\documentclass{article}
\usepackage[a4paper]{geometry}
\usepackage{graphics}

\usepackage{xcolor}
\usepackage[figuresright]{rotating}
\usepackage{amsmath}
\usepackage{amssymb}
\usepackage{booktabs}
\usepackage{subfig}
\usepackage{caption}
\usepackage[colorlinks,citecolor=blue,urlcolor=blue]{hyperref}
\allowdisplaybreaks

\def\bSig\mathbf{\Sigma}

\usepackage[normalem]{ulem}
\usepackage{float}

\usepackage{natbib}

\begin{document}

	{
		\title{\bf An Accelerated Failure Time Regression Model for Illness-Death Data: A Frailty Approach}
		\author{Lea Kats, leatal@mail.tau.ac.il \\
			and \\
			Malka Gorfine, gorfinem@tauex.tau.ac.il \\
			Department of Statistics and Operations Research\\
			Tel Aviv University, Israel}
		\date{}	
		\maketitle
	}

\section*{Abstract}
This work presents a new model and estimation procedure for the illness-death survival data where the hazard functions follow accelerated failure time (AFT) models. A shared frailty variate induces positive dependence among failure times of a subject for handling the unobserved dependency between the non-terminal and the terminal failure times given the observed covariates. Semi-parametric maximum likelihood estimation procedure is developed via a kernel smoothed-aided EM algorithm, and variances are estimated by weighted bootstrap. The model is presented in the context of existing frailty-based illness-death models, emphasizing the contribution of the current work. The breast cancer data of the Rotterdam tumor bank are analyzed using the proposed and existing illness-death models. The results are contrasted and evaluated based on a new graphical goodness-of-fit procedure. Simulation results and data analysis nicely demonstrate the practical utility of the shared frailty variate with the AFT regression model under the illness-death framework. 

\noindent
\textbf{Keywords:} goodness of fit; illness-death model; kernel method; semicompeting risks; shared frailty;

\section{Introduction}\label{s:intro}

The accelerated failure time (AFT) model \citep{kalbfleisch2011statistical} is a well-known alternative to the popular Cox proportional hazard (PH) model (Cox, 1972). The major advantage of AFT over Cox PH is the former being more intuitively interpretable (Wei, 1972; Cox, 1997 among others). This work focuses on the illness-death model which is specified by AFT models for the involved transitions. In illness-death model subjects start at State $0$ (e.g. healthy) and then move to State $2$ (e.g. death) directly or transit first to State $1$ (e.g. the age at diagnosis of the disease under study) and then to State $2$. See Fig.~\ref{fig:fig1}.

\cite{xu2010statistical} suggested an illness-death model with three Cox-based hazard functions. One of their major contributions was the inclusion of a shared gamma-frailty variate, that acts multiplicatively on each of the hazard function,  for incorporating unobserved dependence between the time to disease diagnosis and time to death. \cite{lee2015bayesian} adopted the model of  \cite{xu2010statistical} but replaced their semi-parametric maximum likelihood estimation procedure with a semi-parametric Bayesian estimation approach. \cite{jiang2017semi} developed a class of transformation models for illness-death models that permits a non-parametric specification of the frailty distribution, but to insure identifiabilty,  their model is restricted to parametric transformation and  error distribution. Recently, for a simpler interpretation,  \cite{gorfine2020marginalized} proposed a  frailty-based illness-death model with Cox-type marginalized hazards that also accommodates delayed entry.

Various estimation procedures have been developed for shared-frailty AFT models for clustered data (without competing or semi-competing risks). \cite{pan2001using} considered clustered survival data with an AFT model and a gamma frailty to characterize the unobserved dependence among cluster members. It is assumed that the shared frailty acts multiplicatively on the hazard function of the error term. \cite{zhang2007alternative}, \cite{xu2010like}, \cite{johnson2012smoothing} and \cite{liu2013kernel} adopted Pan's model and provided various estimation methods. All these AFT estimation procedure for clustered data are not directly applicable to our illness-death setting, due to the differences in the likelihood functions, as will be elaborated in Section \ref{ss:estimation_method}.

Unlike Cox-type models with illness-death framework, AFT models are not well developed. The only work which provides an AFT frailty-based model with illness-death setting is of \cite{lee2017accelerated}. \cite{lee2017accelerated}, in contrast to \cite{pan2001using}, used an additive frailty variate in the log-transformed failure time model, and their parametric and semi-parametric estimation methods are based on a Bayesian approach.  Table \ref{tab:methods} summarizes the available frailty-based Cox or AFT models for the illness-death setting.

The current work  fills the gap and provides a gamma-frailty illness-death AFT model where the frailty acts multiplicatively on the hazards of the error terms, in the spirit of \cite{pan2001using}. We extended the estimation approaches of \cite{zeng2007efficient} and \cite{liu2013kernel}, and developed a semi-parametric maximum likelihood estimators (MLEs) based on a kernel-smoothing technique combined with an EM algorithm. Conceptual differences between our model and that of \cite{lee2017accelerated} will be demonstrated in Section \ref{ss:comparison_additive}.

The proposed model and estimation method along with existing methods were applied to the Rotterdam tumor bank of 1,546 breast cancer patients, whom had node-positive disease and underwent a tumor removal surgery between the years 1978--1993. In this example, date at tumor removal surgery is the entrance time to State 0; date at relapse and date at death are the respective entry times to states 1 and 2. Prognostic variables are age at primary surgery, menopausal status, tumor size, tumor grade, number of positive lymph nodes, estrogen and progesterone receptors in the initial biopsy,  hormonal therapy and chemotherapy. For a comparison of the various models, we extended the goodness of fit procedure of \cite{li2021model} to any illness-death model. The results of our proposed goodness of fit visualizing procedure nicely demonstrate the utility of the proposed model and estimation procedure.

The remainder of this article is organized as follows. Section \ref{s:model_methods} describes the proposed gamma frailty-based AFT illness-death regression model, and the estimation method for the regression coefficients, the hazard functions and the parameter of the gamma-frailty distribution. 
The illness-death goodness of fit procedure is provided in Section \ref{sec:visGOF}. In Section \ref{s:simulations} we report the results of extensive simulation study. Section \ref{s:rotterdam} summarises the analyses of the breast cancer data from the Rotterdam tumor bank, while comparing the proposed AFT approach and various existing AFT and Cox models, with and without frailty.  A concluding discussion is provided in Section \ref{s:discussion}. R code for performing the simulations and data analysis in this paper is posted on the following webpage: \url{https://github.com/LeaKats/semicompAFT}.

\section{The Model and Methods}\label{s:model_methods}
\subsection{The Proposed Multiplicative Frailty-Based Model}\label{ss:mutlip}

\begin{figure}
	\centerline{\includegraphics[scale=0.3]{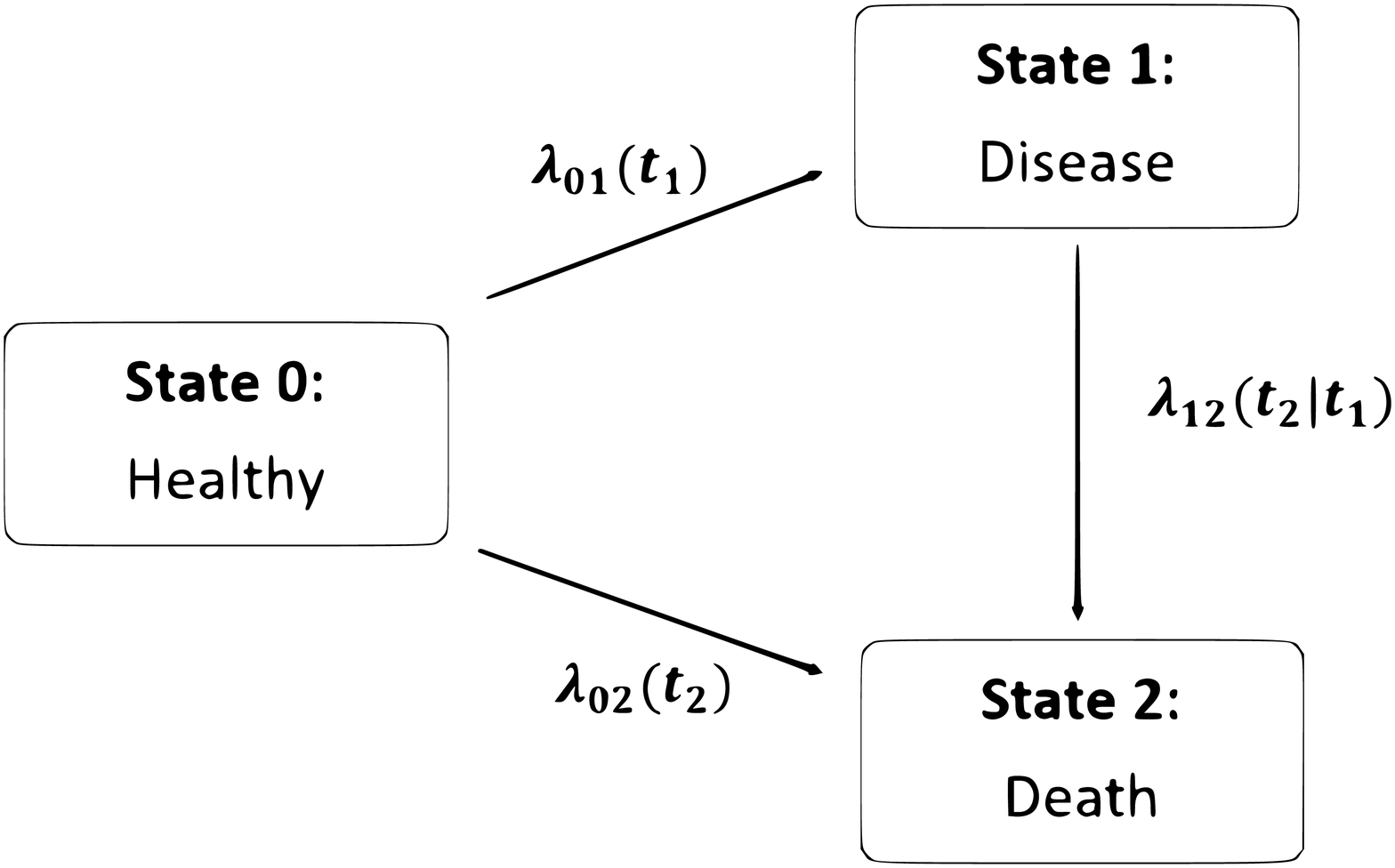}}
	\caption{An illness-death model }
	\label{fig:fig1}
\end{figure}

Assume a sample of $n$ independent observations. Let $T_{1i}$ and $T_{2i}$ be the respective times to the non-terminal and the terminal events of subject $i$, $i=1,\ldots,n$. Let $X_i$ be a time-independent vector of covariates.
The illness-death model (Fig.~\ref{fig:fig1}) is defined by 
\begin{eqnarray}
\log(T_{1i})&=&\beta_{01}^{T} X_{i}+\epsilon_{01i} \, ,  \, T_{1i}>0\label{T1}\\
\log(T_{2i})&=&\beta_{02}^{T} X_{i}+\epsilon_{02i} \, ,  \, T_{2i}>0 \, ,\,\text{given subject $i$ is free of the disease}\label{T2}\\
\log(T_{2i})&=&\beta_{12}^{T} X_{i}+\epsilon_{12i} \, ,  \, T_{2i}>t_{1}>0 \, ,\,\text{given subject $i$ was diagnosed at age $T_{1i}=t_1$}\label{T2|T1}
\end{eqnarray}
where $\beta_{jk}$, $jk \in \{01,02,12\}$, is a vector of regression coefficients of transition $jk$,  and $\epsilon_{jki}$ are random errors with an unspecified distributions. Given that subject $i$ was diagnosed at age $T_{1i}=t_1$, the support of $T_{2i}$ is restricted by $t_1$, so the conditional distribution of $T_{2i}$ is truncated by $t_1$. Model (\ref{T2|T1}) above is not including age at diagnosis, $t_1$, as an additional covariate, but instead, the dependence between $T_{1i}$ and $T_{2i}$ is incorporated via a shared-frailty model. Given the frailty variate of subject $i$, denoted by $\gamma_i$, it is assumed that the respective conditional baseline hazard functions of $\exp(\epsilon_{jki})$, $jk \in \{01,02,12\}$, are given by
\begin{eqnarray}
\lambda^o_{01}(t|\gamma_i) &=& \gamma_i h_{01}^o(t) \, , \, t>0    \label{base_lambda01}\\
\lambda^o_{02}(t|\gamma_i) &=& \gamma_i h_{02}^o(t) \, , \, t>0 \, ,  \text{given subject $i$ is free of the disease} \label{base_lambda02}\\
\lambda^o_{12}(t|t_1,\gamma_i) &=& \gamma_i h_{12}^o(t) \, , \, t>t_1>0 \, , \text{given subject $i$ was diagnosed at age $T_{1i}=t_1$}\label{base_lambda12}
\end{eqnarray}
where each $h^o_{jk}(\cdot)$ is an unspecified baseline hazard function of $\exp(\epsilon_{jk})$ and $\gamma_i$ is an unobservable non-negative random effect, taken to be independent of $X_i$. It is assumed that $\gamma_i$ are gamma distribution with mean 1, unknown variance $\sigma > 0$, and thus with density $f(\gamma;\sigma)=\sigma^{-1/\sigma}\gamma^{1/\sigma-1} e^{-{\gamma}/{\sigma}}/\Gamma({\sigma}^{-1})$. We also assume that $\epsilon_{01i}$, $\epsilon_{02i},$ and $\epsilon_{12i}$ are independent given $(X_i,\gamma_i)$.

Based on Eq.'s~\eqref{T1}--\eqref{base_lambda12}, the conditional hazard functions of the three transitions, given $(X_i,\gamma_i)$,  can be written as
\begin{align}
\hspace{-2mm}\lambda_{0k}(t|X_i, \gamma_i)&=
\lim_{\Delta \to 0}\frac{1}{\Delta}\Pr\left(t\leq T_{ki}<t+\Delta|T_{1i}\geq t, T_{2i}\geq t,X_i,\gamma_i \right)\nonumber\\
&=\gamma_i h^0_{0k}\left(t e^{-\beta^T_{0k}X_i}\right)e^{-\beta^T_{0k}X_i},\quad t>0,\,\,\, k=1,2,\label{lambda0k}
\end{align}
\begin{align}
\lambda_{12}(t|t_1,X_i,\gamma_i)&\hspace{-0.5mm}=\hspace{-1mm}\lim_{\Delta \to 0}\frac{1}{\Delta}\Pr\left(t\leq T_{2i}<t+\Delta|T_{1i}= t_1, T_{2i}\geq t,X_i,\gamma_i \right)\nonumber\\
&\hspace{-0.5mm}=\gamma_i h^0_{12}\left(t e^{-\beta^T_{12}X_i}\right)e^{-\beta^T_{12}X_i},\quad t>t_1>0.\label{lambda12}
\end{align}
For details, see Section S1 of the Web Supplementary Material (WSM).

\subsection{Comparison with the Additive Frailty-Based Model}\label{ss:comparison_additive}
\cite{lee2017accelerated} proposed to directly model the times of the events via the following AFT model specification:
\begin{eqnarray*}
\log(T_{1i})&=&\tilde{\beta}^T_{01} X_i+\tilde{\gamma_i}+\tilde{\epsilon}_{01i} \, ,\, T_{1i}>0\\
\log(T_{2i})&=&\tilde{\beta}^T_{02} X_i+\tilde{\gamma_i}+\tilde{\epsilon}_{02i} \, ,\, T_{2i}>0\, , \,\text{given subject $i$ is free of the disease}\\
\log(T_{2i})&=&\tilde{\beta}^T_{12} X_i+\tilde{\gamma_i}+\tilde{\epsilon}_{12i} \, ,\, T_{2i}>t_{1}>0, \,\text{given subject $i$ was diagnosed at age $T_{1i}=t_1$}
\end{eqnarray*}
where $\tilde{\beta}^T_{jk}$, $jk \in \{01,02,12\}$, are vectors of transition-specific regression coefficients. The errors $\tilde{\epsilon}_{jki}$ are transition-specific random variable with unspecified distributions in the semi-parametric setting, or with a normal distribution in the parametric setting. Also, $\tilde{\gamma_i}$, $i= 1,\ldots,n$, are the unobserved normally distributed frailty variates with mean zero and variance $\theta$, and are assumed to be independent of $X_i$. 

The above additive frailty approach includes the observed covariates $X_i$ and the unobservable $\tilde{\gamma}_i$ in a similar fashion in the models. In contrast, the popular multiplicative frailty approach of \cite{pan2001using} in the context of clustered data and in Section~\ref{ss:mutlip} above, separates the observed covariates $X_i$ and the unobservable component $\gamma_i$. The observed covariates directly affect time to event, and the unobserved affects the hazard functions of the error terms. 

\begin{figure}
	\centering
	\subfloat[]
	[Left: $X_i=0,\mu_{01}=0,\omega_{01}=1$ \quad  Right: $X_i=0,\mu_{01}=-3.5,\omega_{01}=1.2$]
	{{\includegraphics[scale=0.65]{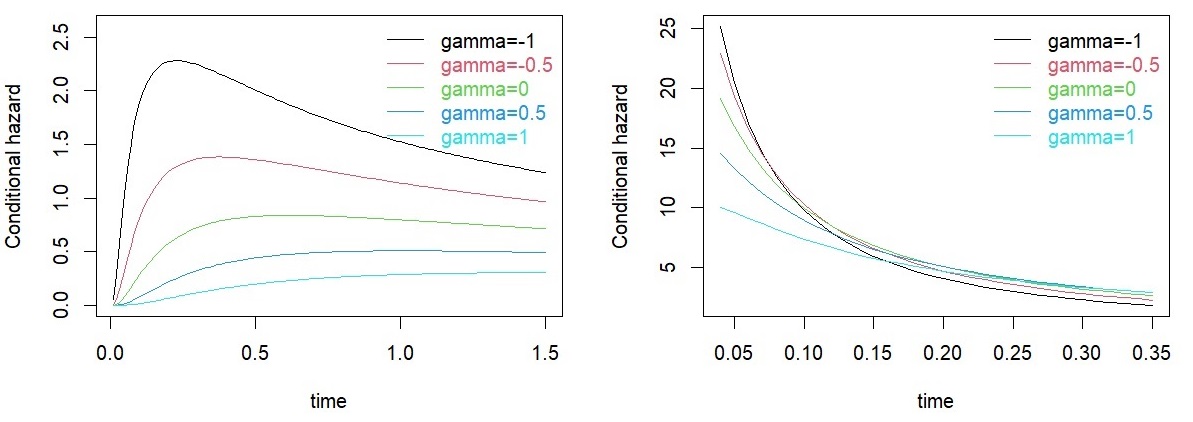} 	}}%
	\qquad
	\subfloat[]
	[$X_i=0,\mu_{01}=-6,\omega_{01}=1.5$]
	{{\includegraphics[scale=0.65]{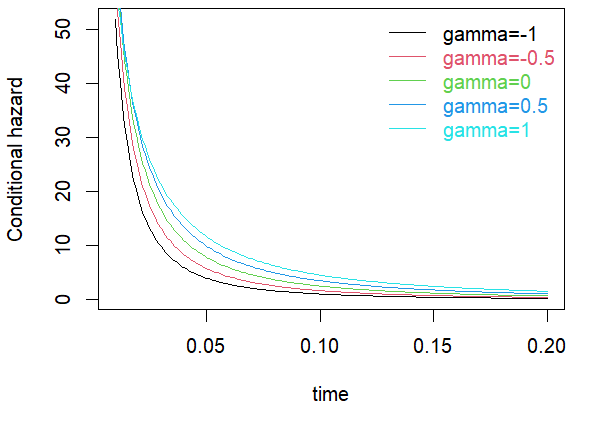}		}}%
	\caption{The additive frailty-based AFT mode of Lee et al. (2017): conditional hazard of transition from healthy state to disease diagnosis.}
\label{fig:lee}
\end{figure}

Assume, for example, that $\exp(\epsilon_{01i})$ and $\tilde{\epsilon}_{01i}$ are normally distributed with  mean $\mu_{01}$ and variance  $\omega^2_{01}$. Then, the respective conditional hazard functions of the multiplicative and additive models are
\begin{equation*}
\lambda_{01}(t|X_i, \gamma_i)
	= \frac{\gamma_i e^{-\beta^T_{01}X_i} }{\omega_{01}}
	{\phi\left(  \frac{t e^{-\beta^T_{01}X_i} -\mu_{01} }{\omega_{01}}   \right)}
\left\{1-\Phi\left(  \frac{t e^{-\beta^T_{01}X_i} -\mu_{01}  }{\omega_{01}}  \right)\right\}^{-1}
\end{equation*}
and
\begin{equation*}\label{additive_cond_hazard_1}
\tilde{\lambda}_{01}(t|X_i,\tilde{\gamma}_i)=\dfrac{1}{\omega_{01} t}\phi\left(\dfrac{\log t-\mu_{01}-\tilde{\beta}^T_1 X_i-\tilde{\gamma}_i}{\omega_{01}}\right)
\left\{1-\Phi\left(\dfrac{\log t-\mu_{01}-\tilde{\beta}^T_1 X_i-\tilde{\gamma}_i}{\omega_{01}}\right)\right\}^{-1}
\end{equation*}
where $\phi(\cdot)$ and $\Phi(\cdot)$ are the density and CDF of the standard normal distribution, respectively. Evidently, $\lambda_{01}$ is of a simpler interpretation than that of $\tilde{\lambda}_{01}$ in terms of the unobserved frailty effect. Fig. \ref{fig:lee} demonstrates $\tilde{\lambda}_{01}(t|X_i,\tilde{\gamma}_i)$ as a function of $t$ for $X_i=0$ and various combinations of $(\tilde{\gamma}_i,\mu_{01},\omega_{01})$. The top-left (bottom) plot of Fig.~\ref{fig:lee} shows that $\tilde{\lambda}_{01}$ decreases (increases) as a function of $\gamma_i$ for any given value of $t$. The top-right plot indicates that  $\tilde{\lambda}_{01}$ could be a non-monotone function of $\gamma_i$ for some values of $t$. In contrast,  ${\lambda}_{01}$ of the multiplicative frailty model, is a monotonic increasing function of the frailty variate $\gamma_i$ for any error distribution 
(see Eq. ~\ref{lambda0k}), and thus the multiplicative-frailty model could be viewed as a model with a simpler interpretation for the unobserved frailty effect.

\subsection{The Proposed Estimation Method}\label{ss:estimation_method}
Out goal is to estimate the unknown set of parameters of the illness death model $\Omega=\{\beta^T_{01},\beta^T_{02},\beta^T_{12},h^o_{01},h^o_{02},h^o_{12},\sigma\}$. 
Let $C_i$ denote the right censoring time of subject $i$, $i=1,\ldots,n$. Then, the observed data consists of $n$ independent observations $\mathcal{O}_i=\left\{V_{i},W_i,\delta_{1i},\delta_{2i},\delta_{3i},{X}_i\right\}$, where $V_{i}=\min(T_{1i},T_{2i},C_i)$, $\delta_{1i}=I(T_{1i}\leq \min(T_{2i},C_i))$, $\delta_{2i}=I(T_{2i}\leq \min(T_{1i},C_i))$, $W_i=\delta_{1i}\min(T_{2i},C_i)$, and $\delta_{3i}=\delta_{1i} I(T_{2i}\leq C_i)$. $V_i$ refers to the first observed time, $\delta_{1i}$ and $\delta_{2i}$ indicate whether the first observed time was age at disease diagnosis $(T_{1i})$, age at death $(T_{2i})$ or age at censoring; $W_i$ is age at death or age at censoring after diagnosis, and  $\delta_{3i}$ indicates whether death was observed after diagnosis. It is assumed that the censoring and the failure times are conditionally independent and non-informative, given $({X}_i,\gamma_i)$, and observations are identically distributed. 

Then, the likelihood function for $\Omega$ is proportional to $L(\Omega) = \prod_{i=1}^n \int L_i(\Omega) d\gamma$
where
\begin{eqnarray*}
L_i(\Omega) &=&\left\{\gamma h^o_{{01}}(V_i e^{-\beta_{01}^{T}X_{i}})e^{-\beta_{01}^{T}X_{i}}\right\}^{\delta_{1i}}
\left\{\gamma h^o_{{02}}(V_i e^{-\beta_{02}^{T}X_{i}})e^{-\beta_{02}^{T}X_{i}}\right\}^{\delta_{2i}}\exp\left\{-\gamma H^o_{01}\left(V_i e^{-\beta_{01}^{T}X_i}\right)\right\}\\
&&\exp\left\{-\gamma H^o_{02}\left(V_i e^{-\beta_{02}^{T}X_i}\right)\right\}\left\{\gamma h^o_{{12}}(W_i e^{-\beta_{12}^{T}X_{i}}) e^{-\beta_{12}^{T}X_{i}} \right\}^{\delta_{3i}}\\
&&\exp\left\{-\delta_{1i}\gamma \left(H^o_{12}(W_i e^{-\beta_{12}^{T}X_{i}})- H^o_{12}(V_i e^{-\beta_{12}^{T}X_{i}})\right)\right\}f(\gamma;\sigma) \, .
\end{eqnarray*}
where $H^o_{jk}(t)=\int_0^t {h}^{o}_{jk}(u)du$, $jk \in \{01,02,12\}$.
For details see Section S2 of WSM. Treating the unobservable frailties as a missing-data problem, calls for the Expectation-Maximization (EM) algorithm for obtaining semi-parametric maximum likelihood estimators \citep{dempster1977maximum}. It can be verified (see S3 of WSM) that the conditional expectation of the complete log-likelihood given the observed data $\mathcal{O}=\{\mathcal{O}_i \, , i=1,\ldots,n\}$ and the parameters' values at the $m$th step, $\widehat{\Omega}^{(m)}$, equals
\begin{equation}\label{eq:estep}
	E\left(l(\sigma)|\mathcal{O},\widehat{\Omega}^{(m)}\right)+E\left(l(\beta_{01},h^o_{01})|\mathcal{O},\widehat{\Omega}^{(m)}\right)
	+E\left(l(\beta_{02},h^o_{02})|\mathcal{O},\widehat{\Omega}^{(m)}\right)+E\left(l(\beta_{12},h^o_{12})|\mathcal{O},\widehat{\Omega}^{(m)}\right)
\end{equation}
where 
\begin{equation}\label{eq:sigma}
E\left(l(\sigma)|\mathcal{O},\widehat{\Omega}^{(m)}\right)=\frac{1}{n}\sum^{n}_{i=1}\left(D_i+\frac{1}{\sigma}\right)\mathcal{E}^{(m)}_{2i}-\frac{1}{n \sigma}\sum^{n}_{i=1}\mathcal{E}^{(m)}_{1i}-\frac{1}{\sigma}\log \sigma -\log\Gamma\left(\frac{1}{\sigma}\right) \, ,
\end{equation}

$$
E\left(l(\beta_{0k},h^o_{0k})|\mathcal{O},\widehat{\Omega}^{(m)}\right)=\frac{1}{n}\sum^{n}_{i=1}\left[\delta_{ki}\left\{\log h^o_{{0k}}(V_i e^{-\beta_{0k}^{T}X_{i}})-\beta_{0k}^{T}X_{i}\right\} -\mathcal{E}^{(m)}_{1i} H^o_{{0k}}(V_i e^{-\beta_{0k}^{T}X_{i}})\right] \, k=1,2 \, ,
$$
\begin{eqnarray*}
		E\left(l(\beta_{12},h^o_{12})|\mathcal{O},\widehat{\Omega}^{(m)}\right)
		&=&\frac{1}{n}\sum^{n}_{i=1}\left[\delta_{3i}\left\{\log h^o_{{12}}(W_i e^{-\beta_{12}^{T}X_{i}})-\beta_{12}^{T}X_{i}\right\} \right.\\
		&&-\delta_{1i}\mathcal{E}^{(m)}_{1i}\left.\left\{H^o_{12}(W_i e^{-\beta_{12}^{T}X_{i}})
		- H^o_{12}(V_i e^{-\beta_{12}^{T}X_{i}})\right\} \right] \, ,
\end{eqnarray*}
$$
	\mathcal{E}^{(m)}_{1i} = E\left( \gamma_i|\mathcal{O},\widehat{\Omega}^{(m)}\right) = \left(D_i+1/\widehat{\sigma}^{(m)} \right) \left\{ 1/\widehat{\sigma}^{(m)} + \widehat{H}^{o(m)} \left(V_i,W_i,\widehat{\beta}^{(m)}\right) \right\}^{-1} \, ,
$$
$$
\mathcal{E}^{(m)}_{2i}=E\left(\log\gamma_i|\mathcal{O},\widehat{\Omega}^{(m)}\right)=\Psi\left(D_i+1/{\widehat{\sigma}^{(m)}}\right)
-\log\left\{1/{\widehat{\sigma}^{(m)}}+\widehat{H}^{o(m)}\left(V_i,W_i, \widehat{\beta}^{(m)}\right) \right\} \, ,
$$
$$
\widehat{H}^{0(m)}(V_i,W_i, \widehat{\beta}^{(m)}) = \sum_{k=1,2} \widehat{H}^{o(m)}_{{0k}}(V_i e^{-\widehat{\beta}_{0k}^{(m)T}X_{i}})+
\delta_{1i}\left\{ \widehat{H}^{o(m)}_{12}(W_i e^{-\widehat{\beta}_{12}^{(m)T}X_{i}})- \widehat{H}^{0(m)}_{12}(V_i e^{-\widehat{\beta}_{12}^{(m)T}X_{i}})\right\} 
$$		
and finally, $D_i=\sum_{k=1}^3\delta_{ji}$, $\Gamma(x)$ is the Gamma function and $\Psi(x)=\Gamma'(x)/\Gamma(x)$ is the digamma function.

The M-step consists of maximization of Eq.~(\ref{eq:estep}). While maximizing $E\left(l(\sigma)|\mathcal{O},\widehat{\Omega}^{(m)}\right)$  as a function of $\sigma$ can be done by gradient-based optimization algorithms, maximizing the other three expectations cannot be done directly with respect to $(\beta_{jk},h^o_{jk})$, $jk\in\{01,02,12\}$,  due to a very non-smooth estimator of the cumulative hazard functions; see \cite{zeng2007efficient} for more details under the standard univariate AFT model. Therefore, we aimed to find a smooth alternative and thus we extended the kernel-smoothing approach of \cite{zeng2007efficient} and \cite{liu2013kernel} to accommodate our semi-competing risks setting.

We start with a simple case of piecewise constant hazard functions
$$
\widetilde{h}^o_{jk} (t) = \sum_{l=1}^{J_{jk,n}}c_{jk,l} I(t_{jk,l-1} \leq t < t_{jk,l}) 
$$
where $0=t_{jk,0}<t_{jk,1}< \cdots <t_{jk,J_{jk,n}}=M_{jk}$, $jk \in \{01,02,12\}$, are equally spaced,  $M_{0k}$ are the respective upper bounds for 
$V_{i}\exp\{-\beta^T_{0k}X_i\}$, $k \in \{1,2\}$, and $M_{12}$ is the upper bound for 
$W_{i}\exp\{-\beta^T_{12}X_i\}$. Then, the cumulative baseline hazard functions are
$$
\widetilde{H}^o_{jk} (t) = \sum_{l=1}^{J_{jk,n}}c_{jk,l}(t-t_{jk,l-1}) I(t_{jk,l-1} \leq t < t_{jk,l}) + (M_{jk}/J_{jk,n})\sum_{l=1}^{J_{jk,n}}c_{jk,l} I(t \geq t_{jk,l}) \, . 
$$
The functions $\widetilde{h}^o_{jk}$ and $\widetilde{H}^o_{jk}$ are plugged in $E\left(l(\beta_{jk},h^o_{jk})|\mathcal{O},\widehat{\Omega}^{(m)}\right)$
and by maximizing the resulting expression with respect to $c_{jk,l}$, $l=1,\ldots,J_{jk,n}$, $jk\in\{01,02,12\}$, for a given $\beta_{jk}$, we left with a closed-form estimator of $c_{jk,l}$, $\widehat{c}^{(m)}_{jk,l}$. Plugging $\widehat{c}^{(m)}_{jk,l}$ in $E\left(l(\beta_{jk},h^o_{jk})|\mathcal{O},\widehat{\Omega}^{(m)}\right)$ provides an approximated profile-likelihood function of $\beta_{jk}$. However, even these profile-likelihood functions are not smooth and have multiple local maxima, and thus additional smoothing step is required. To this end, it can be shown that each of the profile likelihood converges to limiting function of $\beta_{jk}$, $jk \in \{01,02,12\}$, as $n\to\infty$, $J_{jk,n} \to \infty$ and $J_{jk,n}/n \to 0$. Then, for a given kernel function $K$ with bandwidths $a_{jk,n}$, the estimators of $\beta_{jk}$, $jk\in\{01,02,12\}$, are obtained by maximizing a smoothed approximation of the limiting function. In the illness-death setting, the age of death after disease diagnosis is truncated by the age at diagnosis. Therefore, the kernel-smoothing approach is adopted to accommodate left truncation. Given the estimators of the regression coefficients, the proposed smoothing approach also yields estimators of the baseline hazard functions.  Details of the above summary can be found in Section S4 of the WSM. Here we provide the resulting estimation procedure. 

Define $R^V_i(\beta)=\log V_i -\beta^T X_{i}$, $R^W_i(\beta)=\log W_j -\beta^T X_{i}$. Then, $\beta_{0k}$ is estimated by maximization of $l^{s}_{0k}(\beta_{0k})$, $k=1,2$, where
\begin{eqnarray}\label{l_s_beta_01_02}
l^{s}_{0k}(\beta_{0k})&=&-\frac{1}{n}\sum^{n}_{i=1}\delta_{ki}\log V_i+\frac{1}{n}\sum^{n}_{i=1}\delta_{ki}\log \left\{\frac{1}{n a_{0k,n}}\sum^{n}_{j=1}\delta_{kj}K \left(\frac{R^V_j(\beta_{0k})-R^V_i(\beta_{0k})}{a_{0k,n}}\right)\right\} \nonumber \\
&&-\frac{1}{n}\sum^{n}_{i=1}\delta_{ki}\log\left\{ \frac{1}{n}\sum^{n}_{j=1}\mathcal{E}^{(m)}_{1,j}
\int_{-\infty}^{\{ R^V_j(\beta_{0k})-R^V_i(\beta_{0k})\}/a_{0k}}K(s)ds\right\} \, ,
\end{eqnarray}
$\beta_{12}$ is estimated by maximization of 
\begin{eqnarray}\label{l_s_beta_12}
l^{s}_{12}(\beta_{12})&=&-\frac{1}{n}\sum^{n}_{i=1}\delta_{3i}\log W_i +
\frac{1}{n}\sum^{n}_{i=1}\delta_{3i}\log\left\{\frac{1}{n a_{12,n} }\sum^{n}_{j=1}\delta_{3j}K\left(\dfrac{R^W_j(\beta_{12})-R^W_i(\beta_{12})}{a_{12,n}} \right)\right\} \nonumber \\
&&-\frac{1}{n}\sum^{n}_{i=1}\delta_{3i}\log\left\{\frac{1}{n }\sum^{n}_{j=1}\mathcal{E}^{(m)}_{1j}\delta_{1j}\int_{\{R^V_j(\beta_{12})-R^W_i(\beta_{12})\}/a_{12}}^{\{R^W_j(\beta_{12})-R^W_i(\beta_{12})\}/a_{12}}
K\left(s\right)ds\right\} \,
\end{eqnarray}
and given $\widehat{\beta}^{(m)}_{jk}$, $jk \in \{01,02,12\}$, the baseline hazard functions are estimated by
\begin{equation}\label{eq:h0_1_2}
\widehat{h}^{o(m)}_{0k}(t)=\frac{(n a_{0k,n} t)^{-1}\sum^{n}_{i=1} \delta_{ki}K\left(\{R^V_i(\widehat{\beta}^{(m)}_{0k})-\log t\}/
	{a_{0k,n}}\right)}{n^{-1}\sum^{n}_{i=1}\mathcal{E}^{(m)}_{1i}\int^{\{R^V_i(\widehat{\beta}^{(m)}_{0k})-\log t\}/a_{0k,n}}_{-\infty}K(s)ds} \,\,\, 
k=1,2 \, ,
\end{equation}
\begin{equation}\label{eq:h0_12}
\widehat{h}^{o(m)}_{12}(t)=\frac{(n a_{12,n} t)^{-1}
\sum^{n}_{i=1}\delta_{3i}K\left(\{R^W_i(\widehat{\beta}^{(m)}_{12})-\log t\}/{a_{12,n}}\right)}
{n^{-1}\sum^{n}_{i=1}\mathcal{E}^{(m)}_{1i}\delta_{1i}\int_{\{R^V_j(\widehat{\beta}^{(m)}_{12})-\log t\}/{a_{12,n}}}^{\{R^W_i(\widehat{\beta}_{12})-\log t \}/{a_{12,n}}}K(s)ds} \, ,
\end{equation}
and $\widehat{H}^{o(m)}_{jk}(t) = \int_0^t \widehat{h}^{o(m)}_{12}(s)ds$. The following is a summary of our EM-based estimation algorithm:
\begin{description}
\item [{\it Step 0 (Initial values):}] Set $\mathcal{E}^{(0)}_{1i} = 1$, $i=1,\ldots,n$ and $\beta^{(0)}_{12}=0$. Get  $\widehat{\beta}^{(0)}_{0k}$, $k\in\{1,2\}$ by the rank-based estimator of the R package \texttt{aftgee} \citep{aftgee}. Get $\widehat{h}^{o(0)}_{jk}$ using Eq.'s (\ref{eq:h0_1_2})--(\ref{eq:h0_12}) and $\widehat{H}^{o(0)}_{jk}(t)$ by adaptive quadrature (e.g. \texttt{integrate} function of R), $jk \in \{01,02,21\}$. Finally, set  $\widehat{\sigma}^{(0)}$ between 2 to 5 and $m=0$.
\item[{\it Step 1 (E-step):}] Set $m=m+1$ and get $\mathcal{E}^{(m)}_{1i}$ and $\mathcal{E}^{(m)}_{2i}$, $i=1,\ldots,n$.
\item[{\it Step 2 (M-step):}] Obtain $\widehat{\sigma}^{(m)}$ by maximizing  Eq. (\ref{eq:sigma}), and get  $\widehat{\beta}^{(m)}_{jk}$ by maximizing Eq.'s \eqref{l_s_beta_01_02}--\eqref{l_s_beta_12}. Obtain $\widehat{h}^{0(m)}_{jk}(t)$ and $\widehat{H}^{0(m)}_{jk}(t)$, $\,jk\in\{01,02,12\}$. 
\item[{\it Step 3:}] Repeat Steps 1-2 until convergence is reached.
\end{description}

The asymptotic results of \citet[Theorem 1]{liu2013kernel} for clustered data can be extended to establish the asymptotic properties of the proposed illness-death model estimators. Thus, it can be shown that under some regularity conditions, as $n a^2_{jk,n}\to \infty$, $n a^4_{jk,n}\to 0$ and $n \to \infty$, $\sup_{t \in [0,\tau]}|\widehat{H}^o_{jk}(t)-H^o_{jk}(t)| \rightarrow 0$, $\widehat{\beta}_{jk} \rightarrow \beta_{jk}$ almost surely, and $n^{-1/2}(\widehat{\beta}_{jk}-\beta_{jk})$, $jk\in\{01,02,12\}$, converges to a mean-zero multivariate normal distribution.

\subsubsection{Bandwidth Selection}\label{ss;band_slect}
For the bandwidth parameters, $a_{jk,n}$, we recommend a modified version of the optimal bandwidths of \cite{jones1990performance} and \cite{jones1991using}, in the spirit of \cite{liu2013kernel}. The smoothed profile-likelihood function involves the kernel density for uncensored subjects and the cumulative kernel for all subjects. The recommended bandwidth for the kernel density is $\zeta \widehat\tau_{jk}(8\sqrt{2}/3)^{1/5} n^{-1/5}_{jk}$, $jk\in\{01,02,12\}$, where $\widehat\tau_{jk}$ is the sample standard deviation of $\log(T_{jki})$ among the subjects with observed event time of transition $j \rightarrow k$, denoted by $n_{jk}$. The recommended bandwidth for the cumulative kernel is $\zeta \widehat\upsilon_{jk}4^{1/3} n^{-1/3}_{jk}$, where $\widehat\upsilon_{j}$ is the sample standard deviation of  $\log(V_{i})$ or $\log(W_{i})$ among all subjects in state $j$, denoted by $n_j$, $j=0,1$. Based on our extensive simulation study, $\zeta=0.5$ is recommended for $\beta_{jk}$, and $\zeta=0.01$ for $h^o_{jk}$.

\subsubsection{Variance Estimation}\label{ss;variance_est}
In highly censored data, standard bootstrap could produce a low number of distinct event times, which often causes convergence failure. Alternatively, the weighted bootstrap approach can be used \citep{kosorok2004robust}. At each bootstrap step, a random weight from the standard exponential distribution is assigned for each observation, and the estimators are derived by the respective weighted log-likelihood functions or weighted sums (see Section S5 of the WSM for details).

\subsubsection{The Proposed Model and Estimation Method Without Frailty}\label{ss:model_estimation_without_frailty}
The proposed estimation method can also be implemented under the model described by Eq.'s \eqref{T1}--\eqref{base_lambda12} where $\gamma_i \equiv 1$ for all $i=1,\ldots,n$, i.e., with no frailty effect. Namely, AFT models with competing risks, while model \eqref{T2|T1} is an AFT model adjusted for left truncated data, where age at death after disease diagnosis is truncated by the age at diagnosis. If one believes that the functional form of age at diagnosis that effects the time to death after the diagnosis is known, and that frailty is not required, then age at diagnosis should be added as one of the covariates in model \eqref{T2|T1}. In this case, the estimation process is simplified to one step for estimating the vectors of the regression coefficients and to one additional step for the baseline hazard functions' estimators, without the need of iterative process (see Section S6 of the WSM).

\section{Visualizing Goodness of Fit}\label{sec:visGOF}
A goodness-of-fit procedure  aimed to evaluate how closely observed data mirrors expected data under the assumed model. Recently, \cite{li2021model} proposed a goodness of fit (GOF) method for an arbitrary univariate survival model with right censored data which is based on the randomized survival probabilities (RSP). Their key idea is to replace the survival probability of a censored failure time with a uniform random number between 0 and the survival probability at the censored time. They showed that RSPs always uniformly distributed on $(0,1)$ under the true model. Then, graphical methods for comparing the distribution of the RSPs with the standard uniform distribution could be used for detecting a lack of model fit. In contrast,  the distributions of well-known residuals (e.g. Cox-Snell) under the true model are complicated due to censoring and cannot be characterized clearly with a known distribution since their distribution depends on the  censoring distribution. Hence, there is a lack of reference distributions for conducting GOF procedure, and the most widely used diagnostic tool is to apply the Kaplan-Meier estimator on the residual; see \cite{li2021model} and references therein. Here, we extend the RSP approach to the illness-death model.

In our setting of frailty-based AFT illness-death model, the marginal survival functions should be used since the frailties are unobserved. The illness-death model will be examined by two sets of RSPs: (i) the probability of remaining at Stat 0; and (ii) the probability of remaining at State 1 among those who diagnosed with the disease. In particular (for a detailed derivation, see Section S9 of the WSM), 
$$
S^{M}_{0.}(t|X_i) = \Pr(T_{1i}>t, T_{2i}>t|X_i)=\left[1+\sigma  \sum_{j=1,2} H^o_{0j}(t e^{-\beta^T_{0j}X_i})\right]^{-1/\sigma}
$$
and for $t > t_1$, 
\begin{eqnarray*}
	S^M_{12}(t|t_1,X_i)&=&\Pr(T_{2i}>t|T_{1i}=t_1,T_{2i}>t_1,X_i) \\
&=&\left(\frac{1+ \sigma  \sum_{j=1,2} H^o_{0j}(t_1e^{-\beta^T_{0j}X_i})}{1+
\sigma \{ \sum_{j=1,2}  H^o_{0j}(t_1e^{-\beta^T_{0j}X_i}) +
  H^o_{12}(te^{-\beta^T_{12}X_i}) - H^o_{12}(t_1e^{-\beta^T_{12}X_i})
\}
}
\right)^{1/\sigma+1} \, ,
\end{eqnarray*}
where the superscript $M$ denotes the marginal distribution with respect to the frailty variate.
Clearly,  in the absence of frailty (i.e. $\gamma_i \equiv 1$ for all $i=1,\ldots,n$) we get
$$
S_{0.}^M(t|X_i)=\exp\{-H^o_{01}(t e^{-\beta^T_{01}X_i})-H^o_{02}(te^{-\beta^T_{02}X_i})\}
$$
and
$$
		S_{12}^{M}(t|t_1,X_i)
		=\exp\{- H^o_{12}(t e^{-\beta_{12}^{T}X_i})+ H^o_{12}(t_1 e^{-\beta_{12}^{T}X_i})\} \, , \,\,\, t>t_1 \, .
$$

Now we are in a position to define the RSPs:
\begin{equation*}\label{aft_RSP0k}
		S^M_{0.}(V_i,\delta_{1i},\delta_{2i},U_{1i})
		=(\delta_{1i}+\delta_{2i}) S^{M}_{0.}(V_i|X_i)+(1-\delta_{1i}-\delta_{2i})U_{1i} S^{M}_{0.}(V_i|X_i),
\end{equation*}
and
\begin{equation*}\label{aft_RSP12}
		S^{M}_{12}(W_i,V_i,\delta_{3i},U_{2i})
		= \delta_{3i} {S}^{M}_{12}(W_i|V_i,X_i)+(1-\delta_{3i})U_{2i} S^{M}_{12}(W_i|V_i,X_i),
\end{equation*}
where $U_{1i}$ and $U_{2i}$ are independent random samples from the standard uniform distribution $U(0,1)$. Based on \cite{li2021model} it can be shown that if the censoring and failure times are independent, given the observed covariates and the frailty variate, $S_{12}^{M}(t|t_1,X_i)$ and 
$S^M_{0.}(V_i,\delta_{1i},\delta_{2i},U_{1i})$ are  uniformly distributed on $(0,1)$. See the WSM Section S9 for a detailed proof.

Finally, a visualized GOF procedure for any illness-death model could be accomplished by comparing the histograms of $\widehat{S}^M_{0.}(V_i,\delta_{1i},\delta_{2i},U_{1i})$ and of $\widehat{S}^M_{12}(W_i,V_i,\delta_{3i},U_{2i})$ with the expected values under the standard uniform distribution. This procedure will be demonstrated in Section 5.


\section{Simulation Study}
\label{s:simulations}

\subsection{Simulation Setup}
To demonstrate the finite sample properties of the proposed estimation method, we conducted an extensive simulation study. Failure times were generated from models \eqref{T1}-\eqref{T2|T1} with 
$X_i=(X_{1i},X_{2i},X_{3i},X_{4i})^T$,
$\beta_{01}=(1,0.5,0,0)^T$, $\beta_{02}=(0,1,1,0)^T$ and $\beta_{12}=(0.5,0.5,0,1)^T$,
and a sample size of $n=1000$. The baseline hazard functions of $\exp(\epsilon_{jki})$, $jk \in \{01,02,12\}$, were $h^0_{01}(t)=2t,$
$h^0_{02}(t)=3t$, $h^0_{12}(t)=2t$ and $X_{1i},X_{2i},X_{3i},X_{4i}$ were sampled independently such that 
$X_{1i},X_{3i},X_{4i}\sim\text{Uniform(-1,1)}$ and $X_{2i}\sim$ Bernoulli(0.5). Frailty variates $\gamma_i$ were generated from gamma distribution with
various dependence magnitudes $\sigma=0.5,1$ and 2. Failure times, $T_1$ and $T_2$, were generated by solving  $U=\exp\{-\gamma H^0_{0k}(T e^{-\beta^T_{0k}X})\}$, $k\in\{1,2\}$, for $T$, where $U$ is uniformly distributed over $(0,1)$. For those who diagnosed with the disease (i.e. $T_1<T_2$)  new values of $T_2$ were generated from the appropriate left-truncated distribution, by solving  $U=\exp[-\gamma \{H^0_{12}(T e^{-\beta^T_{12}X})-H^0_{12}(T_1 e^{\beta^T_{12}X})\}]$ for $T$ and a new random $U$. Censoring times were sampled from $U(0,15)$, such that for $\sigma=2$ about 16\% of observations were censored prior  to disease diagnosis or death; and among those who diagnosed, about 13\% were censored before death. Under $\sigma=1$ the corresponding censoring rates were 9\% and 10\%, and for $\sigma=0.5$, 7\% and 8\%. The analysis in this work uses the Gaussian kernel with bandwidths values according to Section \ref{ss;band_slect}. A range of values for $\zeta$ were studied, see Section S8 of the WSM. We set $\widehat{\sigma}^{[0]}=2$. Finally, the convergence criteria were
$\max_{1\leq q\leq p_{jk}}\big|\widehat\beta^{(m+1)}_{jkq}-\widehat\beta^{(m)}_{jkq}\big|<0.00001$, $\frac{1}{n_{jk}}\sum_{i=1}^{n_{jk}}\big|\widehat{H}^{0(m+1)}_{jk}(\tilde{t}_{jki})-\widehat{H}^{0(m)}_{jk}(\tilde{t}_{jki})\big|<0.0001$, 
$jk\in\{01,02,12\}$, and
 $\big|\widehat{\sigma}^{(m+1)}-\widehat{\sigma}^{(m)}\big|<0.0001$,
where $p_{jk}$ is the number of components in $\beta_{jk}$, $\beta_{jkq}$ denotes the $q$th component of $\beta_{jk}$, $n_{jk}$ is the number of subjects relevant to transition $j\to k$, $\tilde{t}_{jki}=V_i e^{-\beta_{jk}^{T}X_{i}}$, $jk\in\{01,02\}$ and $\tilde{t}_{12i}=W_i e^{-\beta_{12}^{T}X_{i}}$.

\subsection{Simulation Results}
Tables \ref{tab3}--\ref{tab4} present the performance of the proposed estimation method with frailty. The tables show the empirical mean, empirical standard deviations (SDs), estimated standard errors (SEs) and the empirical coverage rate of 95\% Wald-type confidence interval (CR) of the dependence parameter, the regression coefficients and the cumulative baseline hazard functions at selected time points. Results are based on 100 repetitions. Tables \ref{tab3}--\ref{tab4} indicate that the proposed approach performs well in terms of bias and coverage rates.  Moreover, the empirical standard deviations and the estimated standard errors are reasonably close. 

Tables \ref{tab5}--\ref{tab6} present simulation results of model misspecification. The data were generated with gamma frailty effect but we applied our estimation procedure without frailty effect provided in Section \ref{ss:model_estimation_without_frailty}. Clearly, under high dependence, ignoring the frailty, yield biased estimates and poor coverage rates. For example, with $\sigma=2$ and true coefficients  $\beta_{01}=(1,0.5)^T, \beta_{02}=(1,1)^T$ and $\beta_{12}=(0.5,0.5,1)^T$, the respective mean estimates were $\widehat{\beta}_{01}=(1.27,0.36)^T, \widehat{\beta}_{02}=(1.15,1.27)^T$ and $\widehat{\beta}_{12}=(0.25,0.41,2.09)^T$ with terribly poor coverage rates. In Section S7 and Table S1 of the WSM it is demonstrated that when the true model is free of frailty, our proposed approach of Section \ref{ss:model_estimation_without_frailty} performs very well in term of bias and coverage rate.

\section{Rotterdam Tumor Bank Data}
\label{s:rotterdam}

\subsection{Data and Models}
We analyzed the Rotterdam tumor bank of 1,546 breast cancer patients, whom had node-positive disease and underwent a tumor removal surgery between the years 1978-1993; the dataset is available in the \textbf{survival} R package \citep{survival-package}.   $T_1$ is the time from surgery to relapse and $T_2$ is the time from surgery or relapse to death. Of the 1,546 patients, 924 patients showed a relapse of the disease (63\%), 106 died without a relapse evidence (7\%) and 771 patients died after a relapse (79\% of the patients who showed a relapse of the cancer). Prognostic variables are age at primary surgery (in years), menopausal status (0 = premenopausal, 1 = postmenopausal), tumor size ($\leq 20$ mm, $20 - 50$ mm and $>50$ mm), tumor grade (2 = moderately differentiated, 3 = poorly differentiated), number of positive lymph nodes, estrogen and progesterone receptors in the initial biopsy (fmol/l),  and chemotherapy (0 = no, 1 = yes). 

The following methods were applied: the proposed AFT model and estimation procedure with gamma frailty (SEs are based on 500 bootstrap samples and starting value of $\sigma$ was set to 1),  the gamma-frailty Cox model of \cite{lee2015bayesian}, and the mariginalized gamma-frailty of \cite{gorfine2020marginalized}. The AFT additive-frailty model of \cite{lee2017accelerated} is implemented in the \texttt{R} package \texttt{SemicompRisks} \citep{alvares2019semicomprisks} with sojourn time, $T_2-T_1$, when death occurs after the disease. Applying it to the current data, resulted in convergence failure (the potential scale reduction factors that should be less than 1.05 are much larger for most of the parameters). We hypothesize that the convergence failure is due to the use of sojourn time which could be negatively correlated with time from surgery to relapse, while a gamma frailty model assumes a positive dependence.  
Additionally, the data were analyzed with the proposed AFT model without frailty and a Cox illness-death model (for details, see section S11 in the WSM). 


\subsection{Results}
Table \ref{tab7} presents the estimates of the frailty-based methods with Holm adjusted p-values correction for multiplicity \cite{holm1979simple}. Hereafter, a result is considered significant based on the adjusted p-value and a significance level of 0.05. 
The proposed model suggests that when holding the other covariates constant, higher age, higher progesterone, having chemoteraphy, and having hormone therapy, each goes with a longer time to relapse after surgery. Also, higher number of positive lymph nodes, larger tumor size and poorly differentiated tumor are related to earlier relapse after the surgery. Hormonal treatment and chemotherapy after the surgery are associated with longer time to relapse after surgery. 

In transition from surgery to death,  higher age and higher number of positive lymph nodes are related with earlier death after surgery. For the transition from relapse to death, the proposed AFT model implies that lower number of positive lymph nodes, higher progesterone level associate with longer post-surgery survival time, given that the patient experienced relapse. Finally, the proposed AFT model indicates a strong dependence between time to relapse and time to death. The marginalized Cox model also indicates high level dependence between relapse and death times,  while the conditional Cox model shows somewhat lower dependence. The directions of covariates' effect under the Cox models are similar to each other and to those of the proposed AFT model, but inference results based on these three models somewhat differ. 

\begin{figure}
	\centering
	\subfloat[][\centering  The proposed frailty-based AFT model - $\widehat{S}_{0.}^M$]{\includegraphics[scale=0.36]{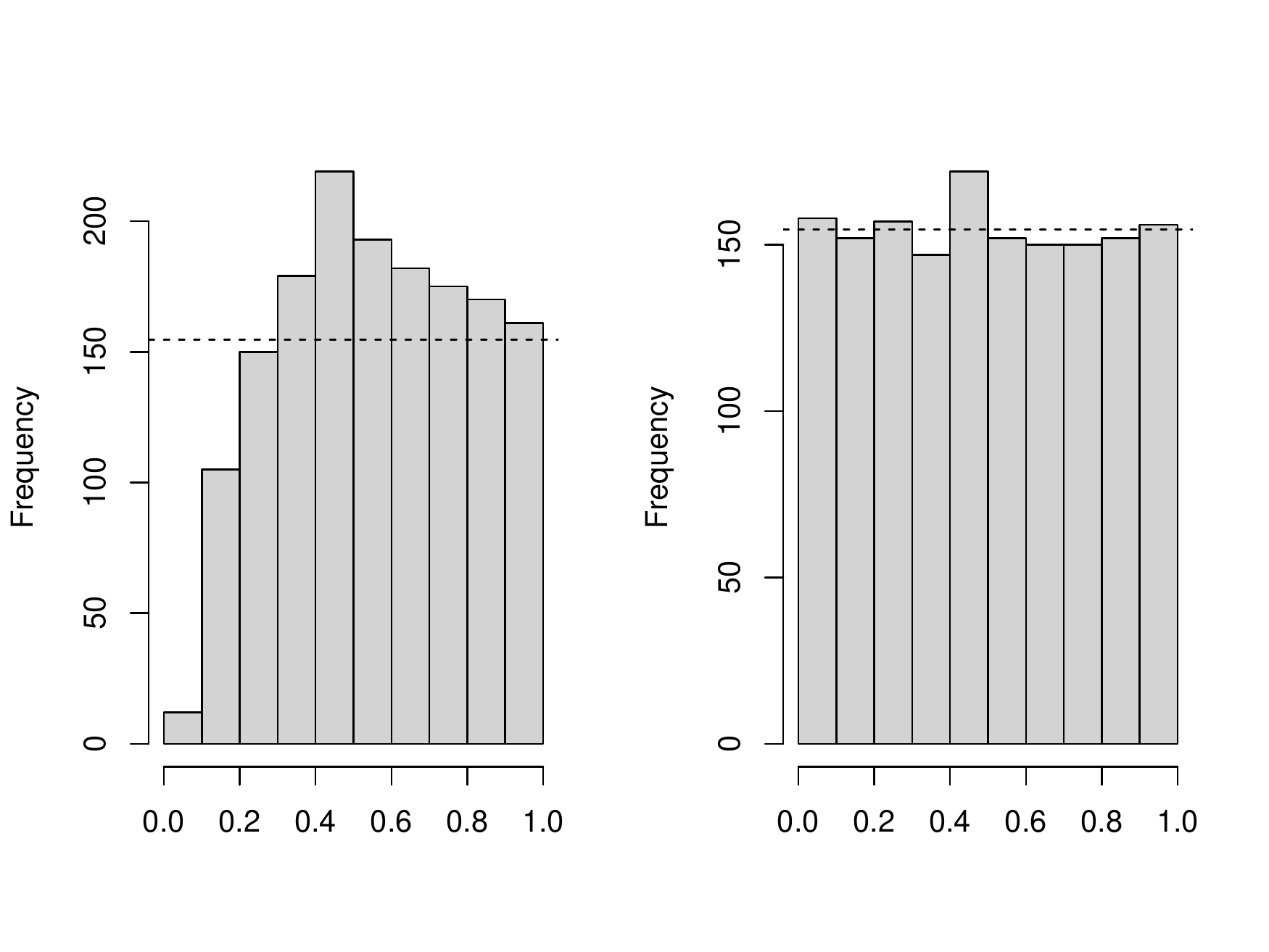}}%
	\qquad
	\subfloat[][\centering The proposed frailty-based AFT model - $\widehat{S}_{12}^M$]{{\includegraphics[scale=0.36]{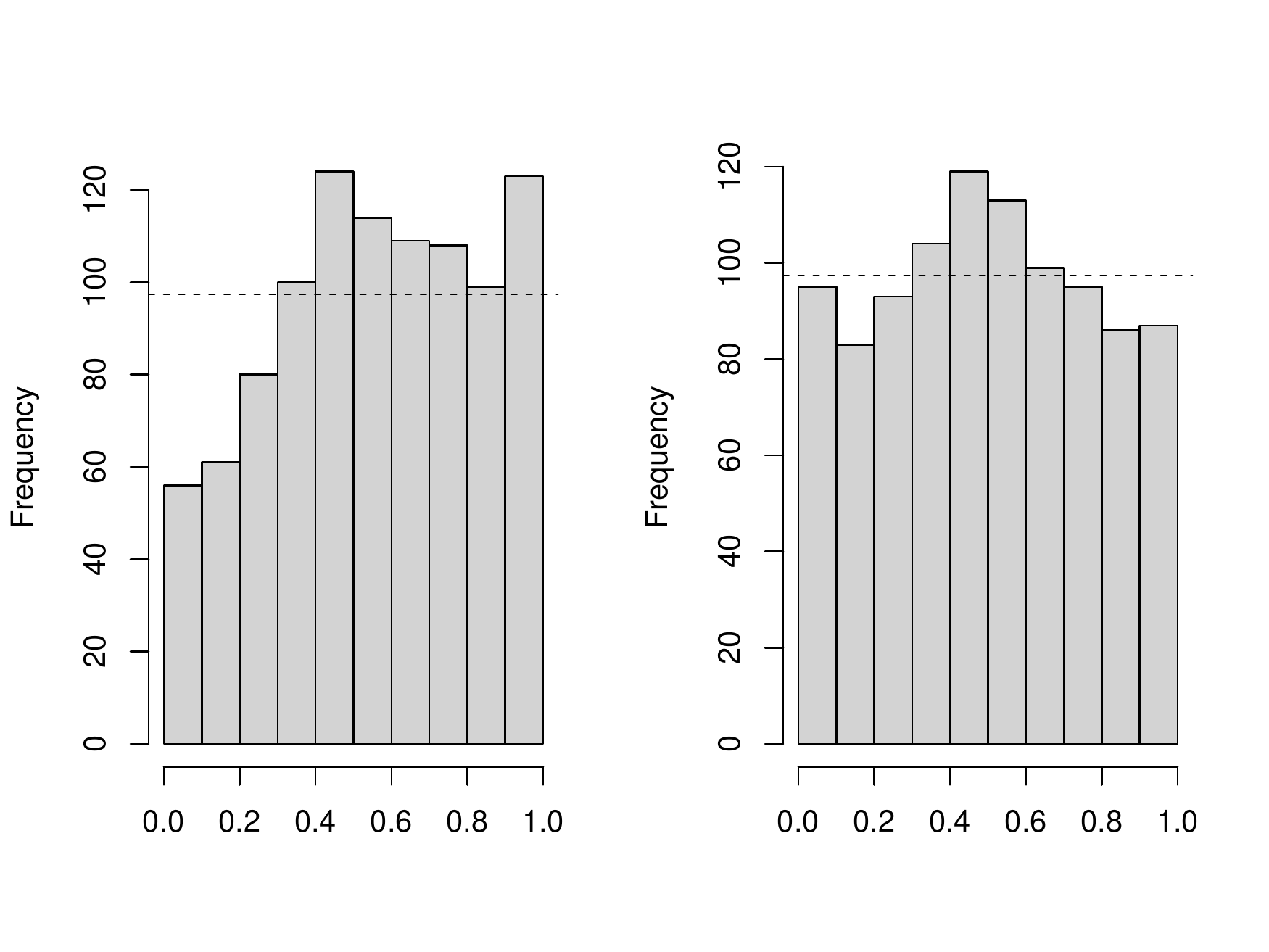} }}%
	\qquad
		\subfloat[][\centering Marginalized Cox model - $\widehat{S}_{0.}^M$]{{\includegraphics[scale=0.36]{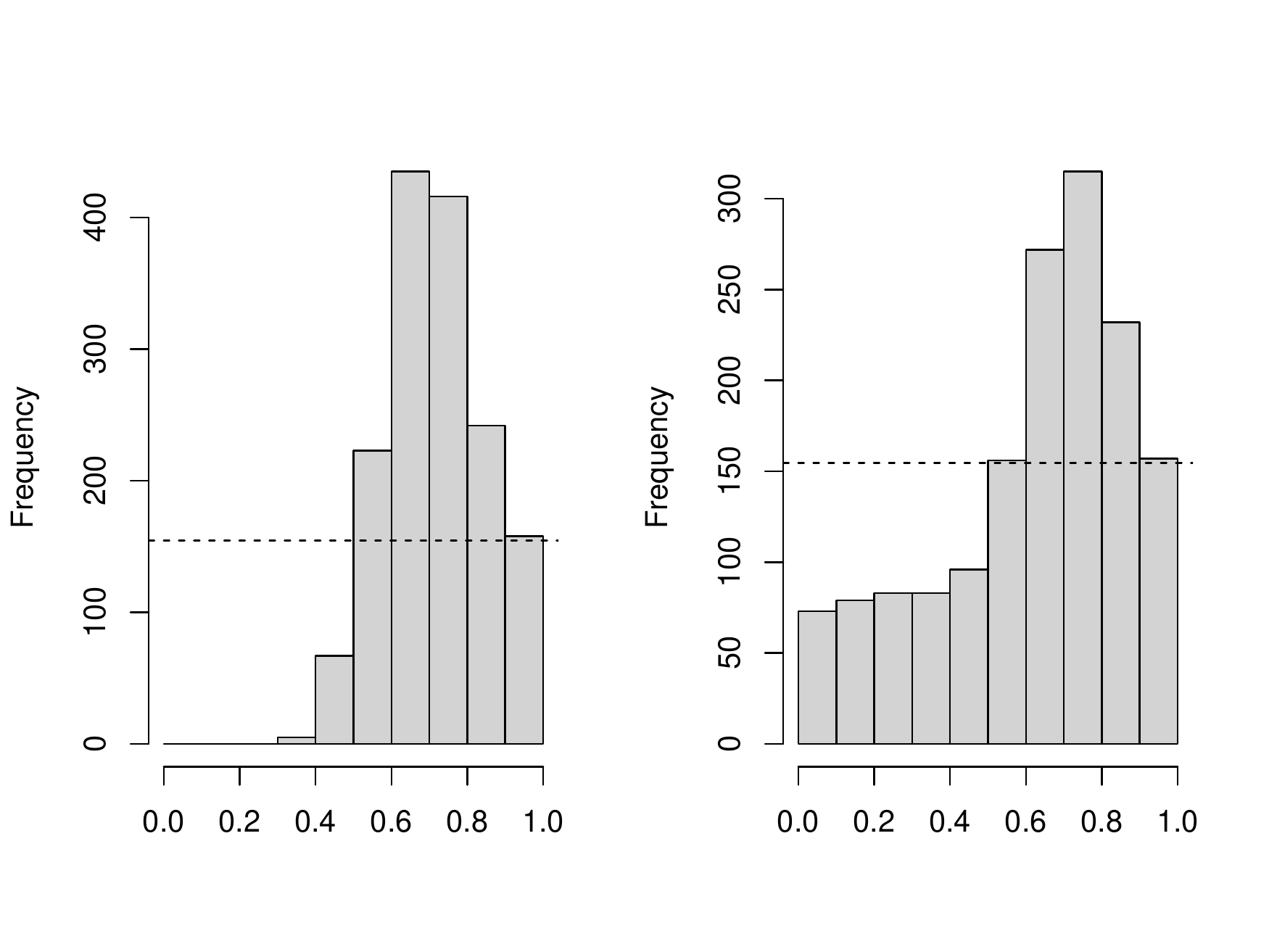} }}%
	\qquad
	\subfloat[][\centering  Marginalized Cox model - $\widehat{S}_{12}^M$]{{\includegraphics[scale=0.36]{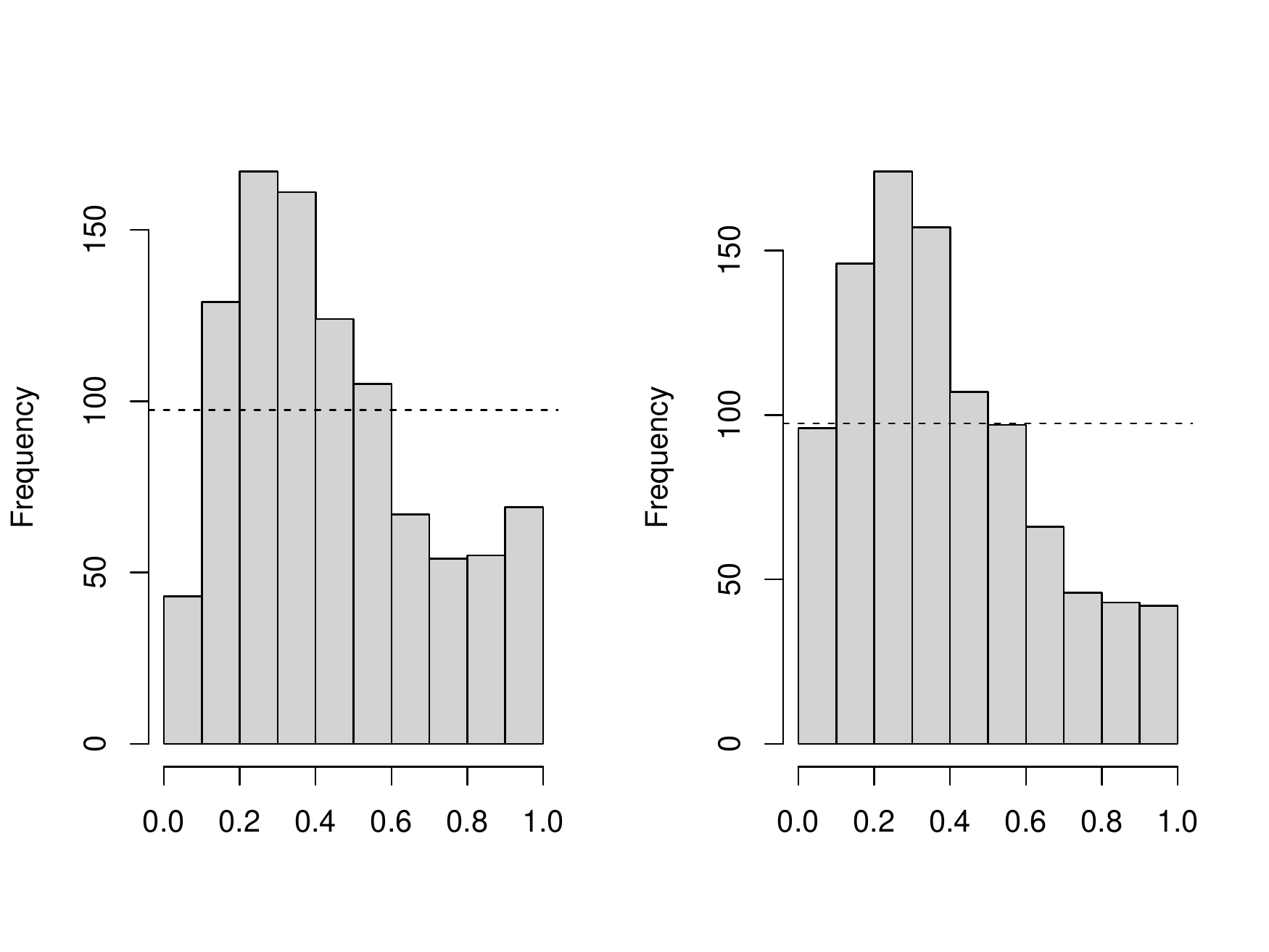} }}%
	\qquad
	\subfloat[][\centering Conditional Cox model - $\widehat{S}_{0.}^M$]{{\includegraphics[scale=0.36]{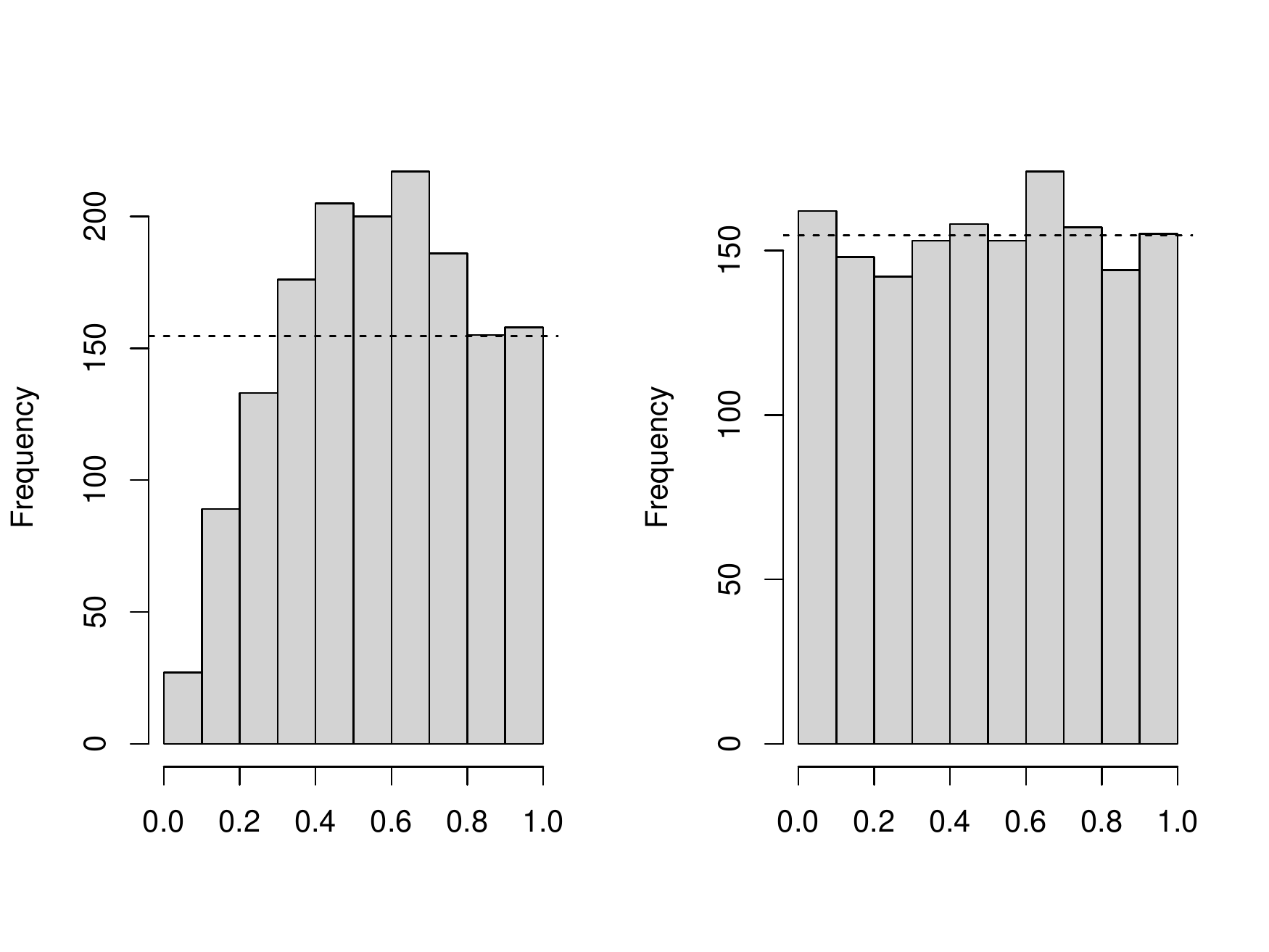} }}%
	\qquad
	\subfloat[][\centering Conditional Cox model - $\widehat{S}_{12}^M$]{{\includegraphics[scale=0.36]{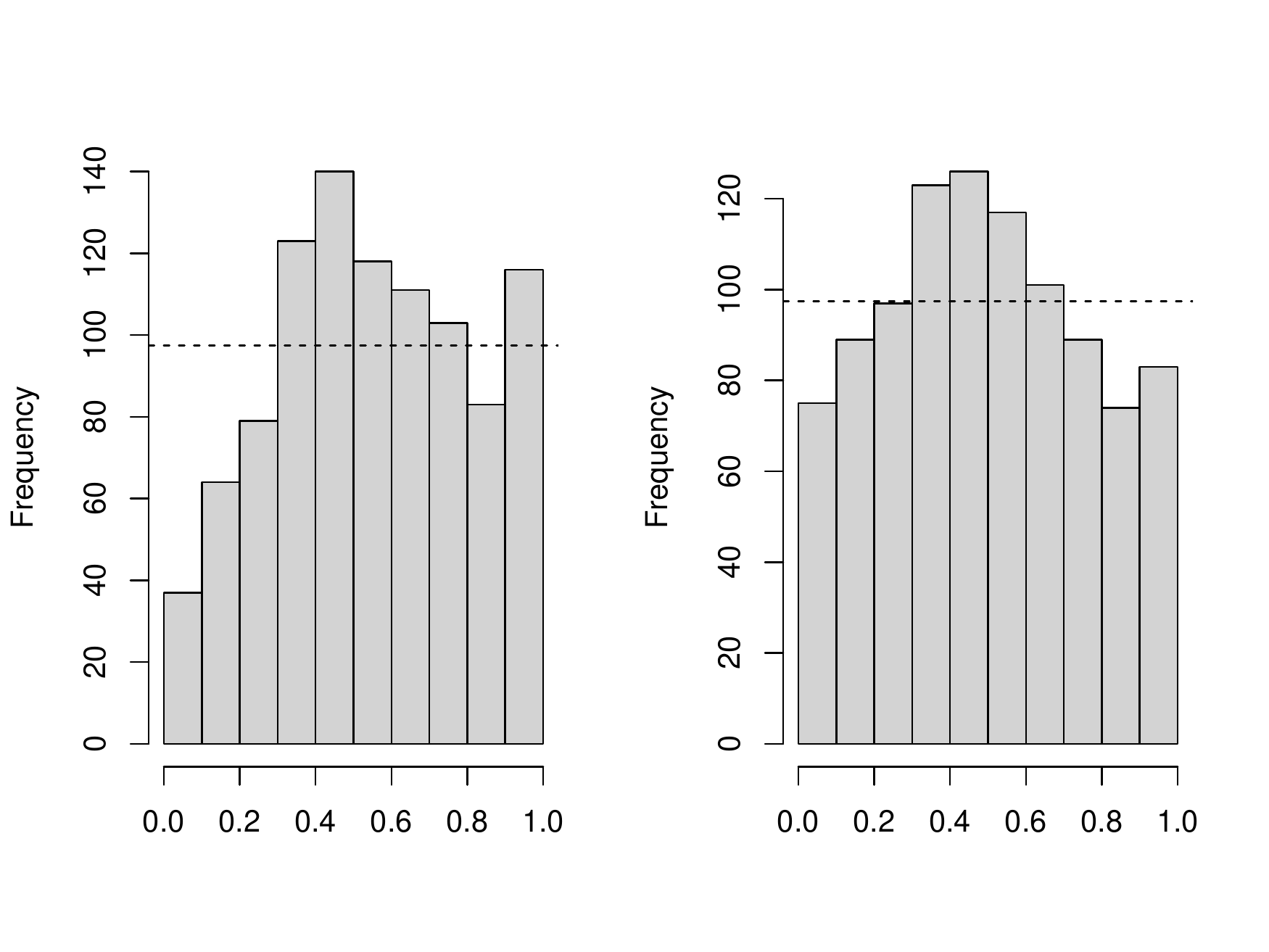} }}%
	\caption{Goodness of fit plots for the illness-death models. Histograms of $\widehat{S}_{0.}^M(V_i|X_i)$ (left of (a), (c), (e)) $\widehat{S}_{0.}^M(V_i,\delta_{1i},\delta_{2i},U_{1i})$ (right of (a), (c), (e)), $\widehat{S}_{12}^M(W_i|V_i,X_i)$ (left of (b), (d), (f)) and $\widehat{S}_{12}^M(W_i,V_i,\delta_{3i},U_{2i})$ (right of (b), (d), (f)). 
		The dashed lines are the expected values under the uniform distribution.}%
	\label{fig_aft_res12}%
\end{figure}

GOF assessment is done by a visual inspection of the histograms of the unmodified estimated survival probabilities and the RSPs histograms, shown in Figure \ref{fig_aft_res12}. As expected, the unmodified histograms are far from  that of a uniform distribution. However, based on the RSPs histogram, it is evident that the proposed AFT model fits well the data and is a better fit to the data, especially in comparison with the marginalized Cox model. The GOF plots of the models without frailty are shown in Figure S1 of the WSM, which indicate that the corresponding models with frailty fit the data much better.

\section{Discussion}
\label{s:discussion}
The main contribution of this work is twofold: (i) New semiparametric model and estimation method for a frailty-based AFT regression within the illness-death framework are provided. (ii) An exploratory method for diagnosis the appropriateness of any illness-death model is presented. 

The proposed model includes covariates and allows to handle a possible residual dependency between the non-terminal  the terminal failure times by incorporating a shared frailty. The estimation method can be applied both with or without frailty, and the simulation study indicates a good finite sample performances of the estimators in terms of bias and variance. 

The proposed model and methods can be extended to other types of multi-state models, for example with multiple non-terminal events and a vector of random effects (frailties) capturing multiple levels of dependence among the event. Furthermore, additional work is required to extend the proposed estimation method for time-dependent covariates.

\section*{Acknowledgements}

This work was supported by the Israel Science Foundation (ISF) grant number 767/21.

\bibliography{biomsample_bib}

\newpage

\begin{table}
	\centering
	\caption{Illness-death Cox and AFT models, methods and software availability.}
	\scalebox{0.8}{
		\begin{tabular}{|p{6em}|p{14em}|p{7.165em}|p{3.5cm}|}
			\toprule
			\textbf{Authors} & \textbf{Model} & \textbf{Estimation Procedure} & \textbf{Software} \\
			\midrule
			\multicolumn{1}{|l|}{\cite{xu2010statistical}} & Cox, gamma frailty, semiparametric & Semi-parametric MLE & none \\
			\midrule
			\multicolumn{1}{|l|}{\cite{lee2015bayesian}} & Cox, gamma frailty, semiparametric & Bayesian & R package \newline{}SemicompRisks \\
			\midrule
			\multicolumn{1}{|l|}{\cite{jiang2017semi}} & Transformation model, known trasformation function, non-parametric frailty  at the price of known error distribution.  & Semiparametric efficient score & none \\
			\midrule
			\multicolumn{1}{|l|}{\cite{lee2017accelerated}} & AFT, additive normal frailty, parametric and semiparametric & Baysian & R package \newline{}SemicompRisks \\
			\midrule
			\multicolumn{1}{|l|}{\cite{gorfine2020marginalized}} & Cox, marginalized gamma frailty, semiparametric & Pseudo-likelihood approach & https://github.com/\newline{}nirkeret/\newline{}frailty-LTRC \\
			\midrule
			Current work & AFT, multiplicative gamma frailty, semiparametric & Semi-parametric MLE & https://github.com/\newline{}LeaKats/\newline{}semicompAFT \\
			\bottomrule
	\end{tabular}}
	\label{tab:methods}%
\end{table}%

\begin{table}
	\centering
	\caption{Simulation results with frailty: means of estimates, empirical standard deviations, bootstrap standard errors and empirical coverage rates.}
	\scalebox{1}{
		\begin{tabular}{llcccccccc}
			\hline
	$\sigma$	&	& $\widehat{\sigma}$ & $\beta_{01,1}$ & $\beta_{01,2}$ & $\beta_{02,1}$ & $\beta_{02,2}$  & $\beta_{12,1}$  &  $\beta_{12,2}$  &  $\beta_{12,3}$  \\
\hline
		&	True values		&       	& 1     & 0.5   & 1     & 1     & 0.5   & 0.5   & 1 \\
			\midrule
	2   &	mean				& 1.84  & 1.05  & 0.48  & 1.03  & 1.05  & 0.51  & 0.49  & 1.06 \\
		&	empirical SD		& 0.27  & 0.10  & 0.13  & 0.12  & 0.10  & 0.11  & 0.12  & 0.10 \\
		&	bootstrap SE		& 0.29  & 0.11  & 0.12  & 0.12  & 0.11  & 0.13  & 0.14  & 0.15 \\
		&	CR    			& 0.95  & 0.95  & 0.94  & 0.93  & 0.96  & 0.99  & 0.99  & 0.96 \\
			\midrule
	1 	&	mean  			& 0.90  & 1.05  & 0.48  & 1.00  & 1.03  & 0.50  & 0.50  & 1.06 \\
		&	empirical SD		& 0.16  & 0.08  & 0.08  & 0.08  & 0.08  & 0.08  & 0.10  & 0.11 \\
		&	bootstrap SE		& 0.16  & 0.09  & 0.10  & 0.10  & 0.09  & 0.10  & 0.11  & 0.12 \\
		&	CR    			& 0.89  & 0.93  & 0.97  & 0.97  & 0.97  & 0.99  & 0.99  & 0.93 \\
			\midrule
	0.5 &	mean  			& 0.44  & 1.02  & 0.51  & 1.02  & 1.02  & 0.51  & 0.51  & 1.05 \\
		&	empirical SD		& 0.10  & 0.06  & 0.07  & 0.07  & 0.07  & 0.07  & 0.08  & 0.08 \\
		&	bootstrap SE		& 0.10  & 0.08  & 0.08  & 0.09  & 0.08  & 0.09  & 0.09  & 0.10 \\
		&	CR    			& 0.93  & 0.98  & 0.98  & 0.96  & 0.97  & 0.96  & 0.96  & 0.99 \\
			\bottomrule
		\end{tabular}%
	}
	\label{tab3}
\end{table}

\begin{table}
	\centering
	\caption{Simulation results of the cumulative baseline hazard functions with frailty: means of estimates, empirical standard deviations, bootstrap standard errors and empirical coverage rates.}
	\label{tab4}
	\scalebox{0.8}{
		\begin{tabular}{clcccccccccc}
			\hline
			$\sigma$		& $t$    		& 0.10  & 0.20  & 0.30  & 0.40  & 0.50  & 0.60  & 0.70  & 0.80  & 0.90  & 1.00 \\
			\midrule
			2     		& $H^0_{01}(t)$ 	& 0.01  & 0.04  & 0.09  & 0.16  & 0.25  & 0.36  & 0.49  & 0.64  & 0.81  & 1.00 \\
						& mean  			& 0.01  & 0.04  & 0.09  & 0.15  & 0.24  & 0.34  & 0.45  & 0.59  & 0.75  & 0.91 \\
						& empirical SD 	& 0.00  & 0.01  & 0.02  & 0.03  & 0.05  & 0.07  & 0.09  & 0.12  & 0.15  & 0.19 \\
						& bootstrap SE 	& 0.00  & 0.01  & 0.02  & 0.03  & 0.04  & 0.07  & 0.09  & 0.13  & 0.17  & 0.21 \\
						& CR    			& 0.96  & 0.93  & 0.95  & 0.93  & 0.97  & 0.96  & 0.95  & 0.97  & 0.98  & 0.97 \\
						\cmidrule{2-12}	
						& $H^0_{02}(t)$ 	& 0.02  & 0.06  & 0.14  & 0.24  & 0.38  & 0.54  & 0.74  & 0.96  & 1.22  & 1.50 \\
						& mean  			& 0.02  & 0.06  & 0.13  & 0.24  & 0.37  & 0.52  & 0.70  & 0.90  & 1.13  & 1.38 \\
						& empirical SD 	& 0.00  & 0.01  & 0.02  & 0.04  & 0.06  & 0.08  & 0.11  & 0.15  & 0.21  & 0.28 \\
						& bootstrap SE 	& 0.01  & 0.01  & 0.02  & 0.04  & 0.06  & 0.09  & 0.13  & 0.19  & 0.25  & 0.32 \\
						& CR    			& 0.99  & 0.97  & 0.96  & 0.97  & 0.98  & 0.97  & 0.99  & 0.99  & 0.98  & 0.98 \\
						\cmidrule{2-12}	
						& $H^0_{12}(t)$ 	& 0.01  & 0.04  & 0.09  & 0.16  & 0.25  & 0.36  & 0.49  & 0.64  & 0.81  & 1.00 \\
						& mean  			& 0.01  & 0.05  & 0.10  & 0.17  & 0.27  & 0.38  & 0.51  & 0.66  & 0.82  & 1.00 \\
						& empirical SD	& 0.03  & 0.04  & 0.05  & 0.05  & 0.07  & 0.09  & 0.11  & 0.13  & 0.15  & 0.18 \\
						& bootstrap SE	& 0.03  & 0.05  & 0.07  & 0.09  & 0.11  & 0.13  & 0.15  & 0.18  & 0.20  & 0.24 \\
						& CR    			& 0.96  & 0.98  & 0.98  & 0.99  & 0.99  & 0.98  & 0.98  & 0.98  & 0.98  & 0.99 \\
			\midrule
			1     		& $H^0_{01}(t)$ 	& 0.01  & 0.04  & 0.09  & 0.16  & 0.25  & 0.36  & 0.49  & 0.64  & 0.81  & 1.00 \\
						& mean  			& 0.01  & 0.04  & 0.09  & 0.15  & 0.24  & 0.34  & 0.46  & 0.60  & 0.75  & 0.91 \\
						& empirical SD	& 0.00  & 0.01  & 0.01  & 0.02  & 0.03  & 0.04  & 0.06  & 0.08  & 0.10  & 0.13 \\
						& bootstrap SE	& 0.00  & 0.01  & 0.01  & 0.02  & 0.04  & 0.05  & 0.07  & 0.10  & 0.13  & 0.17 \\
						& CR    			& 0.97  & 0.99  & 0.98  & 0.97  & 1.00  & 0.99  & 0.98  & 0.98  & 1.00  & 0.99 \\
						\cmidrule{2-12} 	
						& $H^0_{02}(t)$ 	& 0.02  & 0.06  & 0.14  & 0.24  & 0.38  & 0.54  & 0.74  & 0.96  & 1.22  & 1.50 \\
						& mean  			& 0.02  & 0.06  & 0.13  & 0.23  & 0.36  & 0.52  & 0.71  & 0.92  & 1.15  & 1.40 \\
						& empirical SD	& 0.00  & 0.01  & 0.02  & 0.03  & 0.04  & 0.07  & 0.10  & 0.13  & 0.17  & 0.22 \\
						& bootstrap SE	& 0.00  & 0.01  & 0.02  & 0.03  & 0.05  & 0.07  & 0.10  & 0.14  & 0.19  & 0.25 \\
						& CR    			& 0.96  & 0.98  & 0.98  & 0.98  & 0.98  & 0.95  & 0.96  & 0.97  & 0.98  & 0.95 \\
						\cmidrule{2-12} 	
						& $H^0_{12}(t)$ 	& 0.01  & 0.04  & 0.09  & 0.16  & 0.25  & 0.36  & 0.49  & 0.64  & 0.81  & 1.00 \\
						& mean  			& 0.01  & 0.04  & 0.10  & 0.17  & 0.27  & 0.38  & 0.51  & 0.66  & 0.83  & 1.02 \\
						& empirical SD	& 0.03  & 0.03  & 0.04  & 0.05  & 0.07  & 0.08  & 0.10  & 0.11  & 0.13  & 0.15 \\
						& bootstrap SE	& 0.02  & 0.04  & 0.06  & 0.07  & 0.09  & 0.11  & 0.13  & 0.15  & 0.17  & 0.20 \\
						& CR    			& 0.95  & 0.97  & 0.98  & 0.97  & 0.96  & 0.96  & 0.96  & 0.96  & 0.99  & 0.99 \\
			\midrule
			0.5   		& $H^0_{01}(t)$ 	& 0.01  & 0.04  & 0.09  & 0.16  & 0.25  & 0.36  & 0.49  & 0.64  & 0.81  & 1.00 \\
						& mean  			& 0.01  & 0.04  & 0.09  & 0.16  & 0.25  & 0.35  & 0.48  & 0.63  & 0.79  & 0.97 \\
						& empirical SD	& 0.00  & 0.01  & 0.01  & 0.02  & 0.03  & 0.04  & 0.05  & 0.08  & 0.10  & 0.13 \\
						& bootstrap SE	& 0.00  & 0.01  & 0.01  & 0.02  & 0.03  & 0.05  & 0.06  & 0.08  & 0.11  & 0.14 \\
						& CR    			& 0.92  & 0.98  & 0.97  & 0.98  & 0.97  & 0.98  & 0.97  & 0.96  & 0.97  & 0.95 \\
						\cmidrule{2-12}	
						& $H^0_{02}(t)$ 	& 0.02  & 0.06  & 0.14  & 0.24  & 0.38  & 0.54  & 0.74  & 0.96  & 1.22  & 1.50 \\
						& mean  			& 0.02  & 0.06  & 0.13  & 0.24  & 0.37  & 0.53  & 0.72  & 0.93  & 1.18  & 1.44 \\
						& empirical SD	& 0.00  & 0.01  & 0.02  & 0.03  & 0.04  & 0.06  & 0.08  & 0.11  & 0.15  & 0.20 \\
						& bootstrap SE	& 0.00  & 0.01  & 0.02  & 0.03  & 0.04  & 0.06  & 0.09  & 0.12  & 0.15  & 0.20 \\
						& CR    			& 0.96  & 1.00  & 0.98  & 0.97  & 0.97  & 0.95  & 0.94  & 0.94  & 0.93  & 0.96 \\
						\cmidrule{2-12}	
						& $H^0_{12}(t)$ 	& 0.01  & 0.04  & 0.09  & 0.16  & 0.25  & 0.36  & 0.49  & 0.64  & 0.81  & 1.00 \\
						& mean  			& 0.01  & 0.04  & 0.09  & 0.17  & 0.27  & 0.38  & 0.52  & 0.67  & 0.85  & 1.04 \\
						& empirical SD	& 0.02  & 0.03  & 0.04  & 0.05  & 0.06  & 0.07  & 0.09  & 0.11  & 0.13  & 0.14 \\
						& bootstrap SE	& 0.02  & 0.04  & 0.05  & 0.06  & 0.08  & 0.09  & 0.11  & 0.13  & 0.14  & 0.16 \\
						& CR    			& 0.89  & 0.96  & 0.98  & 0.99  & 0.98  & 0.98  & 0.98  & 0.96  & 0.95  & 0.96 \\
			\bottomrule
		\end{tabular}%
	}
\end{table}

\begin{table}
	\centering
	\caption{Simulation results of model misspecification where frailty is ignored: means of estimates, empirical standard deviations and empirical coverage rates.}
	\label{tab5}
	\scalebox{1}{
		\begin{tabular}{rlccccccc}
			\hline
			$\sigma$ 		&       			&  $\beta_{01,1}$ & $\beta_{01,2}$ 
											&  $\beta_{02,1}$ & $\beta_{02,2}$ 
											& $\beta_{12,1}$ & $\beta_{12,2}$ & $\beta_{12,3}$ \\
			\hline
							& true values 	& 1     & 0.5   & 1     & 1     & 0.5   & 0.5   & 1 \\
			\midrule
				2 			& mean  			& 1.27  & 0.36  & 1.15  & 1.27  & 0.25  & 0.41  & 2.09 \\
							& empirical SD 	& 0.10  & 0.14  & 0.15  & 0.15  & 0.32  & 0.31  & 0.50 \\
							& CR 			& 0.25  & 0.86  & 0.82  & 0.53  & 0.90  & 0.94  & 0.45 \\
			\midrule
				1 			& mean  			& 1.19  & 0.42  & 1.08  & 1.17  & 0.37  & 0.46  & 1.63 \\
							& empirical SD 	& 0.09  & 0.09  & 0.10  & 0.09  & 0.16  & 0.16  & 0.19 \\
							& CR 			& 0.48  & 0.86  & 0.87  & 0.56  & 0.86  & 0.93  & 0.08 \\
			\midrule
				0.5 			& mean  			& 1.09  & 0.48  & 1.06  & 1.10  & 0.44  & 0.47  & 1.33 \\
							& empirical SD 	& 0.07  & 0.08  & 0.08  & 0.08  & 0.11  & 0.12  & 0.14 \\
							& CR 			& 0.70  & 0.94  & 0.89  & 0.72  & 0.91  & 0.96  & 0.37 \\
			\bottomrule
	\end{tabular}}
\end{table}

\begin{table}
	\centering
	\caption{Simulation results of the cumulative baseline hazard functions under model misspecification where frailty is ignored: means of estimates, empirical standard deviations, and empirical coverage rates.}
	\label{tab6}
	\scalebox{0.8}{
		\begin{tabular}{clcccccccccc}
			\hline
			$\sigma$	& $t$     			& 0.10  & 0.20  & 0.30  & 0.40  & 0.50  & 0.60  & 0.70  & 0.80  & 0.90  & 1.00 \\
			\midrule
			2     	& $H^{0}_{01}(t)$ 	& 0.01  & 0.04  & 0.09  & 0.16  & 0.25  & 0.36  & 0.49  & 0.64  & 0.81  & 1.00 \\
					& mean  				& 0.01  & 0.03  & 0.06  & 0.10  & 0.14  & 0.19  & 0.23  & 0.28  & 0.32  & 0.36 \\
					& empirical SD 		& 0.00  & 0.01  & 0.01  & 0.02  & 0.02  & 0.03  & 0.03  & 0.04  & 0.04  & 0.04 \\
					& CR 				& 0.96  & 0.73  & 0.31  & 0.06  & 0.00  & 0.00  & 0.00  & 0.00  & 0.00  & 0.00 \\
					\cmidrule{2-12}          
					& $H^{0}_{02}(t)$ 	& 0.02  & 0.06  & 0.14  & 0.24  & 0.38  & 0.54  & 0.74  & 0.96  & 1.22  & 1.50 \\
					& mean  				& 0.02  & 0.06  & 0.11  & 0.17  & 0.23  & 0.29  & 0.35  & 0.40  & 0.46  & 0.51 \\
					& empirical SD 		& 0.00  & 0.01  & 0.02  & 0.02  & 0.03  & 0.03  & 0.03  & 0.04  & 0.04  & 0.04 \\
					& CR 				& 0.91  & 0.92  & 0.67  & 0.12  & 0.00  & 0.00  & 0.00  & 0.00  & 0.00  & 0.00 \\
					\cmidrule{2-12}          
					& $H^{0}_{12}(t)$ 	& 0.01  & 0.04  & 0.09  & 0.16  & 0.25  & 0.36  & 0.49  & 0.64  & 0.81  & 1.00 \\
					& mean  				& 0.19  & 0.44  & 0.66  & 0.88  & 1.07  & 1.26  & 1.43  & 1.58  & 1.73  & 1.89 \\
					& empirical SD 		& 0.24  & 0.36  & 0.42  & 0.45  & 0.48  & 0.50  & 0.52  & 0.52  & 0.55  & 0.56 \\
					& CR 				& 0.90  & 0.86  & 0.82  & 0.75  & 0.72  & 0.69  & 0.68  & 0.68  & 0.72  & 0.77 \\
			\midrule
			1     	& $H^{0}_{01}(t)$ 	& 0.01  & 0.04  & 0.09  & 0.16  & 0.25  & 0.36  & 0.49  & 0.64  & 0.81  & 1.00 \\
					& mean  				& 0.01  & 0.03  & 0.07  & 0.12  & 0.18  & 0.24  & 0.31  & 0.37  & 0.44  & 0.51 \\
					& empirical SD 		& 0.00  & 0.01  & 0.01  & 0.02  & 0.02  & 0.03  & 0.03  & 0.04  & 0.04  & 0.05 \\
					& CR 				& 0.92  & 0.84  & 0.63  & 0.28  & 0.07  & 0.00  & 0.00  & 0.00  & 0.00  & 0.00 \\
					\cmidrule{2-12}          
					& $H^{0}_{02}(t)$ 	& 0.02  & 0.06  & 0.14  & 0.24  & 0.38  & 0.54  & 0.74  & 0.96  & 1.22  & 1.50 \\
					& mean  				& 0.02  & 0.06  & 0.12  & 0.19  & 0.28  & 0.37  & 0.46  & 0.55  & 0.64  & 0.73 \\
					& empirical SD 		& 0.00  & 0.01  & 0.01  & 0.02  & 0.03  & 0.04  & 0.04  & 0.05  & 0.05  & 0.06 \\
					& CR 				& 0.93  & 0.94  & 0.79  & 0.38  & 0.09  & 0.00  & 0.00  & 0.00  & 0.00  & 0.00 \\
					\cmidrule{2-12}          
					& $H^{0}_{12}(t)$ 	& 0.01  & 0.04  & 0.09  & 0.16  & 0.25  & 0.36  & 0.49  & 0.64  & 0.81  & 1.00 \\
					& mean  				& 0.06  & 0.17  & 0.32  & 0.47  & 0.65  & 0.81  & 0.97  & 1.14  & 1.31  & 1.46 \\
					& empirical SD 		& 0.09  & 0.11  & 0.14  & 0.16  & 0.17  & 0.19  & 0.20  & 0.22  & 0.23  & 0.24 \\
					& CR 				& 0.96  & 0.85  & 0.71  & 0.53  & 0.44  & 0.37  & 0.33  & 0.37  & 0.47  & 0.58 \\
			\midrule
			0.5   	& $H^{0}_{01}(t)$ 	& 0.01  & 0.04  & 0.09  & 0.16  & 0.25  & 0.36  & 0.49  & 0.64  & 0.81  & 1.00 \\
					& mean  				& 0.01  & 0.04  & 0.08  & 0.14  & 0.21  & 0.29  & 0.38  & 0.48  & 0.58  & 0.69 \\
					& empirical SD 		& 0.00  & 0.01  & 0.01  & 0.02  & 0.02  & 0.03  & 0.04  & 0.05  & 0.06  & 0.07 \\
					& CR 				& 0.92  & 0.92  & 0.91  & 0.82  & 0.58  & 0.40  & 0.21  & 0.11  & 0.01  & 0.01 \\
					\cmidrule{2-12}          
					& $H^{0}_{02}(t)$ 	& 0.02  & 0.06  & 0.14  & 0.24  & 0.38  & 0.54  & 0.74  & 0.96  & 1.22  & 1.50 \\
					& mean  				& 0.02  & 0.06  & 0.13  & 0.22  & 0.32  & 0.44  & 0.56  & 0.70  & 0.83  & 0.97 \\
					& empirical SD 		& 0.00  & 0.01  & 0.02  & 0.02  & 0.03  & 0.04  & 0.05  & 0.07  & 0.07  & 0.08 \\
					& CR 				& 0.92  & 0.96  & 0.94  & 0.81  & 0.57  & 0.36  & 0.14  & 0.03  & 0.00  & 0.00 \\
					\cmidrule{2-12}          
					& $H^{0}_{12}(t)$ 	& 0.01  & 0.04  & 0.09  & 0.16  & 0.25  & 0.36  & 0.49  & 0.64  & 0.81  & 1.00 \\
					& mean  				& 0.02  & 0.09  & 0.18  & 0.29  & 0.43  & 0.58  & 0.73  & 0.89  & 1.06  & 1.24 \\
					& empirical SD 		& 0.04  & 0.06  & 0.08  & 0.10  & 0.12  & 0.14  & 0.16  & 0.17  & 0.18  & 0.20 \\
					& CR 				& 0.92  & 0.91  & 0.84  & 0.75  & 0.72  & 0.70  & 0.73  & 0.72  & 0.74  & 0.78 \\
			\bottomrule
	\end{tabular}}
\end{table}

\begin{sidewaystable}
	\caption{Rotterdam Tumor Bank Data: Estimates (Est), exponent of estimated regression coefficients (exp), standard errors (SE), p-values and Holm's adjusted p-values. Bold results are significant at 0.05 based on Holm's adjusted p-value. }\label{tab7}
	\centering
	\scalebox{0.75}{
		\begin{tabular}{lccccccccccccccccc}
			& \multicolumn{5}{c}{Proposed Model (zeta=65)}        
			& \multicolumn{5}{c}{Marginalized Cox (Gorfine et al. 2021)}        
			& \multicolumn{5}{c}{Conditional Cox (Lee et al. 2015)} \\
			& Est & exp & SE & p-value & Holm        
			& Est & exp & SE & p-value & Holm       
			& Est & exp & SE & p-value & Holm \\
			\midrule
			$\sigma$ & 2.18 &     & 0.73 & 0.003 & 0.058       
			         & 2.52 &     & 0.54 & 0.000 & \textbf{0.000}        
			         & 1.47 &     & 0.23 & 0.000 & \textbf{0.000} \\
			\midrule
			\textbf{Transition: surgery $\rightarrow$ relapse} \\
			Age at surgery (divided by 10) &  0.14  & 1.15 & 0.06 & 0.012 & 0.185            
			                               &  -0.15 & 0.86 & 0.06 & 0.014 & 0.262 
			                               &  -0.22 & 0.80 & 0.08 & 0.003 & \textbf{0.048} \\
			log of lymph nodes & -0.40 & 0.67 & 0.05 & 0.000 & \textbf{0.000}
			                   & 0.42  & 1.53 & 0.04 & 0.000 & \textbf{0.000}  
			                   & 0.71  & 2.03 & 0.07 & 0.000 & \textbf{0.000}\\
			log of estrogen+1  & 0.07  & 1.07 & 0.03 & 0.030 & 0.390 
			                   & -0.03 & 0.97 & 0.02 & 0.186 & 1.000  
			                   & -0.10 & 0.90 & 0.04 & 0.003 & \textbf{0.047}  \\
			log of progesterone+1 & 0.09  & 1.09 & 0.02 & 0.000 & \textbf{0.005}
			                      & -0.04 & 0.96 & 0.02 & 0.065 & 1.000
			                      & -0.11 & 0.90 & 0.03 & 0.001 & \textbf{0.020} \\
			Postmenopausal (vs. premenopausal) & -0.34 & 0.71 & 0.15 & 0.023 & 0.328
			                                   & 0.13  & 1.14 & 0.13 & 0.296 & 1.000  
			                                   & 0.34  & 1.40 & 0.19 & 0.071 & 0.785 \\
			Tumor size (ref $<20$mm) \\
			\,\,\,\,\, 20-50mm   & -0.32  & 0.73 & 0.09 & 0.001 & \textbf{0.015}
		                                       &  0.20  & 1.22 & 0.07 & 0.006 & 0.116      
		                                       &  0.40  & 1.49 & 0.12 & 0.001 & \textbf{0.019} \\
		     \,\,\,\,\, $>50$mm   & -0.49  & 0.61 & 0.11 & 0.000 & \textbf{0.000}
	                                           &  0.38  & 1.46 & 0.11 & 0.001 & \textbf{0.020}
	                                           &  0.79  & 2.19 & 0.16 & 0.000 & \textbf{0.000} \\
			Hormone therapy &  0.60  & 1.83 & 0.13 & 0.000 & \textbf{0.000}
			                & -0.38  & 0.68 & 0.08 & 0.000 & \textbf{0.000}
			                &  -0.88 & 0.41 & 0.15 & 0.000 & \textbf{0.000} \\
			Chemotherapy &   0.49 & 1.64 & 0.11 & 0.000 & \textbf{0.000}
			             &  -0.37 & 0.69 & 0.11 & 0.001 & \textbf{0.023}
			             &  -0.79 & 0.46 & 0.16 & 0.000 & \textbf{0.000}\\
			Tumor grade 3 (vs. 2) & -0.25 & 0.78 & 0.09 & 0.004 & 0.081
			                      & 0.21  & 1.23 & 0.08 & 0.008 & 0.155
			                      & 0.44  & 1.56 & 0.13 & 0.000 & \textbf{0.010}\\
			\midrule
			\textbf{Transition: surgery $\rightarrow$ death} \\
			Age at surgery (divided by 10) & -0.43 & 0.65 & 0.14 & 0.002 & 0.051 
			                               & 1.32  & 3.74 & 0.37 & 0.000 & \textbf{0.009}
			                               & 1.43  & 4.20 & 0.18 & 0.000 & \textbf{0.000} \\
			log of lymph nodes & -0.14 & 0.87 & 0.08 & 0.091 & 1.000
			                   & 0.13  & 1.14 & 0.12 & 0.298 & 1.000
			                   & 0.44  & 1.54 & 0.15 & 0.004 & \textbf{0.049} \\
			log of estrogen+1  & 0.04  & 1.04 & 0.04 & 0.287 & 1.000
			                   & -0.01 & 0.99 & 0.06 & 0.816 & 1.000
			                   & -0.11 & 0.89 & 0.08 & 0.149 & 1.000 \\
			log of progesterone+1 & 0.01  & 1.01 & 0.04 & 0.827 & 1.000
			                      & 0.08  & 1.08 & 0.06 & 0.205 & 1.000
			                      & 0.01  & 1.01 & 0.07 & 0.857 & 1.000 \\
			Postmenopausal (vs. premenopausal)	& -0.15 & 0.86 & 0.34 & 0.647 & 1.000                               
												& -0.30 & 0.74 & 0.50 & 0.554 & 1.000          
												& -0.35 & 0.70 & 0.70 & 0.613 & 1.000          \\
			Tumor size (ref. $<20$mm)\\
			\,\,\,\,\, 20--50mm 	& -0.13 & 0.88 & 0.15 & 0.376 & 1.000                      
												& -0.16 & 0.85 & 0.25 & 0.526 & 1.000          
												& -0.04 & 0.96 & 0.28 & 0.874 & 1.000 \\
			\,\,\,\,\, $>50$mm  & -0.19 & 0.82 & 0.18 & 0.275 & 1.000                     
											& 0.15  & 1.16 & 0.31 & 0.634 & 1.000
											& 0.58  & 1.79 & 0.35 & 0.096 & 0.956 \\
			Hormone therapy 	& 0.41  & 1.51 & 0.18 & 0.019 & 0.290                      
							& -0.21 & 0.81 & 0.25 & 0.389 & 1.000  
							& -0.69 & 0.50 & 0.29 & 0.018 & 0.220 \\
			Chemotherapy    	& 1.13  & 3.09 & 0.30 & 0.000 & \textbf{0.005}                      
							& -0.22 & 0.81 & 0.81 & 0.789 & 1.000 
							& -0.78 & 0.46 & 0.63 & 0.215 & 1.000 \\
		    Tumor grade 3 (vs. 2) 	& -0.06 & 0.94 & 0.13 & 0.641 & 1.000           
									& -0.01 & 0.99 & 0.28 & 0.961 & 1.000          
									& 0.21  & 1.23 & 0.27 & 0.446 & 1.000\\
			\midrule
		    \textbf{Transition: relapse $\rightarrow$ death}  \\
			Age at surgery (divided by 10)	& 0.00  & 1.00 & 0.07 & 0.956 & 1.000          
											& 0.03  & 1.03 & 0.08 & 0.700 & 1.000          
											& 0.08  & 1.08 & 0.07 & 0.278 & 1.000          \\
			log of lymph nodes 	& -0.25 & 0.78 & 0.07 & 0.000 & \textbf{0.010} 
								& 0.25  & 1.28 & 0.05 & 0.000 & \textbf{0.000} 
								& 0.38  & 1.47 & 0.07 & 0.000 & \textbf{0.000} \\
			log of estrogen+1  	& 0.04  & 1.04 & 0.05 & 0.341 & 1.000          
								& -0.03 & 0.97 & 0.02 & 0.193 & 1.000          
								& -0.10 & 0.90 & 0.04 & 0.008 & 0.102          \\
			log of progesterone+1 	& 0.13  & 1.14 & 0.04 & 0.001 & \textbf{0.021}          
									& -0.08 & 0.92 & 0.02 & 0.000 & \textbf{0.003} 
									& -0.19 & 0.83 & 0.04 & 0.000 & \textbf{0.000} \\
			Postmenopausal (vs. premenopausal)  	& -0.21 & 0.81 & 0.17 & 0.203 & 1.000          
												& -0.05 & 0.95 & 0.13 & 0.731 & 1.000          
												& 0.04  & 1.04 & 0.20 & 0.843 & 1.000          \\
		    Tumor size (ref. $<20$mm)  \\
		    \,\,\,\,\, 20--50mm 	& -0.37 & 0.69 & 0.14 & 0.008 & 0.131          
												& 0.23  & 1.26 & 0.07 & 0.001 & \textbf{0.024} 
												& 0.46  & 1.58 & 0.14 & 0.001 & \textbf{0.015} \\
			\,\,\,\,\,  $>50$mm  & -0.52 & 0.60 & 0.17 & 0.002 & \textbf{0.044}          
											& 0.40  & 1.49 & 0.10 & 0.000 & \textbf{0.002} 
											& 0.67 & 1.96 & 0.18 & 0.000 & \textbf{0.003} \\
			Hormone therapy	& 0.39  & 1.48 & 0.14 & 0.005 & 0.090          
							& -0.18 & 0.84 & 0.09 & 0.037 & 0.633          
							& -0.48 & 0.62 & 0.16 & 0.002 & \textbf{0.037} \\
			Chemotherapy   	& 0.23  & 1.25 & 0.18 & 0.205 & 1.000          
							& -0.16 & 0.85 & 0.13 & 0.227 & 1.000          
							& -0.18 & 0.84 & 0.17 & 0.287 & 1.000          \\
			Tumor grade 3 (vs. 2) 	& -0.26 & 0.77 & 0.13 & 0.047 & 0.569          
									& 0.21  & 1.23 & 0.09 & 0.024 & 0.440          
									& 0.43  & 1.54 & 0.14 & 0.002 & \textbf{0.032}\\
			\midrule
	\end{tabular}}
\end{sidewaystable}

\hspace{15cm}
\newpage
\setcounter{figure}{0}   
\setcounter{table}{0}
\setcounter{section}{0}      

\setlength{\oddsidemargin}{-0.35in}
\setlength{\evensidemargin}{0.0in}
\setlength{\topmargin}{-0.75in}
\setlength{\textheight}{9.25in}
\setlength{\textwidth}{7in}

\setlength{\parskip}{2.3ex}
\setlength{\parindent}{0in}

\renewcommand{\thefigure}{S\arabic{figure}}
\renewcommand{\thetable}{S\arabic{table}}
\renewcommand{\thesection}{S\arabic{section}} 

\bibliographystyle{chicago}

\begin{center}
\textbf{\Large Web Supplementary Material\\
An Accelerated Failure Time Regression Model for\\
Illness-Death Data:\\
\vspace{2mm}
A Frailty Approach}\\
\vspace{5mm}
\large Lea Kats and Malka Gorfine\\
Department of Statistics and Operations Research\\
			Tel Aviv University, Israel
		\date{}	
\end{center}

\section{Conditional Hazards}
\label{ss:conditional_Hazards}
Based on Eq.'s\,(1)-(6) of the main text, we derive the conditional hazard functions of the three transitions, given $(X_i,\gamma_i)$, namely, following conditional hazard functions
\begin{eqnarray*}
	\lambda_{01}(t|X_i, \gamma_i)
	&=&\lim_{\Delta \to 0}\frac{1}{\Delta}\Pr\left(t\leq T_{1i}<t+\Delta|T_{1i}\geq t, T_{2i}\geq t,X_i,\gamma_i \right)\,,\, t>0\,,\\
	\lambda_{02}(t|X_i, \gamma_i)
	&=&\lim_{\Delta \to 0}\frac{1}{\Delta}\Pr\left(t\leq T_{2i}<t+\Delta|T_{1i}\geq t, T_{2i}\geq t,X_i,\gamma_i \right)\,,\, t>0\,,\\
	\lambda_{12}(t|t_1,X_i, \gamma_i)
	&=&\lim_{\Delta \to 0}\frac{1}{\Delta}\Pr\left(t\leq T_{2i}<t+\Delta|T_{1i}= t_1, T_{2i}\geq t,X_i,\gamma_i \right) \, ,\, t>t_1>0 \,.
\end{eqnarray*}
To this end, write
\begin{eqnarray*}
	\lambda_{01}(t|X_i, \gamma_i)
	&=&\frac{\lim_{\Delta\to 0}\frac{1}{\Delta}\Pr\left(T_{1i}\in[t,t+\Delta)\,,T_{2i}\geq t\,|X_i,\gamma_i\right)}{\Pr\left(T_{1i}\geq t,T_{2i}\geq t|X_i,\gamma_i\right)}\\
	&=&\frac{\lim_{\Delta\to 0}\frac{1}{\Delta}\Pr\left(e^{\epsilon_{01i}}\in\,[t e^{-\beta^T_{01}X_i},(t+\Delta)e^{-\beta^T_{01}X_i})\,,\,e^{\epsilon_{02i}}\geq t e^{-\beta^T_{02}X_i}|X_i,\gamma_i\right)}  
	{\Pr\left(e^{\epsilon_{01i}}\geq t e^{-\beta^T_{01}X_i},e^{\epsilon_{02i}}\geq t e^{-\beta^T_{02}X_i}|X_i,\gamma_i\right)}\\
	&=&\frac{\lim_{\Delta\to 0}\frac{1}{\Delta}\Pr\left(e^{\epsilon_{01i}}\in\,[t e^{-\beta^T_{01}X_i},(t+\Delta)e^{-\beta^T_{01}X_i})|X_i,\gamma_i\right)\Pr\left(e^{\epsilon_{02i}}\geq t e^{-\beta^T_{02}X_i}|X_i,\gamma_i\right)}{\Pr\left(e^{\epsilon_{01i}}\geq t e^{-\beta^T_{01}X_i}|X_i,\gamma_i\right)\
		Pr\left(e^{\epsilon_{02i}}\geq t e^{-\beta^T_{02}X_i}|X_i,\gamma_i\right)}\\
	&=&-\frac{\partial}{\partial t}\log S_{\epsilon_{01}}\left(t e^{-\beta^T_{01}X_i}|\gamma_i\right)=\frac{f_{\epsilon_{01}}\left(t e^{-\beta^T_{01}X_i}|\gamma_i\right)e^{-\beta^T_{01}X_i}}{S_{\epsilon_{01}}\left(t e^{-\beta^T_{01}X_i}|\gamma_i\right)}=\lambda_{01}(t e^{-\beta^T_{01}X_i}|\gamma_i) e^{-\beta^T_{01}X_i}\\
	&=&\gamma_i h^o_{01}\left(t e^{-\beta^T_{01}X_i}\right)e^{-\beta^T_{01}X_i}\, ,\, t>0 \,.
\end{eqnarray*}
Due to the similarity between $T_{1i}$ and $T_{2i}$ at transitions $0\to 1$ and $0\to 2$, 
\begin{equation*}
	\lambda_{02}(t|X_i, \gamma_i)=\frac{\lim_{\Delta\to 0}\frac{1}{\Delta}\Pr\big(T_{2i}\in[t,t+\Delta)\,,T_{1i}\geq t\,|X_i,\gamma_i\big)}{\Pr\big(T_{1i}\geq t,T_{2i}\geq t|X_i,\gamma_i\big)}=\gamma_i h^o_{02}\left(t e^{-\beta^T_{02}X_i}\right)e^{-\beta^T_{02}X_i}\, ,\, t>0 \, .
\end{equation*}
Finally, since 
\begin{equation*}
	\lambda_{12}(t|t_1,X_i, \gamma_i)=
	-\dfrac{\partial }{\partial t}\log\Pr\left( T_{2i}>t|T_{1i}=t_1,T_{2i}>t_1,X_i,\gamma_i \right) \, ,
\end{equation*}
we first derive $\Pr\left( T_{2i}>t|T_{1i}=t_1,T_{2i}>t_1,X_i,\gamma_i \right)$, the probability of staying in State 1 until time $t$, after diagnosed with the disease at time $t_1$. Then, for $t > t_1 >0$,
\begin{eqnarray*}
	\lefteqn{\Pr\big(T_{2i}>t|T_{1i}=t_1,T_{2i}>t_1,X_i,\gamma_i \big)}\\
	&&=\frac{\lim_{\Delta\to 0} \frac{1}{\Delta} \Pr(T_{2i}>t,T_{1i}\in [t_1, t_1+\Delta),T_{2i}>t_1|X_i,\gamma_i)}{\lim_{\Delta\to 0} \frac{1}{\Delta} \Pr(T_{1i}\in [t_1, t_1+\Delta),T_{2i}>t_1|X_i,\gamma_i)}\\
	&&=\frac{\lim_{\Delta\to 0} \frac{1}{\Delta} \Pr(T_{2i}>t,T_{1i}\in [t_1, t_1+\Delta)|X_i,\gamma_i)}{\lim_{\Delta\to 0} \frac{1}{\Delta} \Pr(T_{1i}\in [t_1, t_1+\Delta),T_{2i}>t_1|X_i,\gamma_i)}\\
	&&=\frac{\Pr(T_{2i}>t|X_i,\gamma_i) \lim_{\Delta\to 0} \frac{1}{\Delta}  \Pr(T_{1i}\in [t_1 ,t_1+\Delta)|X_i,\gamma_i)}{ \Pr(T_{2i}>t_1|X_i,\gamma_i)\lim_{\Delta\to 0} \frac{1}{\Delta} \Pr(T_{1i}\in [t_1, t_1+\Delta)|X_i,\gamma_i)}\\
	&&=\dfrac{\Pr\left(e^{\epsilon_{12}}>t e^{-\beta_{12}^T X_i}|X_i,\gamma_i\right)}{\Pr\left(e^{\epsilon_{12}}>t_1 e^{-\beta_{12}^T X_i}|X_i,\gamma_i\right)}=\frac{S_{\epsilon_{12}}\left(t e^{-\beta_{12}^{T}X_i}|\gamma_i\right)}{S_{\epsilon_{12}}\left(t_1 e^{-\beta_{12}^{T}X_i}|\gamma_i\right)}\\
	&&=\frac{\exp\left\{- \gamma_i\int_{0}^{t e^{-\beta_{12}^{T}X_i}} h^o_{{12}}(v)dv\right\}}{\exp\left\{- \gamma_i\int_{0}^{t_1 e^{-\beta_{12}^{T}X_i}} h^o_{{12}}(v)dv\right\}}\\
	&&=\exp\left\{-\gamma_i \left[H^o_{12}\left(t e^{-\beta_{12}^{T}X_i}\right)- H^o_{12}\left(t_1 e^{-\beta_{12}^{T}X_i}\right)		\right]\right\}\, .
\end{eqnarray*}
Hence,
\begin{eqnarray*}
	\lambda_{12}(t|t_1,X_i, \gamma_i)
	&=&-\frac{\partial}{\partial t}\log\Big(\exp\left\{-\gamma_i \left[H^o_{12}\left(t e^{-\beta_{12}^{T}X_i}\right)
	- H^o_{12}\left(t_1 e^{-\beta_{12}^{T}X_i}\right)		\right]\right\}\Big)\nonumber\\
	&=&\gamma_i h^o_{12}\left(t e^{-\beta^T_{12}X_i}\right)e^{-\beta^T_{12}X_i} \, , \, t>t_1>0 \, .
\end{eqnarray*}

\section{Likelihood Function}
\label{ss:likelidood_function}
Based on Eq.'s (1)-(6) of the main text, the contribution of individual $i$ to the complete-data likelihood is given by
\begin{eqnarray}\label{likelihood_func}
	\lefteqn{f_{V,W,\delta_{1},\delta_{2},\delta_{3},\gamma}(V_i,W_i,\delta_{1i},\delta_{2i},\delta_{3i},\gamma_i|{X}_i)}  \nonumber\\
	&&=f_{V,W,\delta_{1},\delta_{2},\delta_{3}}(V_i,W_i,\delta_{1i},\delta_{2i},\delta_{3i}|{X}_i,\gamma_i)f_{\gamma| X}(\gamma_i|{X}_i) \nonumber\\
	&&=f_{V,W,\delta_{1},\delta_{2},\delta_{3}}(V_i,W_i,\delta_{1i},\delta_{2i},\delta_{3i}|{X}_i,\gamma_i)f(\gamma_i;\sigma) \nonumber\\
	&&=f(V_i,\delta_{1i},\delta_{2i}|{X}_i,\gamma_i)f(W_i,\delta_{3i}|V_i,W_i>V_i,\delta_{1i}=1,{X}_i,\gamma_i)^{\delta_{1i}}f(\gamma_i;\sigma) \nonumber\\
	&&\propto \lambda_{01}(V_i|{X}_i,\gamma_i)^{\delta_{1i}}\lambda_{02}(V_i|{X}_i,\gamma_i)^{\delta_{2i}}\exp\left\{-\gamma_i \left[H^o_{01}(V_i e^{-\beta_{01}^{T}X_i})+H^o_{02}(V_i e^{-\beta_{02}^{T}X_i})\right]\right\} \nonumber\\
	&& \,\,\,\,  \Bigg({\lambda_{12}(W_i|V_i,{X}_i,\gamma_i)^{\delta_{3i}}}\exp\left\{-\gamma_i \left[H^o_{12}(W_i e^{-\beta_{12}^{T}X_i})- H^o_{12}(V_i e^{-\beta_{12}^{T}X_i}) \right]\right\}\bigg)^{\delta_{1i}}f(\gamma_i;\sigma) \nonumber\\
	&&=\left\{\gamma_i h^o_{{01}}(V_i e^{-\beta_{01}^{T}X_{i}})e^{-\beta_{01}^{T}X_{i}}\right\}^{\delta_{1i}} \left\{\gamma_i h^o_{{02}}(V_i e^{-\beta_{02}^{T}X_{i}})e^{-\beta_{02}^{T}X_{i}}\right\}^{\delta_{2i}}\exp\left\{-\gamma_i H^o_{01}(V_i e^{-\beta_{01}^{T}X_i})\right\} \nonumber\\
	&&\,\,\,\,  \exp\left\{-\gamma_i H^o_{02}(V_i e^{-\beta_{02}^{T}X_i})\right\} \left[\left\{\gamma_i h^o_{{12}}(W_i e^{-\beta_{12}^{T}X_{i}}) e^{-\beta_{12}^{T}X_{i}} \right\}^{\delta_{3i}}\right. \nonumber\\
	&&\,\,\,\,   \left.\exp\left\{-\gamma_i \left(H^o_{12}(W_i e^{-\beta_{12}^{T}X_{i}})- H^o_{12}(V_i e^{-\beta_{12}^{T}X_{i}})\right)\right\}\right]^{\delta_{1i}}f(\gamma_i;\sigma) \, .
\end{eqnarray}

\section{Conditional Expectations in the EM Algorithm}
\label{ss:conditional_expectations} 
Given Eq. \eqref{likelihood_func}, the complete-data log-likelihood function can be written as
\begin{equation*}
	l(\Omega)=\log L(\Omega)=l(\sigma)+l(\beta_{01},h^o_{01})+l(\beta_{02},h^o_{02})+l(\beta_{12},h^o_{12})
\end{equation*}
where
\begin{equation*}
	l(\sigma)=\frac{1}{n}\sum^{n}_{i=1}\log f(\gamma_i;\sigma)+\frac{1}{n}\sum^{n}_{i=1}(\delta_{1i}+\delta_{2i}+\delta_{3i})\log\gamma_i \, ,
\end{equation*}
\begin{equation*}
	l(\beta_{0k},h^o_{0k})=\frac{1}{n}\sum^{n}_{i=1}\left\{\delta_{ki}\left[\log h^o_{{0k}}(V_i e^{-\beta_{0k}^{T}X_{i}})-\beta_{0k}^{T}X_{i}\right]-\gamma_i H^o_{{0k1}}(V_i e^{-\beta_{0k}^{T}X_{i}})\right\} \, , \, k=1,2 \,  ,
\end{equation*}
and
\begin{eqnarray*}
	l(\beta_{12},h^o_{12})
	&=&\frac{1}{n}\sum^{n}_{i=1}\left\{\delta_{3i}\left[\log h^o_{{12}}(W_i e^{-\beta_{12}^{T}X_{i}})
	-\beta_{12}^{T}X_{i}\right]\right.\\
	&&\left.-\delta_{1i}\gamma_i\left[ H^o_{12}(W_i e^{-\beta_{12}^{T}X_{i}})
	- H^o_{12}(V_i e^{-\beta_{12}^{T}X_{i}})\right] \right\} \, .
\end{eqnarray*}
In the E-step of the EM algorithm, the expectation of the complete log-likelihood function given the observed data, $\mathcal{O}=\{(V_i,W_i,\delta_{1i},\delta_{2i},\delta_{3i},{X}_i):i=1,\ldots,n\}$, and the current values of the estimated parameters is required. The conditional density of $\gamma_i$, given the observed data $\mathcal{O}_i={(V_i,W_i,\delta_{1i},\delta_{2i},\delta_{3i},{X}_i)}$ and the parameter estimates at step $m$, denoted by $\widehat{\Omega}^{(m)}=(\widehat{{\beta}}^{T(m)}_{01},\widehat{{\beta}}^{T(m)}_{12},\widehat{{\beta}}^{T(m)}_{12},\widehat{{h}}^{o,(m)}_{01},\widehat{{h}}^{o,(m)}_{02},\widehat{{h}}^{o,(m)}_{12},\widehat{\sigma}^{(m)})$, is obtained by
\begin{equation*}
	f(\gamma_i|\mathcal{O}_i,\widehat{\Omega}^{(m)})=
	\frac{f(\mathcal{O}_i|\gamma_i,\widehat{\Omega}^{(m)})f(\gamma_i;\widehat{\sigma}^{(m)})}{\int_{0}^{\infty}f(\mathcal{O}_i|\gamma_i,\widehat{\Omega}^{(m)})f(\gamma_i;\widehat{\sigma}^{(m)})d\gamma_i} \, ,
\end{equation*}
where
\begin{eqnarray*}
	f(\mathcal{O}_i|\gamma_i,\widehat{\Omega}^{(m)})
	&=&\big\{\gamma_i \widehat{h}^{o(m)}_{{01}}(V_i e^{-\widehat{\beta}_{01}^{(m)T}X_{i}})e^{-\widehat{\beta}_{01}^{(m)T}X_{i}}\big\}^{\delta_{1i}}
	\big\{\gamma_i \widehat{h}^{o(m)}_{{02}}(V_i e^{-\widehat{\beta}_{02}^{(m)T}X_{i}})e^{-\widehat{\beta}_{02}^{T(m)}X_{i}}\big\}^{\delta_{2i}} \\
	&&\exp\big\{-\gamma_i \big[ \widehat{H}^{o(m)}_{{01}}(V_i e^{-\widehat{\beta}_{01}^{(m)T}X_{i}})
	-\widehat{H}_{02}^{0(m)}(V_i e^{-\widehat{\beta}_{02}^{(m)T}X_{i}})\big]\big\}\\
	&&\big[\big\{ \gamma_i \widehat{h}^{o(m)}_{{12}}(W_i e^{-\widehat{\beta}_{12}^{(m)T}X_{i}}) 
	e^{-\widehat{\beta}_{12}^{(m)T}X_{i}} \big\}^{\delta_{3i}}\\
	&&\exp\big\{-\gamma_i \big( \widehat{H}^{o(m)}_{12}(W_i e^{-\widehat{\beta}_{12}^{(m)T}X_{i}})
	- \widehat{H}^{o(m)}_{12}(V_i e^{-\widehat{\beta}_{12}^{(m)T}X_{i}})\big)\big\}\big]^{\delta_{1i}}.
\end{eqnarray*}
Under the gamma frailty model, the respective conditional expectations of $\gamma_i$, and $\log \gamma_i$ at $m$th step are
\begin{equation*}
	\mathcal{E}^{(m)}_{1i} 
	= E\left( \gamma_i|\mathcal{O},\widehat{\Omega}^{(m)}\right) 
	= \left(D_i+1/\widehat{\sigma}^{(m)} \right) \left\{ 1/\widehat{\sigma}^{(m)}  + \widehat{H}^{o(m)}(V_i,W_i,\widehat{\beta}^{(m)}) \right\}^{-1} 
\end{equation*}
and
\begin{equation*}
	\mathcal{E}^{(m)}_{2i} 
	= E\left(\log\gamma_i|\mathcal{O},\widehat{\Omega}^{(m)}\right)
	=\Psi\left(D_i+1/{\widehat{\sigma}^{(m)}}\right)
	-\log\left\{1/{\widehat{\sigma}^{(m)}}+\widehat{H}^{o(m)}(V_i,W_i, 					\widehat{\beta}^{(m)}) \right\} \, ,
\end{equation*}
where
\begin{equation*}
	\widehat{H}^{0(m)}(V_i,W_i, \widehat{\beta}^{(m)}) 
	= \sum_{k=1,2} \widehat{H}^{o(m)}_{{0k}}(V_i e^{-\widehat{\beta}_{0k}^{(m)T}X_{i}})
	+\delta_{1i}\left\{ \widehat{H}^{o(m)}_{12}(W_i e^{-\widehat{\beta}_{12}^{(m)T}X_{i}})
	-\widehat{H}^{0(m)}_{12}(V_i e^{-\widehat{\beta}_{12}^{(m)T}X_{i}})\right\} \, ,
\end{equation*}		
$D_i=\sum_{k=1}^3\delta_{ji}$, $\Gamma(x)$ is the Gamma function and $\Psi(x)=\Gamma'(x)/\Gamma(x)$ is the digamma function. Therefore, the conditional expectation of the complete-data log-likelihood  equals
\begin{equation*}
	E\left(l(\sigma)|\mathcal{O},\widehat{\Omega}^{(m)}\right)+E\left(l(\beta_{01},h^o_{01})|\mathcal{O},\widehat{\Omega}^{(m)}\right)
	+E\left(l(\beta_{02},h^o_{02})|\mathcal{O},\widehat{\Omega}^{(m)}\right)+E\left(l(\beta_{12},h^o_{12})|\mathcal{O},\widehat{\Omega}^{(m)}\right)
\end{equation*}
where 
\begin{equation*}
	E\left(l(\sigma)|\mathcal{O},\widehat{\Omega}^{(m)}\right)
	=\frac{1}{n}\sum^{n}_{i=1}\left(D_i+\frac{1}{\sigma}\right)\mathcal{E}^{(m)}_{2i}
	-\frac{1}{n \sigma}\sum^{n}_{i=1}\mathcal{E}^{(m)}_{1i}
	-\frac{1}{\sigma}\log \sigma
	-\log\Gamma\left(\frac{1}{\sigma}\right) \, ,
\end{equation*}
\begin{equation*}
	E\left(l(\beta_{0k},h^o_{0k})|\mathcal{O},\widehat{\Omega}^{(m)}\right)=\frac{1}{n}\sum^{n}_{i=1}\left[\delta_{ki}\left\{\log h^o_{{0k}}(V_i e^{-\beta_{0k}^{T}X_{i}})-\beta_{0k}^{T}X_{i}\right\} -\mathcal{E}^{(m)}_{1i} H^o_{{0k}}(V_i e^{-\beta_{0k}^{T}X_{i}})\right] \, , \, k=1,2 \, ,
\end{equation*}
and
\begin{eqnarray*}
	E\left(l(\beta_{12},h^o_{12})|\mathcal{O},\widehat{\Omega}^{(m)}\right)
	&=&\frac{1}{n}\sum^{n}_{i=1}\left[\delta_{3i}\left\{\log h^o_{{12}}(W_i e^{-\beta_{12}^{T}X_{i}})-\beta_{12}^{T}X_{i}\right\} \right.\\
	&&-\delta_{1i}\mathcal{E}^{(m)}_{1i}
	\left.\left\{H^o_{12}(W_i e^{-\beta_{12}^{T}X_{i}})
	- H^o_{12}(V_i e^{-\beta_{12}^{T}X_{i}})\right\} \right] \, .
\end{eqnarray*}


\section{Derivation of the Estimators} 
\label{ss:estimators}
We start with a simple case of piecewise constant hazard function, 
$$\tilde{h}^o_{jk}(t)=\sum^{J_{jk,n}}_{l=1}c_{jk,l} I(t_{jk,l-1}\leq t<t_{jk,l})$$ 
on $[0,M_{jk}]$, where $0=t_{jk,0}<t_{jk,1}< \dots <t_{jk,J_{jk,n}}=M_{jk}$ are equally spaced, $M_{0k}$ denotes an upper bound for $V_i e^{-\beta^T_{jk}X_i}$, $\,k\in\{1,2\}$, and $M_{12}$ denotes an upper bounds for  $W_i e^{-\beta^T_{12}X_i}$. Then, the cumulative piecewise hazard function are
\begin{equation*}
	\tilde{H}^o_{{jk}}(t)=\sum^{J_{jk,n}}_{l=1}c_{jk,l}(t-t_{jk,l-1})I(t_{jk,l-1}\leq t<t_{jk,l})+(M_{jk}/J_{jk,n})\sum^{J_{jk,n}}_{l=1}c_{jk,l} I(t\geq t_{jk,l}) \, .
\end{equation*}
In the following, these piecewise functions will be used to acquire an approximated profile-likelihood function of $\beta_{jk}$, which converges to a limiting function of $\beta_{jk}$, $\,jk\in\{01,02,12\}$, as $n\to\infty$, $J_{jk,n}\to\infty$ and $J_{jk,n}/n\to 0$. The estimators of $\beta_{jk}$, $jk\in\{01,02,12\}$, are defined as the maximizing values of respective smooth approximations of the limiting functions, with respect to $\beta_{jk}$. Given the estimates of the regression coefficients, the proposed smoothing approach also provides estimators of the baseline hazard functions. In the next subsections we provide a detailed derivation of the proposed estimations.

\subsection{Estimating \texorpdfstring{$(\beta^T_{01},h^o_{01})$}{bh} and \texorpdfstring{$(\beta^T_{02},h^o_{02})$}{bh2}}\label{ss:estimating_01}
We begin with the estimation method of $(\beta^T_{01},h^o_{01})$ by  substituting $\tilde{h}^{o}_{{01}}(t)$ and $\tilde{H}^{o}_{{01}}(t)$ in $E\left(l(\beta_{01},h^o_{01})|\mathcal{O},\widehat{\Omega}^{(m)}\right)$:
\begin{eqnarray}\label{eq:E_post_l_beta01}
	\lefteqn{\widetilde{E}\Big(l(\beta_{01},h^o_{01})|\mathcal{O},\widehat{\Omega}^{(m)}\Big)=} \nonumber \\
	&=&\frac{1}{n}\sum_{i=1}^{n}\left\{\delta_{1i}\left[\log\sum^{J_{01,n}}_{l=1}c^{01}_l I(t_{01,l-1}\leq V_i e^{-\beta_{01}^{T}X_{i}}<t_{01,l})-\beta_{01}^{T}X_{i}\right]\right. \nonumber \\
	&&\hspace{14mm} -\mathcal{E}^{(m)}_{1,i}\left.\left[\sum^{J_{01,n}}_{l=1}c_{01,l}(V_i e^{-\beta_{01}^{T}X_{i}}-t_{01,l-1})I(t_{01,l-1}\leq V_i e^{-\beta_{01}^{T}X_{i}}<t_{01,l})\right.\right. \nonumber \\
	&&\hspace{25mm} \left.\left.+\frac{M_{01}}{J_{01,n}}\sum^{J_{01,n}}_{l=1}c_{01,l} I(V_i e^{-\beta_{01}^{T}X_{i}}\geq t_{01,l})\right]\right\} \nonumber \\
	&=&-\frac{1}{n}\sum_{i=1}^{n}\delta_{1i}\beta_{01}^{T}X_{i}+\frac{1}{n}\sum_{i=1}^{n}\delta_{1i}\log\sum^{J_{01,n}}_{l=1}c_{01,l} I(t_{01,l-1}\leq V_i e^{-\beta_{01}^{T}X_{i}}<t_{01,l}) \nonumber \\
	&&-\frac{1}{n}\sum_{i=1}^{n}\mathcal{E}^{(m)}_{1,i}
	\left[\sum^{J_{01,n}}_{l=1}c_{01,l}(V_i e^{-\beta_{01}^{T}X_{i}}-t_{01,l-1})I(t_{01,l-1}\leq V_i e^{-\beta_{01}^{T}X_{i}}<t_{01,l})\right. \nonumber \\
	&&\hspace{25mm} \left.+\frac{M_{01}}{J_{01,n}}\sum^{J_{01,n}}_{l=1}c_{01,l} I(V_i e^{-\beta_{01}^{T}X_{i}}\geq t_{01,l})\right] \nonumber \\
	&=&-\frac{1}{n}\sum_{i=1}^{n}\delta_{1i}\beta_{01}^{T}X_{i}+\frac{1}{n}\sum^{J_{01,n}}_{l=1}\log c_{01,l} \sum_{i=1}^{n}\delta_{1i}I(t_{01,l-1}\leq V_i e^{-\beta_{01}^{T}X_{i}}<t_{01,l}) \nonumber \\
	&&-\frac{1}{n}\sum^{J_{01,n}}_{l=1}c_{01,l}\left[\sum_{i=1}^{n}\mathcal{E}^{(m)}_{1,i}(V_i e^{-\beta_{01}^{T}X_{i}}-t_{01,l-1})I(t_{01,l-1}\leq V_i e^{-\beta_{01}^{T}X_{i}}<t_{01,l})\right. \nonumber \\
	&&\hspace{25mm} \left.+\frac{M_{01}}{J_{01,n}}\sum_{i=1}^{n}\mathcal{E}^{(m)}_{1,i} I(V_i e^{-\beta_{01}^{T}X_{i}}\geq t_{01,l})\right] \, .
\end{eqnarray}
For a given value of  $\beta_{01}$, $c^{(m)}_{01,l}$ that maximizes Eq. (\ref{eq:E_post_l_beta01}), $l={1,2,\ldots,J_{01,n}}$, is given by
\begin{equation}\label{eq:cl}
	\widehat{c}_{01,l}^{(m)}=\frac{\sum_{i=1}^{n}\delta_{1i} I(t_{01,l-1}\leq V_i e^{-\beta_{01}^{T}X_{i}}<t_{01,l})}{\sum_{i=1}^{n}\mathcal{E}^{(m)}_{1,i}(V_i e^{-\beta_{01}^{T}X_{i}}-t_{01,l-1})I(t_{01,l-1}\leq V_i e^{-\beta_{01}^{T}X_{i}}<t_{01,l})+\frac{M_{01}}{J_{01,n}}
		\sum_{i=1}^{n}\mathcal{E}^{(m)}_{1,i} I(V_i e^{-\beta_{01}^{T}X_{i}}\geq t_{01,l})} \, .
\end{equation}
An approximated profile-likelihood function of $\beta_{01}$, based on the piecewise hazard, is obtained by plugging (\ref{eq:cl}) into (\ref{eq:E_post_l_beta01}),
\begin{eqnarray*}
	\lefteqn{l^{*p}_{01}(\beta_{01})=}\\
	&=&-\frac{1}{n}\sum_{i=1}^{n}\delta_{1i}\beta_{01}^{T}X_{i}+\frac{1}{n}\sum^{J_{01,n}}_{l=1}\log(\widehat{c}_{01,l})\sum_{i=1}^{n}\delta_{1i}I(t_{01,l-1}\leq V_i e^{-\beta_{01}^{T}X_{i}}<t_{01,l})\\
	&&-\frac{1}{n}\sum^{J_{01,n}}_{l=1}\widehat{c}_{01,l}\left[\sum_{i=1}^{n}\mathcal{E}^{(m)}_{1,i}(V_i e^{-\beta_{01}^{T}X_{i}}-t_{01,l-1})I(t_{01,l-1}\leq V_i e^{-\beta_{01}^{T}X_{i}}<t_{01,l})\right.\\
	&&\hspace{30mm}\left.+\dfrac{M_{01}}{J_{01,n}}\sum_{i=1}^{n}\mathcal{E}^{(m)}_{1,i} I(V_i e^{-\beta_{01}^{T}X_{i}}\geq t_{01,l})\right]\\
	&=&-\frac{1}{n}\sum_{i=1}^{n}\delta_{1i}\beta_{01}^{T}X_{i}-\frac{1}{n}\sum_{l=1}^{J_{01,n}}\sum_{i=1}^{n}\delta_{1i}I(t_{01,l-1}\leq V_i e^{-\beta_{01}^{T}X_{i}}<t_{01,l})\\
	&&+\frac{1}{n}\sum^{J_{01,n}}_{l=1}\sum_{i=1}^{n}\delta_{1i} I(t_{01,l-1}\leq V_i e^{-\beta_{01}^{T}X_{i}}<t_{01,l})\left[{\log\left\{{\sum_{j=1}^{n}\delta_{1j} I(t_{01,l-1}\leq V_j e^{-\beta_{01}^{T}X_{j}}<t_{01,l})}\right\}}\right.\\
	&&-\log\left\{\sum_{j=1}^{n}\mathcal{E}^{(m)}_{1,j}(V_j e^{-\beta_{01}^{T}X_{j}}-t_{01,l-1})I(t_{01,l-1}\leq V_j e^{-\beta_{01}^{T}X_{j}}<t_{01,l})\right.\\
	&&\hspace{30mm}\left.\left.+\frac{M_{01}}{J_{01,n}}\sum_{j=1}^{n}\mathcal{E}^{(m)}_{1,j} I(V_j e^{-\beta_{01}^{T}X_{j}}\hspace{-1mm}\geq\hspace{-1mm} t_{01,l})\right\}\right] \, .
\end{eqnarray*}
By multiplying the numerator and the denominator by $\frac{J_{01,n}}{nM_{01}}$ we obtain
\begin{eqnarray*}
	\lefteqn{l^{*p}_{01}(\beta_{01})=}\\
	&=& -\frac{1}{n}\sum_{i=1}^{n}\delta_{1i}\beta_{01}^{T}X_{i}-\frac{1}{n}\sum_{l=1}^{J_{01,n}}\sum_{i=1}^{n}\delta_{1i}I(t_{01,l-1}\leq V_i e^{-\beta_{01}^{T}X_{i}}<t_{01,l})\\
	&&+\frac{1}{n}\sum^{J_{01,n}}_{l=1}\left(\sum_{i=1}^{n}\delta_{1i} I(t_{01,l-1}\leq V_i e^{-\beta_{01}^{T}X_{i}}<t_{01,l})\right.
	\\
	&&\hspace{15mm} \left[\log\left\{{\frac{J_{01,n}}{nM_{01}}\sum_{j=1}^{n}\delta_{1j} I(t_{01,l-1}\leq V_j e^{-\beta_{01}^{T}X_{j}}<t_{01,l})}\right\}\right.\\
	&&\hspace{17mm} -\left.\left.\log\left\{\frac{J_n}{nM_{01}}\sum_{j=1}^{n}\mathcal{E}^{(m)}_{1,j}(V_j e^{-\beta_{01}^{T}X_{j}}-t_{01,l-1})I(t_{01,l-1}\leq V_j e^{-\beta_{01}^{T}X_{j}}<t_{01,l})\right.\right.\right.\\
	&&\hspace{30mm} \left.\left.\left.+\frac{1}{n}
	\sum_{j=1}^{n}\mathcal{E}^{(m)}_{1,j} I(V_j e^{-\beta_{01}^{T}X_{j}}\geq t_{01,l})\right\}\right]\right)\\
	&=& -\frac{1}{n}\sum_{i=1}^{n}\delta_{1i}\beta_{01}^{T}X_{i}-\frac{1}{n}\sum_{l=1}^{J_{01,n}}\sum_{i=1}^{n}\delta_{1i}I(t_{01,l-1}\leq V_i e^{-\beta_{01}^{T}X_{i}}<t_{01,l})\\
	&& +\sum^{J_{01,n}}_{l=1}\left[\frac{1}{n}\sum_{i=1}^{n}\delta_{1i}I(t_{01,l-1}\leq V_i e^{-\beta_{01}^{T}X_{i}}<t_{01,l}) \log\left\{\frac{J_{01,n}}{nM_{01}}\sum_{j=1}^{n}\delta_{1j}I(t_{01,l-1}\leq V_j e^{-\beta_{01}^{T}X_{j}}<t_{01,l})\right\}\right.\\
	&&\hspace{15mm} -\frac{1}{n}\sum_{i=1}^{n}\delta_{1i}I(t_{01,l-1}\leq V_i e^{-\beta_{01}^{T}X_{i}}<t_{01,l})\log\left\{\frac{1}{n}\sum_{j=1}^{n}\mathcal{E}^{(m)}_{1,j} I(V_j e^{-\beta_{01}^{T}X_{j}}\geq t_{01,l})\right.\\
	&&\hspace{40mm} \left.\left.+\frac{J_{01,n}}{nM_{01}}\sum_{j=1}^{n}\mathcal{E}^{(m)}_{1,j}(V_j e^{-\beta_{01}^{T}X_{j}}-t_{01,l-1})I(t_{01,l-1}\leq V_j e^{-\beta_{01}^{T}X_{j}}<t_{01,l})\right\}\right] \, .
\end{eqnarray*}
The term $-\frac{1}{n}\sum_{l=1}^{J_{01,n}}\sum_{i=1}^{n}\delta_{1i}I(t_{01,l-1}\leq V_i e^{-\beta_{01}^{T}X_{i}}<t_{01,l})$  equals  $-\frac{1}{n}\sum_{i=1}^{n}\delta_{1i},$  because for subject $i$, $V_i e^{-\beta_{01}^{T}X_{i}}$ must belong to one of the intervals. Therefore, for maximizing $l^{*p}_{01}(\beta_{01})$ with respect to $\beta_{01}$ it is enough to consider
\begin{eqnarray}
	\lefteqn{l^{p}_{01}(\beta_{01})=} \nonumber\\
	&=& -\frac{1}{n}\sum_{i=1}^{n}\delta_{1i}\beta_{01}^{T}X_{i}\nonumber\\
	&&	+\sum^{J_{01,n}}_{l=1}\left[\frac{1}{n}\sum_{i=1}^{n}\delta_{1i}I(t_{01,l-1}\leq V_i e^{-\beta_{01}^{T}X_{i}}<t_{01,l}) \log\left\{\frac{J_{01,n}n}{nM_{01}}\sum_{j=1}^{n}\delta_{1j}I(t_{01,l-1}\leq V_j e^{-\beta_{01}^{T}X_{j}}<t_{01,l})\right\}\right.\nonumber\\
	&&\hspace{14mm} -\frac{1}{n}\sum_{i=1}^{n}\delta_{1i}I(t_{01,l-1}\leq V_i e^{-\beta_{01}^{T}X_{i}}<t_{01,l})\log\left\{\frac{1}{n}\sum_{j=1}^{n}\mathcal{E}^{(m)}_{1,j} I(V_j e^{-\beta_{01}^{T}X_{j}}\geq t_{01,l})\right.\nonumber\\
	&&\hspace{3cm} +\left.\left.\frac{J_{01,n}}{nM_{01}}\sum_{j=1}^{n}\mathcal{E}^{(m)}_{1,j}(V_j e^{-\beta_{01}^{T}X_{j}}-t_{01,l-1})I(t_{01,l-1}\leq V_j e^{-\beta_{01}^{T}X_{j}}<t_{01,l})\right\}\right] \, .
\end{eqnarray}

In the spirit of  \cite{zeng2007efficient} and \cite{liu2013kernel}, as $n \rightarrow \infty$, $J_{01,n} \rightarrow \infty$ and $J_{01,n}/n \rightarrow 0$, it can be shown that $l^{p}_{01}(\beta_{01})$ converges uniformly in $\beta_{01}$ to a limiting function $l_{01}(\beta_{01})$, where
\begin{equation*}
	l_{01}(\beta_{01})=E\left.\left[-\delta_1\beta_{01}^{T}X+\delta_1\log\left\{\frac{\frac{\partial}{\partial t} P(\delta_1=1,V\exp\{-\beta_{01}^T X\}\leq t)|_{t=v\exp\{-\beta_{01}^T X\}}}{E\left[\widehat{\gamma}\,I\left(V\exp\{-\beta_{01}^T X\}\geq v\exp\{-\beta_{01}^T X\}\right)\right]} \right\}\right.\right] \, ,
\end{equation*}
and $\widehat{\gamma}=E(\gamma_1|\mathcal{O}_1,\widehat{\Omega})$.

Moreover, $\frac{\partial}{\partial t} P\left(\delta_1=1,V\exp\{-\beta_{01}^T X\}\leq t\right)$ can be approximated by
\begin{equation*}
	\frac{1}{t n a_{01}}\sum_{j=1}^{n}\delta_{1j}K\left(\frac{\log\{V_j\}-\beta_{01}^T X_{j}-\log t\}}{a_{01}} \right) \, ,
\end{equation*}
and $E\left[\widehat{\gamma}\,I(V\exp\{-\beta_{01}^T X\}\geq t)\right]$ can be approximated by
\begin{equation*}
	\frac{1}{n}\sum_{j=1}^{n}\mathcal{E}^{(m)}_{1,j}\int_{\log t}^{\infty}\frac{1}{a_{01}}K\left(\frac{\log\{V_j\}-\beta_{01}^T X_{j}-s}{a_{01}} \right)ds,
\end{equation*}
where $K(\cdot)$ is a kernel function and $a_{01}$ is the bandwidth. By setting $x=\frac{\log\{V_j\}-\beta_{01}^T X_{j}-s}{a_{01}}$
we get
\begin{equation*}
	\frac{1}{n}\sum_{j=1}^{n}\mathcal{E}^{(m)}_{1,j}\int_{-\infty}^{(\log\{V_j\}-\beta_{01}^T X_{j}-\log t)/{a_{01}}}K\left(x\right)dx \, .
\end{equation*}
Thus, the term $E\left.\left[\delta_1\log\left\{\dfrac{\frac{\partial}{\partial t} P(\delta_1=1,V\exp\{-\beta_{01}^T X\}\leq t)|_{t=v\exp\{-\beta_{01}^T X\}}}{E[\widehat{\gamma}\,I(V\exp\{-\beta_{01}^T X\}\geq v\exp\{-\beta_{01}^T X\})]} \right\}\right.\right]$ can be approximated by
\begin{equation*}
	\frac{1}{n}\sum^{n}_{i=1}\delta_{1i}\log\left\{\dfrac{ \left[V_i\exp\{-\beta_{01}^T X_{i}\}n a_{01}\right]^{-1}\sum_{j=1}^{n}\delta_{1j}K\left(\dfrac{\log\{V_j\}-\beta_{01}^T X_{j}-\left(\log\{V_i\}-\beta_{01}^T X_{i}\right)}{a_{01}} \right)}{(n)^{-1}\sum_{j=1}^{n}\mathcal{E}^{(m)}_{1,j}\int_{-\infty}^{[\log\{V_j\}-\beta_{01}^T X_{j}-(\log\{V_i\}-\beta_{01}^T X_{i})]/{a_{01}}}K\left(x\right)dx}\right\} \, .
\end{equation*}

Based on the approximations above, $l^{p}_{01}(\beta_{01})$ 
can be approximated by 
\begin{eqnarray*}
	\lefteqn{l^{s}_{01}(\beta_{01})} \\
	&=& -\frac{1}{n}\sum^{n}_{i=1}\delta_{1i}\beta_{01}^{T}X_{i} \\
	&& +\frac{1}{n}\sum^{n}_{i=1}\delta_{1i}\log\left\{\frac{ \left[\exp(\log\{V_i\}-\beta_{01}^T X_{i})n a_{01}\right]^{-1}\sum_{j=1}^{n}\delta_{1j}K\left(\frac{\log\{V_j\}-\beta_{01}^T X_{j}-\log\{V_i\}+\beta_{01}^T X_{i}}{a_{01}} \right)}{(n)^{-1}\sum_{j=1}^{n}\mathcal{E}^{(m)}_{1,j}\int_{-\infty}^{[\log\{V_j\}-\beta_{01}^T X_{j}-\log\{V_i\}+\beta_{01}^T X_{i}]/{a_{01}}}K\left(x\right)dx}\right\}\\
	&=& -\frac{1}{n}\sum^{n}_{i=1}\delta_{1i}\log\{V_i\}+\frac{1}{n}\sum^{n}_{i=1}\delta_{1i}\log\left\{\frac{1}{na_{01}}\sum^{n}_{j=1}\delta_{1j}K\left(\frac{\log\{V_j\}-\beta_{01}^T X_{j}-\log\{V_i\}+\beta_{01}^T X_{i}}{a_{01}}\right)\right\} \\
	&& -\frac{1}{n}\sum^{n}_{i=1}\delta_{1i}\log\left\{\frac{1}{n}\sum^{n}_{j=1}\mathcal{E}^{(m)}_{1,j}\int_{-\infty}^{\left[\log\{V_j\}-\beta_{01}^T X_{j}-\log\{V_i\}+\beta_{01}^T X_{i}\right]/a_{01}}K(s)ds\right\} \, .
\end{eqnarray*}
Since $K(\cdot)$ is a smooth function, the maximizer of $l^{s}_{01}(\beta_{01})$, denoted by $\widehat{\beta}_{01}$, can be found by gradient-based algorithms. Let $\widehat{\beta}_{01}^{(m)}$ denote the maximizer of $l^{s}_{01}(\beta_{01})$ at step $m$. Given $\widehat{\beta}_{01}^{(m)}$, a smooth estimator of $h^o_{{01}}(t)$ can be obtained by
\begin{equation*}
	\widehat{h}^{o(m)}_{01}\left(t;\widehat{\beta}^{(m)}_{01} \right)=\dfrac{(n a_{01} t)^{-1}\sum^{n}_{j=1}\delta_{1j}K\left(\dfrac{\log\{V_j\}-\widehat{\beta}_{01}^{T(m)} X_{j}-\log(t)}{a_{01}}\right)}{(n)^{-1}\sum^{n}_{j=1}\mathcal{E}^{(m)}_{1,j}\int^{[\log\{V_j\}-\widehat{\beta}_{01}^{T(m)} X_{j}-\log(t)]/a_{01}}_{-\infty}K(u)du} \, ,\, t>0 \, ,
\end{equation*}
and the estimator of ${H}_{01}^{o}(t)$, is obtained by 
\begin{equation*}
	\widehat{H}_{01}^{o(m)}\left(t;\widehat{\beta}_{01}^{(m)}\right)=\int^{t}_{0}\widehat{h}^{o(m)}_{{01}}\left(s;\widehat{\beta}_{01}^{(m)}\right)ds \, .
\end{equation*} 

Similarly, due to the symmetry between transition $0\to1$ and $0\to 2$, we get
\begin{equation*}
	\widehat{h}^{o(m)}_{{02}}\left(t;\widehat{\beta}^{(m)}_{02}\right)=\dfrac{ (n a_{02} t)^{-1}\sum^{n}_{j=1}\delta_{2j}K\left(\dfrac{\log\{V_j\}-\widehat{\beta}_{02}^{T(m)} X_{j}-\log(t)}{a_{02}}\right)}{(n)^{-1}\sum^{n}_{j=1}\mathcal{E}^{(m)}_{1,j}\int^{[\log\{V_j\}-\widehat{\beta}_{02}^{T(m)} X_{j}-\log(t)]/a_{02}}_{-\infty}K(u)du}\, ,\, t>0 \, ,
\end{equation*}
where $\widehat{\beta}_{02}^{(m)}$ is the maximizer of $l^{s}_{02}(\beta_{02})$ at step $m$ and
\begin{eqnarray*}
	\lefteqn{l^{s}_{02}(\beta_{02})} \\
	&=& -\frac{1}{n}\sum^{n}_{i=1}\delta_{2i}\log\{V_i\}+\frac{1}{n}\sum^{n}_{i=1}\delta_{2i}\log\left\{\dfrac{1}{n a_{02}}\sum^{n}_{j=1}\delta_{2j}K\left(\frac{\log\{V_j\}-{\beta}_{02}^T X_{j}-\log\{V_i\}+{\beta}_{02}^T X_{i}}{a_{02}}\right)\right\}\\
	&& -\frac{1}{n}\sum^{n}_{i=1}\delta_{2i}\log\left\{\frac{1}{n}\sum^{n}_{j=1}\mathcal{E}^{(m)}_{1,j}\int_{-\infty}^{[\log\{V_j\}-{\beta}_{02}^T X_{j}-\log\{V_i\}+{\beta}_{02}^T X_{i}]/a_{02}}K(s)ds\right\} \, .
\end{eqnarray*}

\subsection{Estimating \texorpdfstring{$(\beta^T_{12},h^o_{12})$}{bh}}
\label{ss:estimating_12}
In the illness-death setting, the death time after disease diagnosis is truncated by the age at diagnosis. Therefore, the kernel smoothing method described previously should be adapted to accommodate left truncation. To this end, we start with substituting $\tilde{h}^{o}_{{12}}(t)$ and $\tilde{H}^{o}_{{12}}(t)$ into $E\left(l(\beta_{12},h^o_{12})|\mathcal{O},\widehat{\Omega}^{(m)}\right)$ and get
\begin{eqnarray}\label{eq:E_post_l_beta12}
	\lefteqn{\tilde{E}\left(l_n(\beta_{12},h^o_{12})|\mathcal{O},\widehat{\Omega}^{(m)}\right)} \nonumber\\
	&=& -\frac{1}{n}\sum_{i=1}^{n}\delta_{3i}\beta_{12}^{T}X_{i}+\frac{1}{n}\sum_{i=1}^{n}\delta_{3i}\log\left\{\sum^{J_{12,n}}_{l=1}c_{12,l} I(t_{12,l-1}\leq W_i e^{-\beta_{12}^{T}X_{i}}<t_{12,l})\right\} \nonumber\\
	&& -\frac{1}{n}\sum_{i=1}^{n}\delta_{1i}\mathcal{E}^{(m)}_{1,i}\left\{\sum^{J_{12,n}}_{l=1}c_{12,l}(W_i e^{-\beta_{12}^{T}X_{i}}-t_{12,l-1})I(t_{12,l-1}\leq W_i e^{-\beta_{12}^{T}X_{i}}<t_{12,l})\right. \nonumber\\
	&&\hspace{32mm} -\left.\sum^{J_{12,n}}_{l=1}c_{12,l}(V_i e^{-\beta_{12}^{T}X_{i}}-t_{12,l-1})I(t_{12,l-1}\leq V_i e^{-\beta_{12}^{T}X_{i}}<t_{12,l})\right. \nonumber\\
	&&\hspace{32mm} \left.+\frac{M_{12}}{J_{12,n}}\sum^{J_{12,n}}_{l=1}c_{12,l} I(W_i e^{-\beta_{12}^{T}X_{i}}\geq t_{12,l})-\frac{M_{12}}{J_{12,n}}\sum^{J_{12,n}}_{l=1}c_{12,l} I(V_i e^{-\beta_{12}^{T}X_{i}}\geq t_{12,l})\right\} \nonumber\\
	&=& -\frac{1}{n}\sum_{i=1}^{n}\delta_{3i}\beta_{12}^{T}X_{i}+\frac{1}{n}\sum^{J_{12,n}}_{l=1}\log(c_{12,l})\sum_{i=1}^{n}\delta_{3i}I(t_{12,l-1}\leq W_i e^{-\beta_{12}^{T}X_{i}}<t_{12,l}) \nonumber\\
	&& -\frac{1}{n}\sum^{J_{12,n}}_{l=1}c_{12,l}\left[\sum_{i=1}^{n}\delta_{1i}\mathcal{E}^{(m)}_{1,i}(W_i e^{-\beta_{12}^{T}X_{i}}-t_{12,l-1})I(t_{12,l-1}\leq W_i e^{-\beta_{12}^{T}X_{i}}<t_{12,l})\right. \nonumber\\
	&&\hspace{24mm} \left.-\sum_{i=1}^{n}\delta_{1i}\mathcal{E}^{(m)}_{1,i}(V_i e^{-\beta_{12}^{T}X_{i}}-t_{12,l-1})I(t_{12,l-1}\leq V_i e^{-\beta_{12}^{T}X_{i}}<t_{12,l})\right. \nonumber\\
	&&\hspace{24mm} \left.+\frac{M_{12}}{J_{12,n}}\sum_{i=1}^{n}\delta_{1i}\mathcal{E}^{(m)}_{1,i} I(W_i e^{-\beta_{12}^{T}X_{i}}\geq t_{12,l})-\frac{M_{12}}{J_{12,n}}\sum_{i=1}^{n}\delta_{1i}\mathcal{E}^{(m)}_{1,i} I(V_i e^{-\beta_{12}^{T}X_{i}}\geq t_{12,l}) \right] \, . 
\end{eqnarray}
For  a given value of $\beta_{12}$, maximizing Eq. (\ref{eq:E_post_l_beta12}) with respect to $c_{12,l}$, $l=1,2,\ldots,J_n$, gives the following maximizers
\begin{eqnarray}\label{eq:cl12}
	\widehat{c}_{12,l}^{(m)}
	&=&\left\{{\sum_{i=1}^{n}\delta_{3i} I(t_{12,l-1}\leq W_i e^{-\beta_{12}^{T}X_{i}}<t_{12,l})}\right\} \nonumber\\
	&&\hspace{-1mm}\left[\sum_{i=1}^{n}\delta_{1i}\mathcal{E}^{(m)}_{1,i}(W_i e^{-\beta_{12}^{T}X_{i}}-t_{12,l-1})I(t_{12,l-1}\leq W_i e^{-\beta_{12}^{T}X_{i}}<t_{12,l})\right. \nonumber\\
	&&\left.-\sum_{i=1}^{n}\delta_{1i}\mathcal{E}^{(m)}_{1,i}(V_i e^{-\beta_{12}^{T}X_{i}}-t_{12,l-1})I(t_{12,l-1}\leq V_i e^{-\beta_{12}^{T}X_{i}}<t_{12,l})\right. \nonumber\\
	&&\left.+\frac{M_{12}}{J_{12,n}}
	\sum_{i=1}^{n}\delta_{1i}\mathcal{E}^{(m)}_{1,i} I(W_i e^{-\beta_{12}^{T}X_{i}}\geq t_{12,l})-\frac{M_{12}}{J_{12,n}}
	\sum_{i=1}^{n}\delta_{1i}\mathcal{E}^{(m)}_{1,i} I(V_i e^{-\beta_{12}^{T}X_{i}}\geq t_{12,l}) \right]^{-1} \, .
\end{eqnarray}
The piecewise-based profile-likelihood function of $\beta_{12}$ is obtained  by plugging Eq. (\ref{eq:cl12}) into Eq. (\ref{eq:E_post_l_beta12}),
\begin{eqnarray*}
	\lefteqn{l^{*p}_{12}(\beta_{12})}\\
	&=& -\frac{1}{n}\sum_{i=1}^{n}\delta_{3i}\beta_{12}^{T}X_{i}-\frac{1}{n}\sum^{J_{12,n}}_{l=1}\sum_{i=1}^{n}\delta_{3i} I(t_{12,l-1}\leq W_i e^{-\beta_{12}^{T}X_{i}}<t_{12,l})\\
	&& +\frac{1}{n}\sum^{J_{12,n}}_{l=1}\left(\left[\sum_{i=1}^{n}\delta_{3i}I(t_{12,l-1}\leq W_i e^{-\beta_{12}^{T}X_{i}}<t_{12,l})\right]\log(\widehat{c}^{12}_l)\right)\\
	&=& -\frac{1}{n}\sum_{i=1}^{n}\delta_{3i}\beta_{12}^{T}X_{i}-\frac{1}{n}\sum_{i=1}^{n}\delta_{3i}\\
	&& +\frac{1}{n}\sum^{J_{12,n}}_{l=1}\left( \sum_{i=1}^{n}\delta_{3i} I(t_{12,l-1}\leq W_i e^{-\beta_{12}^{T}X_{i}}<t_{12,l})\right.\\
	&&\hspace{8mm} {\left[\log\left\{{\frac{J_{12,n}}{nM_{12}}{\sum_{j=1}^{n}\delta_{3j} I(t_{12,l-1}\leq W_j e^{-\beta_{12}^{T}X_{j}}<t_{12,l})}}\right\}\right.}\\
	&&\hspace{8mm} \left.\left.-\log\left\{\frac{J_{12,n}}{nM_g}\sum_{j=1}^{n}\mathcal{E}^{(m)}_{1,j}\delta_{1j}(W_j e^{-\beta_{12}^{T}X_{j}}-t_{12,l-1})I(t_{12,l-1}\leq W_j e^{-\beta_{12}^{T}X_{j}}<t_{12,l})\right.\right.\right.\\
	&&\hspace{21mm} \left.\left.-\left.\frac{J_{12,n}}{nM_{12}}\sum_{j=1}^{n}\mathcal{E}^{(m)}_{1,j}\delta_{1j}(V_j e^{-\beta_{12}^{T}X_{j}}-t_{12,l-1})I(t_{12,l-1}\leq V_j e^{-\beta_{12}^{T}X_{j}}<t_{12,l})\right.\right.\right.\\
	&&\hspace{21mm} \left.\left.\left.+\frac{1}{n}
	\sum_{j=1}^{n}\mathcal{E}^{(m)}_{1,j}\delta_{1j} I(W_j e^{-\beta_{12}^{T}X_{j}}\geq t_{12,l})-\frac{1}{n}
	\sum_{j=1}^{n}\mathcal{E}^{(m)}_{1,j}\delta_{1j} I(V_j e^{-\beta_{12}^{T}X_{j}}\geq t_{12,l})\right\} \right]\right) \, .
\end{eqnarray*}
The term $ I(W_j e^{-\beta_{12}^{T}X_{j}}\geq t_{12,l})-I(V_j e^{-\beta_{12}^{T}X_{j}}\geq t_{12,l})$ consists of the following  possible cases. If $V_j e^{-\beta_{12}^{T}X_{j}}<W_j e^{-\beta_{12}^{T}X_{j}}< t_{12,l}$, then  $I(W_j e^{-\beta_{12}^{T}X_{j}}\geq t_{12,l})-I(V_j e^{-\beta_{12}^{T}X_{j}}\geq t_{12,l})=0-0=0$.  If $t_{12,l}\leq V_j e^{-\beta_{12}^{T}X_{j}}< W_j e^{-\beta_{12}^{T}X_{j}} $, then $I(W_j e^{-\beta_{12}^{T}X_{j}}\geq t_{12,l})-I(V_j e^{-\beta_{12}^{T}X_{j}}\geq t_{12,l})=1-1=0\,.$ If $V_j e^{-\beta_{12}^{T}X_{j}}<t_{12,l}\leq W_j e^{-\beta_{12}^{T}X_{j}} $, then  $I(W_j e^{-\beta_{12}^{T}X_{j}}\geq t_{12,l})-I(V_j e^{-\beta_{12}^{T}X_{j}}\geq t_{12,l})=1-0=1$.
Thus, we can write 
\begin{equation*}
	I(W_j e^{-\beta_{12}^{T}X_{j}}\geq t_{12,l})-I(V_j e^{-\beta_{12}^{T}X_{j}}\geq t_{12,l})=I(V_j e^{-\beta_{12}^{T}X_{j}}<t_{12,l}\leq W_j e^{-\beta_{12}^{T}X_{j}})
\end{equation*}
and as a consequence,
\begin{eqnarray*}
	\lefteqn{\frac{1}{n}\sum_{j=1}^{n}\mathcal{E}^{(m)}_{1,j}\delta_{1j} I(W_j e^{-\beta_{12}^{T}X_{j}}\geq t_{12,l})-\frac{1}{n}
		\sum_{j=1}^{n}\mathcal{E}^{(m)}_{1,j}\delta_{1j} I(V_j e^{-\beta_{12}^{T}X_{j}}\geq t_{12,l})}\\
	&&= \frac{1}{n}\sum_{j=1}^{n}\mathcal{E}^{(m)}_{1,j}\delta_{1j}I(V_j e^{-\beta_{12}^{T}X_{j}}<t_{12,l}\leq W_j e^{-\beta_{12}^{T}X_{j}}).
\end{eqnarray*}
After plugging in this term into $l^{p}_{12}(\beta_{12})$, the profile log-likelihood becomes
\begin{eqnarray*}
	\lefteqn{l^{*p}_{12}(\beta_{12})}\\
	&=& -\frac{1}{n}\sum_{i=1}^{n}\delta_{3i}\beta_{12}^{T}X_{i}-\frac{1}{n}\sum_{i=1}^{n}\delta_{3i}\\[-1.5\jot]
	&& +\frac{1}{n}\sum^{J_{12,n}}_{l=1}\left( \sum_{i=1}^{n}\delta_{3i} I(t_{12,l-1}\leq W_i e^{-\beta_{12}^{T}X_{i}}<t_{12,l})\right.\\
	&&\hspace{20mm} {\left[\log\left\{{\frac{J_{12,n}}{nM_g}{\sum_{j=1}^{n}\delta_{3j} I(t_{12,l-1}\leq W_j e^{-\beta_{12}^{T}X_{j}}<t_{12,l})}}\right\}\right.}\\
	&&\hspace{21mm} \left.\left.-\log\left\{\frac{J_{12,n}}{nM_{12}}\sum_{j=1}^{n}\mathcal{E}^{(m)}_{1,j}\delta_{1j}(W_j e^{-\beta_{12}^{T}X_{j}}-t_{12,l-1})I(t_{12,l-1}\leq W_j e^{-\beta_{12}^{T}X_{j}}<t_{12,l})\right.\right.\right.\\
	&&\hspace{35mm}-\frac{J_{12,n}}{nM_{12}}\sum_{j=1}^{n}\mathcal{E}^{(m)}_{1,j}\delta_{1j}(V_j e^{-\beta_{12}^{T}X_{j}}-t_{12,l-1})I(t_{12,l-1}\leq V_j e^{-\beta_{12}^{T}X_{j}}<t_{12,l})\\
	&&\hspace{35mm} \left.\left.\left.+\frac{1}{n}
	\sum_{j=1}^{n}\mathcal{E}^{(m)}_{1,j}\delta_{1j} I(V_j e^{-\beta_{12}^{T}X_{j}}<t_{12,l}\leq W_j e^{-\beta_{12}^{T}X_{j}})\right\}\right]\right) \, .
\end{eqnarray*}
By dropping the term $-\frac{1}{n}\sum_{i=1}^{n}\delta_{3i}$, and using similar arguments given earlier, as $n \rightarrow \infty$, $J_{12,n} \rightarrow \infty$ and $J_{12,n}/n \rightarrow 0$, $l^{*p}_{12}(\beta_{12})$ converges uniformly in $\beta_{12}$ to a limiting function $l_{12}(\beta_{12})$,
\begin{equation*}
	l_{12}(\beta_{12})=E\left[-\delta_3\beta^T_{12}X_{} +\delta_3\log\left\{\frac{\frac{\partial}{\partial t}P(\delta_3=1,W e^{-\beta^T_{12}X_{}}\leq t)|_{t=w e^{-\beta^T_{12}X_{}}}}{E[\widehat{\gamma}\delta_1 I(V e^{-\beta^T_{12}X_{}} < w e^{-\beta^T_{12}X_{}}\leq W e^{-\beta^T_{12}X_{}})]} \right\}\right] \, .
\end{equation*}

The term $\frac{\partial}{\partial t}P(\delta_3=1,W e^{-\beta^T_{12}X_{}}\leq t)$ can be approximated by
\begin{equation*}
	\frac{1}{t n a_{12}}\sum_{j=1}^{n}\delta_{3j}K\left(\frac{\log\{W_j\}-\beta_{12}^T X_{j}-\log t}{a_{12}} \right) \, ,
\end{equation*}
and the term 
$E\left[\widehat{\gamma}\delta_1 I(V e^{-\beta^T_{12}X_{}} < t\leq W e^{-\beta^T_{12}X_{}})\right]$
can be approximated by 
\begin{equation*}
	\frac{1}{n}\sum_{j=1}^{n}\mathcal{E}^{(m)}_{1,j}\delta_{1j}\int_{\left(\log\{V_j\}-\beta_{12}^T X_{j}-\log t\right)/{a_{12}}}^{\left(\log\{W_j\}-\beta_{12}^T X_{j}-\log t\right)/{a_{12}}}K\left(x\right)dx \, .
\end{equation*}
Clearly, this approximation takes into account the left truncation of the survival time by age at disease diagnosis.
Thus, the term 
\begin{equation*}
	E\left[\delta_3\log\left\{\frac{\frac{\partial}{\partial t}P(\delta_3=1,W e^{-\beta^T_{12}X_{}}\leq t)|_{t=w e^{-\beta^T_{12}X_{}}}}{E[\widehat{\gamma}\delta_1 I(V e^{-\beta^T_{12}X_{}} < w e^{-\beta^T_{12}X_{}}\leq W e^{-\beta^T_{12}X_{}})]} \right\}\right]
\end{equation*}
can be approximated by
\begin{equation*}
	\frac{1}{n}\sum^{n}_{i=1}\delta_{3i}\log\left\{\dfrac{ \left[W_i\exp\{-\beta_{12}^T X_{i}\}n a_{12}\right]^{-1}\sum_{j=1}^{n}\delta_{3j}K\left(\dfrac{\log\{W_j\}-\beta_{12}^T X_{j}-\log\{W_i\}+\beta_{12}^T X_{i}}{a_{12}} \right)}{n^{-1}\sum_{j=1}^{n}\mathcal{E}^{(m)}_{1,j}\delta_{1j}\int_{\left(\log\{V_j\}-\beta_{12}^T X_{j}-\log \{W_i\}+\beta_{12}^T X_{i}\right)/{a_{12}}}^{\left(\log\{W_j\}-\beta_{12}^T X_{j}-\log \{W_i\}+\beta_{12}^T X_{i}\right)/{a_{12}}}K\left(x\right)dx}\right\} \, .
\end{equation*}
Combining the approximations above, $l^{p}_{12}(\beta_{12})$ is approximated by the following smooth function
\begin{eqnarray*}
	\lefteqn{l^{s}_{12}(\beta_{12})}\\
	&=& -\frac{1}{n}\sum^{n}_{i=1}\delta_{3i}\beta^T_{12}X_{i}\\
	&& +\frac{1}{n}\sum^{n}_{i=1}\delta_{3i}\log\left\{\dfrac{ \left[W_i\exp\{-\beta_{12}^T X_{i}\}n a_{12}\right]^{-1}\sum_{j=1}^{n}\delta_{3j}K\left(\frac{\log\{W_j\}-\beta_{12}^T X_{j}-\log\{W_i\}+\beta_{12}^T X_{i}}{a_{12}} \right)}{n^{-1}\sum_{j=1}^{n}\mathcal{E}^{(m)}_{1,j}\delta_{1j}\int_{\left(\log\{V_j\}-\beta_{12}^T X_{j}-\log \{W_i\}+\beta_{12}^T X_{i}\right)/{a_{12}}}^{\left(\log\{W_j\}-\beta_{12}^T X_{j}-\log \{W_i\}+\beta_{12}^T X_{i}\right)/{a_{12}}}K\left(x\right)dx}\right\}\\
	&=& -\frac{1}{n}\sum^{n}_{i=1}\delta_{3i}\log\{W_i\}+\frac{1}{n}\sum^{n}_{i=1}\delta_{3i}\log\left\{\frac{1}{n a_{12} }\sum^{n}_{j=1}\delta_{3j}K\left(\frac{\log\{W_j\}-\beta_{12}^T X_{j}-\log\{W_i\}+\beta_{12}^T X_{i}}{a_{12}} \right)\right\}\\
	&& -\frac{1}{n}\sum^{n}_{i=1}\delta_{3i}\log\left\{\frac{1}{n }\sum^{n}_{j=1}\mathcal{E}^{(m)}_{1,j}\delta_{1j}\int_{\left(\log\{V_j\}-\beta_{12}^T X_{j}-\log \{W_i\}+\beta_{12}^T X_{i}\right)/{a_{12}}}^{\left(\log\{W_j\}-\beta_{12}^T X_{j}-\log \{W_i\}+\beta_{12}^T X_{i}\right)/{a_{12}}}K\left(x\right)dx\right\} \, .
\end{eqnarray*}
Given the maximizer of $l^{s}_{12}(\beta_{12})$ at step $m$, denoted by $\widehat{\beta}^{(m)}_{12}$, $h^{0}_{12}(t)$ is estimated by
\begin{equation*}
	\widehat{h}^{o(m)}_{12}\left(t;\widehat{\beta}^{(m)}_{12} \right)=\frac{\dfrac{1}{n a_{12} t}\sum^{n}_{j=1}\delta_{3j}K\left(\dfrac{\log\{W_j\}-\widehat{\beta}^{T(m)}_{12}X_{j}-\log(t)}{a_{12}}\right)}{\dfrac{1}{n}\sum^{n}_{j=1}\mathcal{E}^{(m)}_{1,j}\delta_{1j}\int_{\left(\log\{V_j\}-\widehat{\beta}^{T(m)}_{12} X_{j}-\log (t)\right)/{a_{12}}}^{\left(\log\{W_j\}-\widehat{\beta}^{T(m)}_{12} X_{j}-\log (t)\right)/{a_{12}}}K(u)du} \, ,\, t>0 \,.
\end{equation*}

\section{Variance Estimation by weighted bootstrap}
\label{ss:variance_estimation}
Let $G_{i},\,i=1,\ldots,n,$ be $n$ independent copies of a positive random variable from a known distribution with mean and variance 1. Let 
\begin{equation*}
	l^{*}_{n}(\Omega)=l^{*}_{n}(\sigma)+l^{*}_{n}(\beta_{01},h^o_{01})+l^{*}_{n}(\beta_{02},h^o_{02})+l^{*}_{n}(\beta_{12},h^o_{12})
\end{equation*}
be a weighted log-likelihood function of the complete data, where
\begin{eqnarray*}
	l^{*}_{n}(\sigma)
	&=& \frac{1}{n}\sum^{n}_{i=1}D_i \mathcal{E}_{2,i}G_i+\frac{1}{n}\sum^{n}_{i=1}\mathcal{E}_{3,i}G_i\\
	&=& \frac{1}{n}\sum^{n}_{i=1}\left(D_i+\frac{1}{\sigma}-1\right)\mathcal{E}_{2,i}G_i-\frac{1}{n\sigma}\sum^{n}_{i=1}\mathcal{E}_{1,i}G_i+\frac{1}{n}\left[\frac{1}{\sigma}\log\left\{\frac{1}{\sigma}\right\}-\log\Gamma\left(\frac{1}{\sigma}\right)\right]\sum^{n}_{i=1}G_i,\\
	l^{*}_{n}(\beta_{01},h^o_{01})
	&=&\frac{1}{n}\sum^{n}_{i=1}\left\{\delta_{1i}\left[\log h^o_{{01}}(V_i e^{-\beta_{01}^{T}X_{i}})-\beta_{01}^{T}X_{i}\right]-\mathcal{E}_{1,i} H^o_{{01}}(V_i e^{-\beta_{01}^{T}X_{i}})\right\}G_i,\\
	l^{*}_{n}(\beta_{02},h^o_{02})
	&=&\frac{1}{n}\sum^{n}_{i=1}\left\{ \delta_{2i}\left[\log h^o_{{02}}(V_i e^{-\beta_{02}^{T}X_{i}})-\beta_{02}^{T}X_{i}\right]-\mathcal{E}_{1,i} H^o_{{02}}(V_i e^{-\beta_{02}^{T}X_{i}})\right\}G_i,\\
	l^{*}_{n}(\beta_{12},h^o_{12})
	&=& \frac{1}{n}\sum^{n}_{i=1}\left\{\delta_{3i}\left[\log h^o_{{12}}(W_i e^{-\beta_{12}^{T}X_{i}})-\beta_{12}^{T}X_{i}\right] \right.\\
	&& -\delta_{1i}\mathcal{E}_{1,i}\left.\left[H^o_{12}(W_i e^{-\beta_{12}^{T}X_{i}}) - H^o_{12}(V_i e^{-\beta_{12}^{T}X_{i}})\right] \right\}G_i \, .
\end{eqnarray*}
Similarly, the weighted versions of $l^{s}_{jk}(\beta_{jk})$, $jk \in {01,02,12}$, are  given by
\begin{eqnarray}\label{eq:l_s*_0k}
	\lefteqn{l^{s*}_{0k}(\beta_{0k})} \nonumber\\
	&=& -\frac{1}{n}\sum^{n}_{i=1}\delta_{ki}V_i G_i+\frac{1}{n}\sum^{n}_{i=1}\delta_{ki}G_i\log\left\{\dfrac{1}{na_{0k}}\sum^{n}_{j=1}\delta_{kj}G_i K\left(\dfrac{\log\{V_j\}-\beta^T_{0k}X_j-\log\{V_i\}+\beta^T_{0k}X_i}{a_{0k}}\right)\right\} \nonumber\\
	&& -\frac{1}{n}\sum^{n}_{i=1}\delta_{ki}G_i\log\left\{\frac{1}{n}\sum^{n}_{j=1}\mathcal{E}_{1,j} G_i\int_{-\infty}^{\left[\log\{V_j\}-\beta^T_{0k}X_j-\log\{V_i\}+\beta^T_{0k}X_i\right]/a_{0k}}K(s)ds\right\}, \, \text{for}\, k=1,2\, ,
\end{eqnarray}
and
\begin{eqnarray}\label{eq:l_s*_12}
	\lefteqn{l^{s*}_{12}(\beta_{12})} \nonumber\\
	&=& -\frac{1}{n}\sum^{n}_{i=1}\delta_{3i}w_{i}G_i
	+\frac{1}{n}\sum^{n}_{i=1}\delta_{3i}G_i\log\left\{ \dfrac{1}{n a_{12}}\sum_{j=1}^{n}\delta_{3j}G_i K\left(\dfrac{\log\{W_j\}-\beta_{12}^T X_{j}-\log\{W_i\}+\beta_{12}^T X_{i}}{a_{12}} \right) \right\} \nonumber\\
	&& -\frac{1}{n}\sum^{n}_{i=1}\delta_{3i}G_i\log\left\{ \dfrac{1}{n}\sum_{j=1}^{n}\mathcal{E}_{1,j} \delta_{1j}G_i \int_{\left(\log\{V_j\}-\beta_{12}^T X_{j}-\log \{W_i\}+\beta_{12}^T X_{i}\right)/{a_{12}}}^{\left(\log\{W_j\}-\beta_{12}^T X_{j}-\log \{W_i\}+\beta_{12}^T X_{i}\right)/{a_{12}}}K\left(x\right)dx\right\} \,.
\end{eqnarray}
The estimators of each bootstrap sample are derived by maximizing of Eq.\, (\ref{eq:l_s*_0k}) and (\ref{eq:l_s*_12}).

Given the bootstrap estimates of the resression coefficients, the corresponding bootstrap estimators of the baseline hazard functions are given by
\begin{equation*}
	\widehat{h}^{o*}_{0k}\left(t;\widehat{\beta}_{0k} \right)=\dfrac{(n a_{0k} t)^{-1}\sum^{n}_{j=1}\delta_{kj}G_i K\left(\dfrac{\log\{V_j\}-\widehat{\beta}^{*T}_{0k}X_j-\log(t)}{a_{0k}}\right)}{(n)^{-1}\sum^{n}_{j=1}\mathcal{E}_{1,j}G_i \int^{[\log\{V_j\}-\widehat{\beta}^{*T}_{0k}X_j-\log(t)]/a_{0k}}_{-\infty}K(u)du} \, ,\, t>0 \,, \, k=1,2 \,,
\end{equation*}
and
\begin{equation*}
	\widehat{h}^{o*}_{12}\left(t;\widehat{\beta}_{12} \right)=\frac{(n a_{12} t)^{-1}\sum^{n}_{j=1}\delta_{3j}G_i K\left(\dfrac{\log\{W_j\}-\widehat{\beta}^{*T}_{12}X_{j}-\log(t)}{a_{12}}\right)}{(n)^{-1}\sum^{n}_{j=1}\mathcal{E}_{1,j} \delta_{1j}G_i\int_{\left(\log\{V_j\}-\widehat{\beta}^{*T}_{12} X_{j}-\log t\right)/{a_{12}}}^{\left(\log\{W_j\}-\widehat{\beta}^{*T}_{12} X_{j}-\log t\right)/{a_{12}}}K(u)du} \, ,\, t>0 \, .
\end{equation*}

\section{Estimation Procedure When Frailty is Not Required}\label{ss:etimation_no_frailty}
Define $R^V_i(\beta)=\log V_i -\beta^T X_{i}$, $R^W_i(\beta)=\log W_j -\beta^T X_{i}$. Then, $\beta_{0k}$ is estimated by maximization of $l^{s}_{0k}(\beta_{0k})$, $k=1,2$, where
\begin{eqnarray*}
	l^{s}_{0k}(\beta_{0k})&=&-\frac{1}{n}\sum^{n}_{i=1}\delta_{ki}\log V_i+\frac{1}{n}\sum^{n}_{i=1}\delta_{ki}\log \left\{\frac{1}{n a_{0k,n}}\sum^{n}_{j=1}\delta_{kj}K \left(\frac{R^V_j(\beta_{0k})-R^V_i(\beta_{0k})}{a_{0k,n}}\right)\right\} \nonumber \\
	&&-\frac{1}{n}\sum^{n}_{i=1}\delta_{ki}\log\left\{ \frac{1}{n}\sum^{n}_{j=1}
	\int_{-\infty}^{\{ R^V_j(\beta_{0k})-R^V_i(\beta_{0k})\}/a_{0k}}K(s)ds\right\} \, ,
\end{eqnarray*}
$\beta_{12}$ is estimated by maximization of 
\begin{eqnarray*}
	l^{s}_{12}(\beta_{12})&=&-\frac{1}{n}\sum^{n}_{i=1}\delta_{3i}\log W_i +
	\frac{1}{n}\sum^{n}_{i=1}\delta_{3i}\log\left\{\frac{1}{n a_{12,n} }\sum^{n}_{j=1}\delta_{3j}K\left(\dfrac{R^W_j(\beta_{12})-R^W_i(\beta_{12})}{a_{12,n}} \right)\right\} \nonumber \\
	&&-\frac{1}{n}\sum^{n}_{i=1}\delta_{3i}\log\left\{\frac{1}{n }\sum^{n}_{j=1}\delta_{1j}\int_{\{R^V_j(\beta_{12})-R^W_i(\beta_{12})\}/a_{12}}^{\{R^W_j(\beta_{12})-R^W_i(\beta_{12})\}/a_{12}}
	K\left(s\right)ds\right\} \, ,
\end{eqnarray*}
and given $\widehat{\beta}^{(m)}_{jk}$, $jk \in \{01,02,12\}$, the baseline hazard functions are estimated by
\begin{equation*}
	\widehat{h}^{o(m)}_{0k}(t)=\frac{(n a_{0k,n} t)^{-1}\sum^{n}_{i=1} \delta_{ki}K\left(\{R^V_i(\widehat{\beta}^{(m)}_{0k})-\log t\}/
		{a_{0k,n}}\right)}{n^{-1}\sum^{n}_{i=1}\int^{\{R^V_i(\widehat{\beta}^{(m)}_{0k})-\log t\}/a_{0k,n}}_{-\infty}K(s)ds} \,\,\, 
	k=1,2 \, ,
\end{equation*}
\begin{equation*}
	\widehat{h}^{o(m)}_{12}(t)=\frac{(n a_{12,n} t)^{-1}
		\sum^{n}_{i=1}\delta_{3i}K\left(\{R^W_i(\widehat{\beta}^{(m)}_{12})-\log t\}/{a_{12,n}}\right)}
	{n^{-1}\sum^{n}_{i=1}\delta_{1i}\int_{\{R^V_j(\widehat{\beta}^{(m)}_{12})-\log t\}/{a_{12,n}}}^{\{R^W_i(\widehat{\beta}_{12})-\log t \}/{a_{12,n}}}K(s)ds} \, ,
\end{equation*}
and $\widehat{H}^{o(m)}_{jk}(t) = \int_0^t \widehat{h}^{o(m)}_{12}(s)ds$. No recursive algorithm is required.

\section{Simulation Results Without Frailty}
\label{ss:simulation_no_frailty}
To demonstrate the performance of the estimation method without frailty,  data were generated similar to Section 4 but with $\mathcal{E}_{1i}\equiv 1$, $i=1,\ldots,1000$. The regression coefficients and the hazard functions were estimated based on the procedure of Sections S5 above. In this simulation about 17\% of the observations were censored at healthy state, and about 19\% of those who diagnosed with the disease were censored before death. The results are summarized in Tables \ref{tab2}. As expected, the proposed estimators performs well in terms of bias and coverage rates.

\section{Additional Simulations Results}
\label{ss:additional_simulations}
Table \ref{tab_add_sim} presents simulation results for several baseline hazard functions with different values of $\zeta$ in the bandwidths used in the estimation of the regression coefficients.

\section{Marginal Survival Functions under the AFT Model}
\label{ss:marginal_surv_aft}
We begin with the probability
$S^M_{0.}(t|X_i)=Pr(T_{1i}>t,T_{2i}>t|X_i)$. 
\begin{eqnarray*}
	S^M_{0.}(t|X_i)
	&=&Pr(T_{1i}>t,T_{2i}>t|X_i)\\
	&=&\int_{0}^{\infty}Pr(T_{1i}>t,T_{2i}>t|X_i,\gamma_i)f(\gamma_i|X_i)d\gamma_i\\
	&=&\int_{0}^{\infty}Pr(T_{1i}>t,T_{2i}>t|X_i,\gamma_i)f(\gamma_i)d\gamma_i\\
	&=&\int_{0}^{\infty}Pr(T_{1i}>t|X_i,\gamma_i)Pr(T_{2i}>t|X_i,\gamma_i)f(\gamma_i)d\gamma_i\\
	&=&\int_{0}^{\infty}\exp\{-\gamma_i [H_{01}^o(t e^{-\beta^T_{01}X_i})+H_{02}^o(t e^{-\beta^T_{02}X_i})]\}f(\gamma_i)d\gamma_i\\
	&=&\left\{1+\sigma [H^o_{01}(t e^{-\beta^T_{01}X_i})+H_{02}^o(t e^{-\beta^T_{02}X_i})]\right\}^{-1/\sigma} \, .
\end{eqnarray*}
The marginal survival function $S^M_{12}(t|t_1,X_i)=Pr(T_{2i}>t|T_{1i}=t_1,T_{2i}>t_1,X_i)$ for $t>t_1>0$ equals
\begin{equation*}
	S^M_{12}(t|t_1,X_i)=
	\int_{0}^{\infty}\Pr(T_{2i}>t|T_{1i}=t_1,T_{2i}>t_1,X_i,\gamma_i)f(\gamma_i|T_{1i}=t_1,T_{2i}>t_1,X_i)d\gamma_i \, .
\end{equation*}
For the conditional density of $\gamma_i$, we write
\begin{equation*}
	f(\gamma_i|T_{1i}=t_1,T_{2i}>t_1,X_i)=\dfrac{\lim_{\Delta\to 0}\frac{1}{\Delta}\Pr(T_{1i}\in [t_1,t_1+\Delta),T_{2i}>t_1|X_i,\gamma_i)f(\gamma_i|X_i)}{\lim_{\Delta\to 0}\frac{1}{\Delta}\Pr(T_{1i}\in [t_1,t_1+\Delta),T_{2i}>t_1|X_i)}.
\end{equation*}
where $f(\gamma_i|X_i)=f(\gamma_i)$ and under the gamma frailty model
\begin{eqnarray*}
	\lefteqn{\lim_{\Delta\to 0}\dfrac{1}{\Delta}\Pr(T_{1i}\in [t_1,t_1\Delta),T_{2i}>t_1|X_i,\gamma_i)=}\\
	&=&\Pr(T_{1i}>t_1,T_{2i}>t_1|X_i,\gamma_i)\lambda_{01}(t_1|X_i,\gamma_i)\\
	&=&\exp\{-\gamma_i [H_{01}^o(t_1 e^{-\beta^T_{01}X_i})+H_{02}^o(t_1 e^{-\beta^T_{02}X_i})]\}\gamma_i h_{01}^o(t_1 e^{-\beta^T_{01}X_i}) e^{-\beta^T_{01}X_i} \, .
\end{eqnarray*}
Therefore, 
\begin{eqnarray*}
	\lefteqn{\lim_{\Delta\to 0}\dfrac{1}{\Delta}\Pr(T_{1i}\in [t_1,t_1+\Delta),T_{2i}>t_1|X_i)=}\\
	&=& \int_{0}^{\infty}\lim_{\Delta\to 0}\frac{1}{\Delta}\Pr(T_{1i}\in [t_1,t_1+\Delta),T_{2i}>t_1|X_i,\gamma_i)f(\gamma_i|X_i)d\gamma_i\\
	&=& \int_{0}^{\infty}\exp\{-\gamma_i [H_{01}^o(t_1 e^{-\beta^T_{01}X_i})+H_{02}^o(t_1 e^{-\beta^T_{02}X_i})]\}\gamma_i h_{01}^o(t_1 e^{-\beta^T_{01}X_i}) e^{-\beta^T_{01}X_i}f(\gamma_i)d\gamma_i\\
	&=& h_{01}^o(t_1 e^{-\beta^T_{01}X_i}) e^{-\beta^T_{01}X_i}\int_{0}^{\infty}\exp\{-\gamma_i [H_{01}^o(t_1 e^{-\beta^T_{01}X_i})+H_{02}^o(t_1 e^{-\beta^T_{02}X_i})]\}\gamma_i f(\gamma_i)d\gamma_i\\
	&=& h_{01}^o(t_1 e^{-\beta^T_{01}X_i}) e^{-\beta^T_{01}X_i}(1+\sigma){\left\{1+\sigma[H_{01}^o(t_1 e^{-\beta^T_{01}X_i})+H_{02}^o(t_1 e^{-\beta^T_{02}X_i})]\right\}^{-1/\sigma+1}} \, .
\end{eqnarray*}
By combining the above expressions, we get
\begin{eqnarray*}
	\lefteqn{S^M_{12}(t|t_1,X_i)=
		Pr(T_{2i}>t|T_{1i}=t_1,T_{2i}>t_1,X_i)=}\\
	&=& \int_{0}^{\infty}Pr(T_{2i}>t|T_{1i}=t_1,T_{2i}>t_1,X_i,\gamma_i)f(\gamma_i|T_{1i}=t_1,T_{2i}>t_1,X_i)d\gamma_i\\
	&=& \int_{0}^{\infty}\frac{\exp\left\{-\gamma_i H^o_{12}\left(t e^{-\beta_{12}^{T}X_i}\right)\right\}}{\exp\left\{-\gamma_i H^o_{12}\left(t_1 e^{-\beta_{12}^{T}X_i}\right)\right\}}\dfrac{\lim_{\Delta\to 0}\frac{1}{\Delta}Pr(T_{1i}\in [t_1,t_1+\Delta),T_{2i}>t_1|X_i,\gamma_i)f(\gamma_i|X_i)}{\lim_{\Delta\to 0}\dfrac{1}{\Delta}Pr(T_{1i}\in [t_1,t_1+\Delta),T_{2i}>t_1|X_i)}d\gamma_i\\
	&=& \left\{\dfrac{1+\sigma[H_{01}^o(t_1 e^{-\beta^T_{01}X_i})+H_{02}^o(t_1 e^{-\beta^T_{02}X_i})]}{1+\sigma[H_{01}^o(t_1 e^{-\beta^T_{01}X_i})+H_{02}^o(t_1 e^{-\beta^T_{02}X_i})+H^o_{12}\left(t e^{-\beta_{12}^{T}X_i}\right)- H^o_{12}\left(t_1 e^{-\beta_{12}^{T}X_i}\right)]} \right\}^{1/\sigma+1} \, .
\end{eqnarray*}

\section{Marginal Survival under the Conditional Cox Model}
\label{ss:marginal_surv_cox}
\begin{eqnarray*}
	S^{M,Cox}_{0.}(t|X_i)&=&\Pr(T_{1i}>t,T_{2i}>t|X_i)\\
	&=& \int_{0}^{\infty}\Pr(T_{1i}>t,T_{2i}>t|X_i)f(\gamma_i|X_i,\gamma_i)d\gamma_i\\
	&=& \int_{0}^{\infty}\exp\left\{\gamma_i\left[H^o_{01}(t)e^{\beta^T_{01}X_i}+H^o_{02}(t)e^{\beta^T_{02}X_i}\right]\right\}f(\gamma_i)d\gamma_i\\
	&=& \left\{1+\sigma\left[H^o_{01}(t)e^{\beta^T_{01}X_i}+H^o_{02}(t)e^{\beta^T_{02}X_i}\right]\right\}^{-1/\sigma} \, .
\end{eqnarray*}
and
\begin{eqnarray*}
	S^{M,Cox}_{12}(t|t_1,X_i)
	&=& Pr(T_{2i}>t|T_{1i}=t_1,T_{2i}>t_1,X_i)\\
	&=& \left\{\dfrac{1+\sigma[H_{01}^o(t_1)e^{\beta^T_{01}X_i}+H_{02}^o(t_1) e^{\beta^T_{02}X_i}]}{1+\sigma[H_{01}^o(t_1)e^{\beta^T_{01}X_i}+H_{02}^o(t_1)e^{\beta^T_{02}X_i}+H^o_{12}\left(t\right) e^{\beta_{12}^{T}X_i}- H^o_{12}\left(t_1\right) e^{\beta_{12}^{T}X_i}]} \right\}^{1/\sigma+1}.
\end{eqnarray*}

\section{Analysis of the Rotterdam Breast Cancer Data Without Frailty}
\label{ss:breats_results_no_frailty}
\begin{figure}%
\centering
\subfloat[][\centering The proposed AFT without frailty]{\includegraphics[scale=0.47]{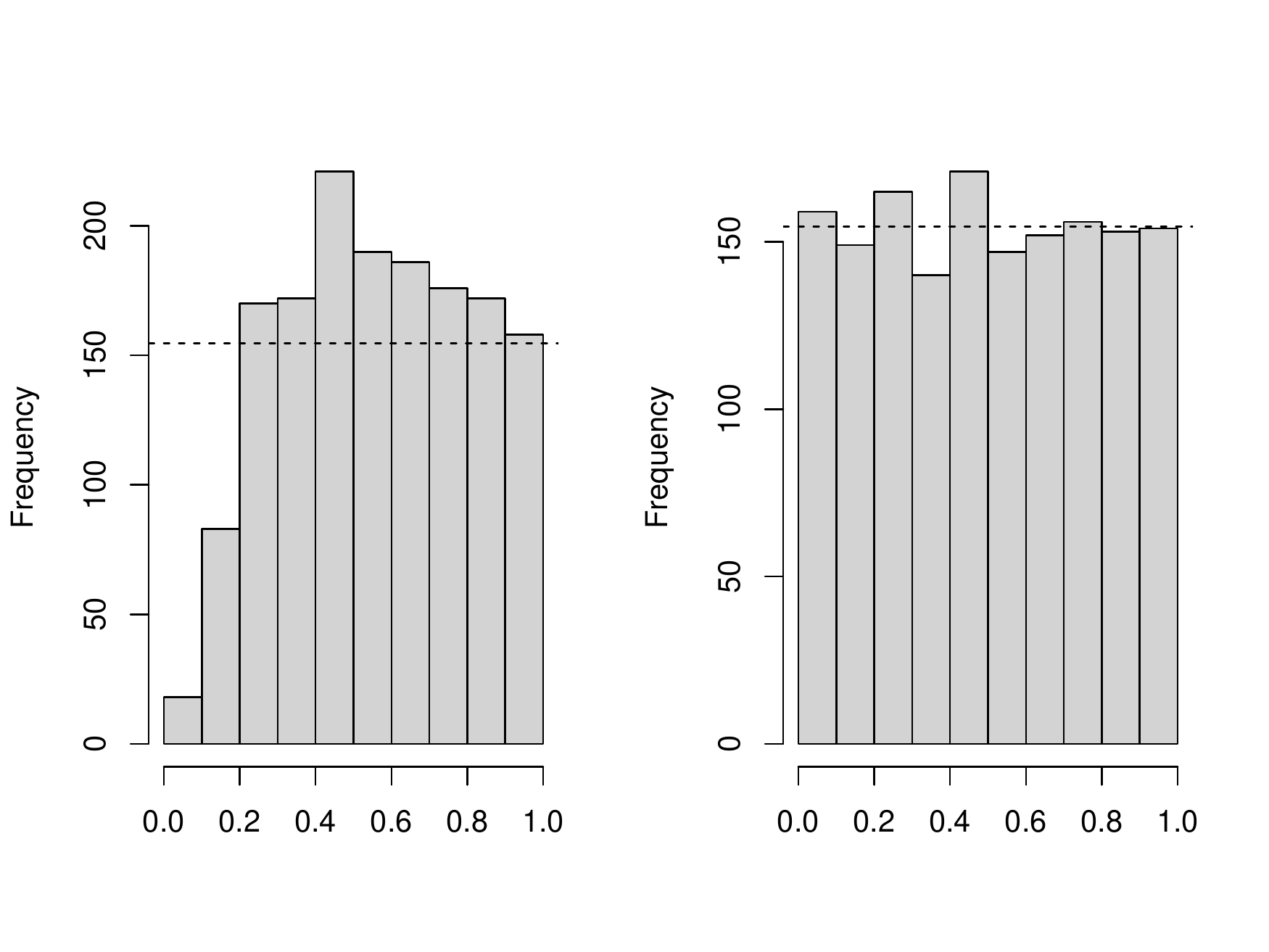}}
	\qquad	
\subfloat[][\centering The proposed AFT without frailty]{\includegraphics[scale=0.47]{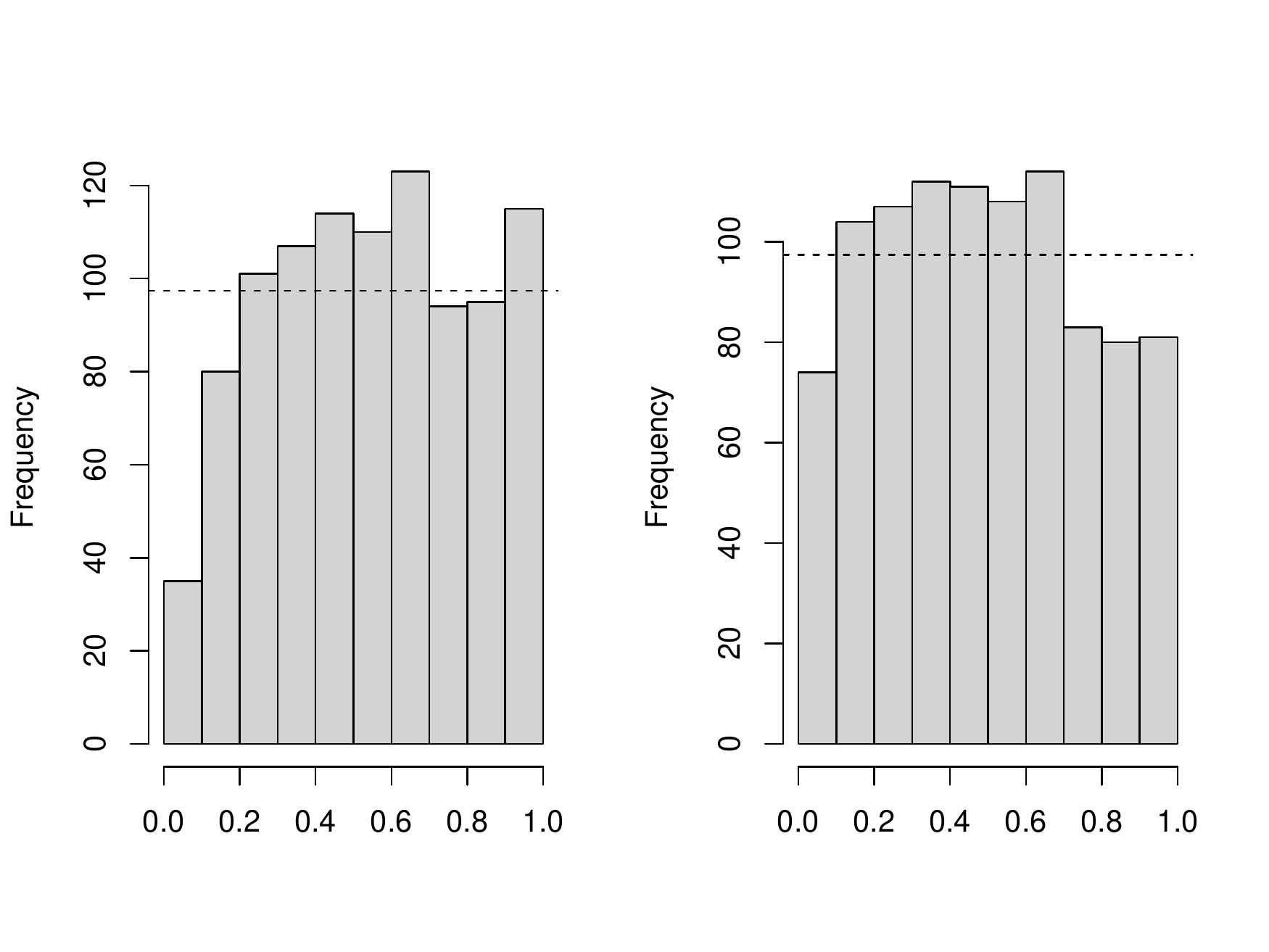}}
	\qquad
\subfloat[][\centering Cox model without frailty]{\includegraphics[scale=0.47]{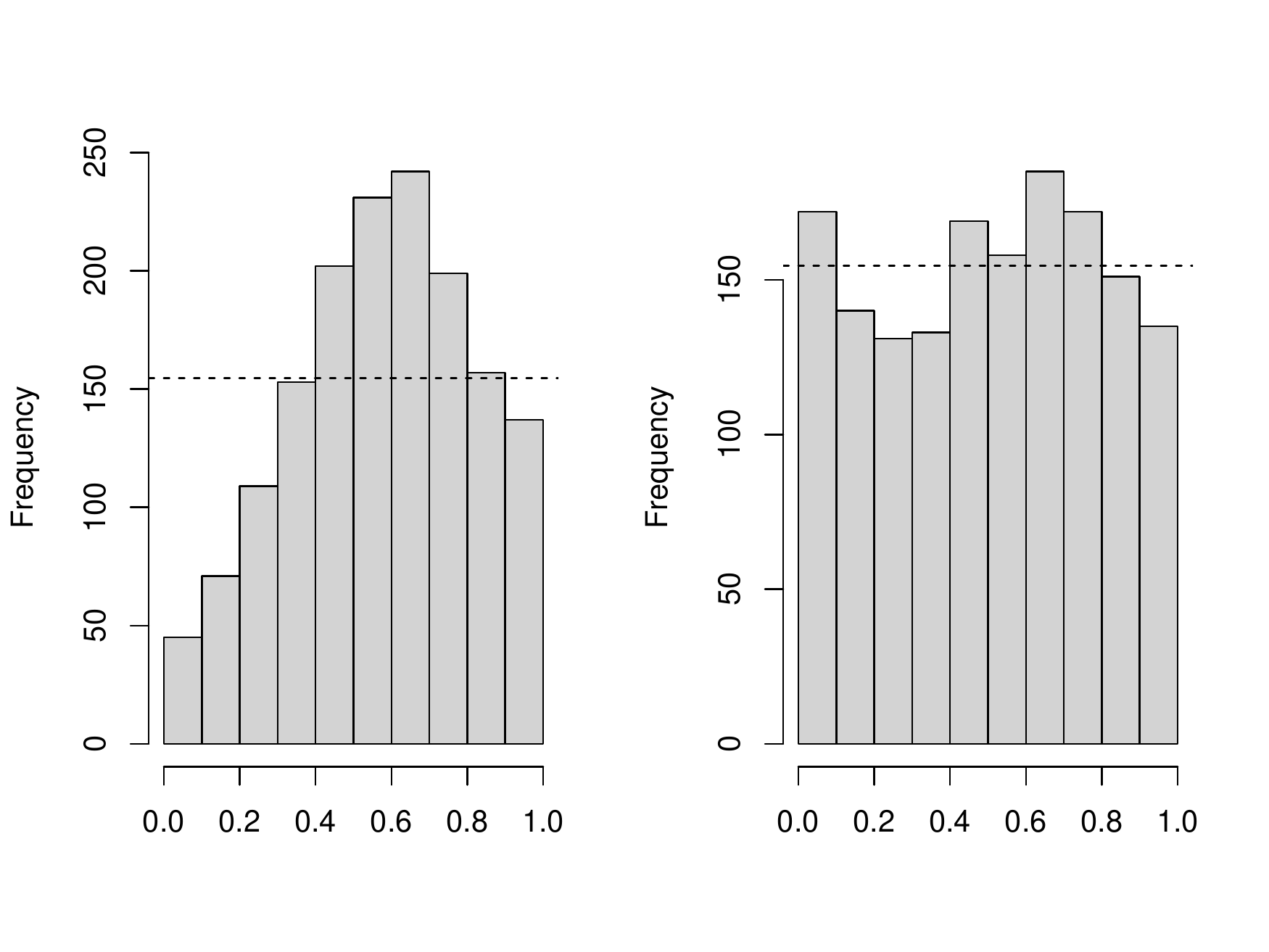}}
	\qquad
\subfloat[][\centering Cox model without frailty]{\includegraphics[scale=0.47]{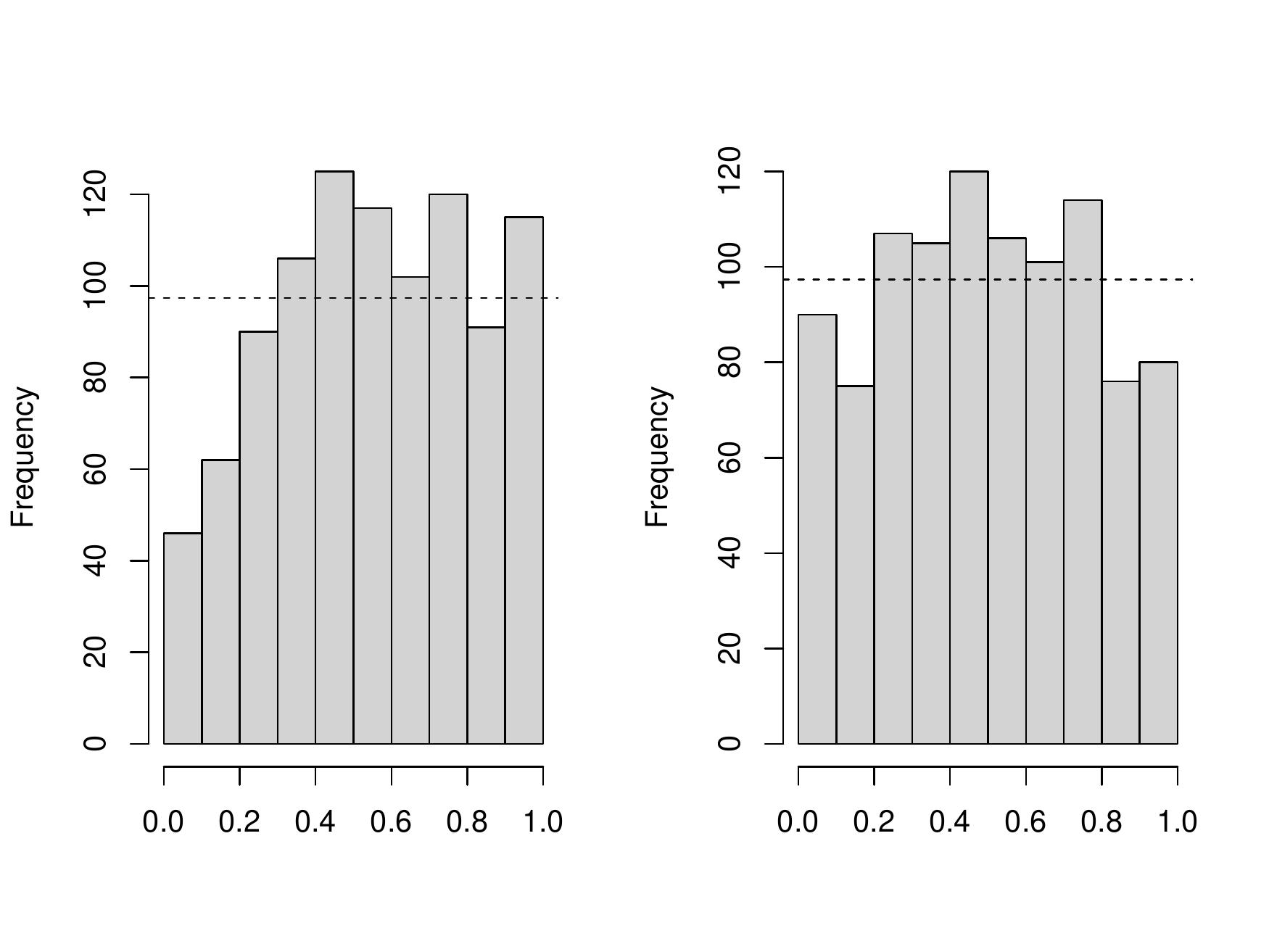}}
	
%
	\caption{Goodness of fit plots for the illness-death models. Histograms of $\widehat{S}_{0.}^M(V_i|X_i)$ (left of (a), (c)) $\widehat{S}_{0.}^M(V_i,\delta_{1i},\delta_{2i},U_{1i})$ (right of (a), (c)), $\widehat{S}_{12}^M(W_i|V_i,X_i)$ (left of (b), (d)) and $\widehat{S}_{12}^M(W_i,V_i,\delta_{3i},U_{2i})$ (right of (b), (d)). 
		The dashed lines are the expected values under the uniform distribution.}%
	\label{fig:fig_aft_res12}%
\end{figure}

\begin{figure}
	\begin{center}
		\includegraphics[scale=0.85]{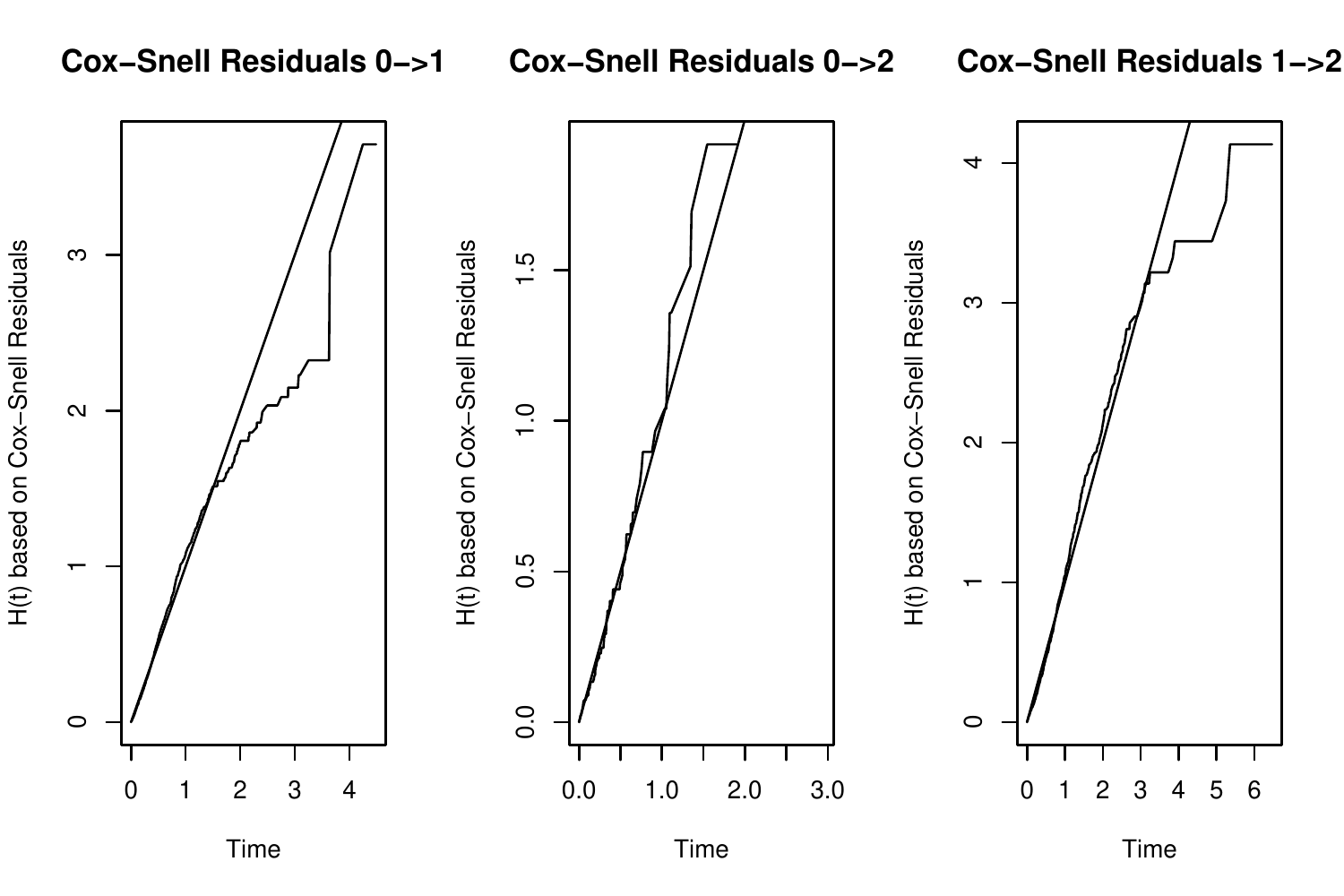}
		\caption{\label{fig:fig_coxsnell} Cox-Snell residuals of each transition, Cox model\label{RSP_cox_cs}}
	\end{center}
\end{figure}

The analysis of the Rotterdam breast cancer data without frailty  is presented in Table \ref{reg_no_frailty}. 
Additionally, the RSPs of the proposed AFT model and the RSPs of the Cox model are presented in Figures \ref{fig:fig_aft_res12}-\ref{fig:fig_coxsnell}. 

\begin{table}
	\centering
	\caption{Simulation results without frailty}\label{tab2}
	\scalebox{1}{
		\begin{tabular}{lccccccc}
		\multicolumn{8}{c}{(a) The regression coefficients.}\\
		&\\
		[-0.75em]
			\midrule
			& $\widehat{\beta}_{01,1}$ & $\widehat{\beta}_{01,2}$ & $\widehat{\beta}_{02,1}$ & $\widehat{\beta}_{02,2}$ & $\widehat{\beta}_{12,1}$ & $\widehat{\beta}_{12,2}$ & $\widehat{\beta}_{12,3}$  \\
			\midrule
			True values  & 1     & 0.5   & 1     & 1     & 0.5   & 0.5   & 1 \\
			\midrule
			mean         & 1.01  & 0.50  & 1.01  & 1.01  & 0.49  & 0.49  & 1.03 \\
			empirical SD & 0.06  & 0.06  & 0.06  & 0.05  & 0.08  & 0.07  & 0.07 \\
			CR           & 0.95  & 0.97  & 0.93  & 0.92  & 0.93  & 0.96  & 0.90 \\
			\bottomrule
	\end{tabular}}
	\bigskip
	\centering
	\scalebox{1}{
		\begin{tabular}{lcccccccccc}
		&\\
		\multicolumn{11}{c}{(b) The cumulative baseline hazard functions.}\\
		&\\
		[-0.75em]
			\midrule
			$t$             & 0.10  & 0.20  & 0.30  & 0.40  & 0.50  & 0.60  & 0.70  & 0.80  & 0.90  & 1.00 \\
			\midrule
			$H^{o}_{01}(t)$ & 0.01  & 0.04  & 0.09  & 0.16  & 0.25  & 0.36  & 0.49  & 0.64  & 0.81  & 1.00 \\
			mean            & 0.01  & 0.04  & 0.09  & 0.16  & 0.25  & 0.36  & 0.49  & 0.64  & 0.81  & 1.00 \\
			empirical SD    & 0.00  & 0.01  & 0.01  & 0.02  & 0.02  & 0.03  & 0.04  & 0.06  & 0.08  & 0.10 \\
			CR              & 0.95  & 0.96  & 0.96  & 0.96  & 0.95  & 0.97  & 0.96  & 0.95  & 0.96  & 0.96 \\
			\midrule
			$H^{o}_{02}(t)$ & 0.02  & 0.06  & 0.14  & 0.24  & 0.38  & 0.54  & 0.74  & 0.96  & 1.22  & 1.50 \\
			mean            & 0.02  & 0.06  & 0.13  & 0.24  & 0.37  & 0.54  & 0.73  & 0.96  & 1.22  & 1.50 \\
			empirical SD    & 0.00  & 0.01  & 0.02  & 0.02  & 0.04  & 0.04  & 0.06  & 0.07  & 0.10  & 0.14 \\
			CR              & 0.94  & 0.97  & 0.95  & 0.95  & 0.97  & 0.94  & 0.95  & 0.97  & 0.96  & 0.96 \\
			\midrule
			$H^{o}_{12}(t)$ & 0.01  & 0.04  & 0.09  & 0.16  & 0.25  & 0.36  & 0.49  & 0.64  & 0.81  & 1.00 \\
			mean            & 0.01  & 0.04  & 0.09  & 0.16  & 0.26  & 0.37  & 0.50  & 0.65  & 0.83  & 1.01 \\
			empirical SD    & 0.03  & 0.04  & 0.05  & 0.06  & 0.06  & 0.07  & 0.08  & 0.09  & 0.10  & 0.13 \\
			CR              & 0.88  & 0.92  & 0.95  & 0.96  & 0.95  & 0.94  & 0.94  & 0.94  & 0.94  & 0.95 \\
			\bottomrule
	\end{tabular}}
\end{table}

\begin{table}
	\small
	\centering
	\caption{Simulation results with frailty, various bandwidths $\zeta$ and various baseline hazard functions. \label{tab_add_sim}}
	\scalebox{1}{
		\begin{tabular}{lllllllllll}
			\midrule
			\multicolumn{11}{c}{$h^o_{01}(t)=2$, $h^o_{02}(t)=3$, $h^o_{12}(t)=2$} \\
			\midrule
			$\sigma$ &$\zeta$&            & $\widehat{\sigma}$ & $\widehat{\beta}_{01,1}$ & $\widehat{\beta}_{01,2}$ & $\widehat{\beta}_{02,1}$ & $\widehat{\beta}_{02,2}$ & $\widehat{\beta}_{12,1}$ & $\widehat{\beta}_{12,2}$ & $\widehat{\beta}_{12,3}$ \\
			\midrule
			&       & True values  &      & 1.0  & 0.5  & 1.0  & 1.0  & 0.5  & 0.5  & 1.0 \\
			\midrule
			2    & 35    & mean         & 1.98 & 0.98 & 0.45 & 0.97 & 1.00 & 0.46 & 0.41 & 0.92 \\
			&       & empirical SD & 0.34 & 0.18 & 0.27 & 0.28 & 0.18 & 0.25 & 0.30 & 0.21 \\
			&       & CR           & 0.96 & 0.95 & 0.95 & 0.96 & 0.97 & 0.94 & 0.90 & 0.94 \\
			\cmidrule{2-11}
			& 50    & mean         & 1.91 & 1.04 & 0.51 & 1.03 & 1.03 & 0.53 & 0.51 & 1.04 \\
			&       & empirical SD & 0.33 & 0.17 & 0.27 & 0.26 & 0.17 & 0.21 & 0.25 & 0.22 \\
			&       & CR           & 0.94 & 0.94 & 0.93 & 0.96 & 0.96 & 0.97 & 0.92 & 0.96 \\
			\cmidrule{2-11}
			& 75    & mean         & 1.88 & 1.06 & 0.51 & 1.05 & 1.05 & 0.54 & 0.54 & 1.08 \\
			&       & empirical SD & 0.30 & 0.15 & 0.24 & 0.23 & 0.15 & 0.20 & 0.22 & 0.17 \\
			&       & CR           & 0.93 & 0.91 & 0.91 & 0.95 & 0.95 & 0.94 & 0.90 & 0.93 \\
			\cmidrule{2-11}
			& 100   & mean         & 1.81 & 1.10 & 0.52 & 1.08 & 1.08 & 0.56 & 0.54 & 1.14 \\
			&       & empirical SD & 0.30 & 0.15 & 0.23 & 0.23 & 0.14 & 0.20 & 0.23 & 0.16 \\
			&       & CR           & 0.91 & 0.89 & 0.93 & 0.95 & 0.94 & 0.92 & 0.93 & 0.87 \\
			\midrule
			\multicolumn{11}{c}{$h^o_{jk}(t)=a_{jk}/(1+t)$, $a_{01}=2$, $a_{02}=3$, $a_{12}=2$} \\
			\midrule
			2  & 35    & mean         & 2.05 & 0.93 & 0.48 & 0.96 & 0.96 & 0.37 & 0.40 & 0.73 \\
			&       & empirical SD & 0.40 & 0.24 & 0.33 & 0.29 & 0.20 & 0.33 & 0.42 & 0.35 \\
			&       & CR           & 0.98 & 0.92 & 0.95 & 0.95 & 0.96 & 0.93 & 0.94 & 0.86 \\
			\cmidrule{2-11}
			& 50    & mean         & 1.95 & 1.01 & 0.50 & 1.01 & 1.01 & 0.48 & 0.51 & 0.94 \\
			&       & empirical SD & 0.40 & 0.20 & 0.29 & 0.27 & 0.19 & 0.34 & 0.40 & 0.36 \\
			&       & CR           & 0.96 & 0.93 & 0.94 & 0.96 & 0.95 & 0.93 & 0.95 & 0.96 \\
			\cmidrule{2-11}
			& 75    & mean         & 1.92 & 1.04 & 0.48 & 1.01 & 1.02 & 0.52 & 0.50 & 1.09 \\
			&       & empirical SD & 0.37 & 0.18 & 0.26 & 0.24 & 0.17 & 0.32 & 0.38 & 0.31 \\
			&       & CR   		& 0.95 & 0.93 & 0.95 & 0.97 & 0.93 & 0.92 & 0.93 & 0.94 \\
			\cmidrule{2-11}
			& 100   & mean         & 1.90 & 1.05 & 0.48 & 1.02 & 1.04 & 0.53 & 0.49 & 1.13 \\
			&       & empirical SD & 0.36 & 0.16 & 0.25 & 0.24 & 0.16 & 0.29 & 0.36 & 0.28 \\
			&       & CR           & 0.96 & 0.94 & 0.94 & 0.97 & 0.94 & 0.93 & 0.93 & 0.92 \\
			\midrule
			\multicolumn{11}{c} {$h^o_{jk}(t)=t^{-1}\phi\{\log(t)\}/[1-\Phi\{\log(t)\}]$, $jk \in {01,02,12}$}\\
			\midrule
			&       & True values  &      & 1.0  & 0.5  & 1.0  & 1.0  & 0.5  & 0.5  & 1.0 \\
			\midrule
			2  & 35    & mean         & 2.02 & 0.98 & 0.47 & 0.95 & 0.96 & 0.44 & 0.38 & 0.82 \\
			&       & empirical SD & 0.35 & 0.14 & 0.18 & 0.19 & 0.15 & 0.24 & 0.25 & 0.26 \\
			&       & CR           & 0.96 & 0.97 & 0.95 & 0.95 & 0.95 & 0.95 & 0.91 & 0.88 \\
			\cmidrule{2-11}
			& 50    & mean         & 1.90 & 1.02 & 0.50 & 1.00 & 1.01 & 0.46 & 0.46 & 0.98 \\
			&       & empirical SD & 0.32 & 0.12 & 0.16 & 0.18 & 0.14 & 0.26 & 0.25 & 0.30 \\
			&       & CR           & 0.94 & 0.98 & 0.95 & 0.96 & 0.95 & 0.96 & 0.95 & 0.94 \\
			\cmidrule{2-11}
			& 75    & mean         & 1.86 & 1.03 & 0.49 & 1.01 & 1.02 & 0.50 & 0.52 & 1.13 \\
			&       & empirical SD & 0.33 & 0.10 & 0.14 & 0.16 & 0.12 & 0.27 & 0.27 & 0.32 \\
			&       & CR           & 0.90 & 0.94 & 0.96 & 0.96 & 0.92 & 0.97 & 0.94 & 0.93 \\
			\cmidrule{2-11}
			& 100   & mean         & 1.84 & 1.04 & 0.49 & 1.01 & 1.03 & 0.50 & 0.50 & 1.18 \\
			&       & empirical SD & 0.32 & 0.10 & 0.12 & 0.15 & 0.11 & 0.22 & 0.25 & 0.31 \\
			&       & CR           & 0.92 & 0.91 & 0.96 & 0.95 & 0.92 & 0.97 & 0.96 & 0.93 \\
			\bottomrule
	\end{tabular}}
\end{table}

\begin{table}
	\centering
	\caption{Rotterdam Tumor Bank Data Analysis: AFT model with our proposed estimation procedure without frailty and Cox illness-death model without frailty. Adjusted p-values in bold are less than 0.05.\newline{}  }
	\scalebox{0.75}{
		\begin{tabular}{lcccccccccc}
			\midrule
			& \multicolumn{5}{c}{Proposed Model ($\zeta=65$)} 
			& \multicolumn{5}{c}{Cox Model} \\
			&Est &exp &SE &P-value &Holm
			&Est &exp &SE &P-value &Holm \\
			\midrule
			\textbf{Transition: surgery $\rightarrow$ relapse} \\
			\midrule
			Age at surgery*	& 0.13  & 1.14 & 0.05 & 0.016 & 0.052          
			& -0.16 & 0.85 & 0.04 & 0.000 & \textbf{0.002} \\
			log of lymph nodes	& -0.41 & 0.67 & 0.06 & 0.000 & \textbf{0.000} 
			& 0.43  & 1.54 & 0.04 & 0.000 & \textbf{0.000} \\
			log of estrogen+1	& 0.08  & 1.08 & 0.03 & 0.024 & 0.052          
			& -0.04 & 0.96 & 0.02 & 0.027 & 0.082          \\
			log of progesterone+1	& 0.08  & 1.08 & 0.03 & 0.007 & \textbf{0.035} 
			& -0.02 & 0.98 & 0.02 & 0.206 & 0.260          \\
			Postmenopausal (vs. premenopausal)	& -0.30 & 0.74 & 0.16 & 0.059 & 0.059          
			& 0.18  & 1.19 & 0.12 & 0.130 & 0.260          \\
			Tumor size (vs. $<20$mm)\\
			\,\,\,\,\, 20-50mm 	& -0.30 & 0.74 & 0.10 & 0.004 & \textbf{0.021} 
			& 0.21  & 1.24 & 0.08 & 0.006 & \textbf{0.026} \\
			\,\,\,\,\, $>50$mm 	& -0.48 & 0.62 & 0.12 & 0.000 & \textbf{0.000} 
			& 0.43  & 1.54 & 0.10 & 0.000 & \textbf{0.000} \\
			Hormone therapy	& 0.56  & 1.76 & 0.13 & 0.000 & \textbf{0.000} 
			& -0.42 & 0.66 & 0.09 & 0.000 & \textbf{0.000} \\
			Chemotherapy		& 0.47  & 1.60 & 0.10 & 0.000 & \textbf{0.000} 
			& -0.47 & 0.63 & 0.09 & 0.000 & \textbf{0.000} \\
			Tumor grade 3 (vs. 2)	& -0.25 & 0.78 & 0.10 & 0.013 & 0.052          
			& 0.24  & 1.27 & 0.08 & 0.003 & \textbf{0.016}\\
			
			\midrule
			\textbf{Transition: surgery $\rightarrow$ death}\\
			\midrule
			Age at surgery*	& -0.98 & 0.37 & 0.13 & 0.000 & \textbf{0.000} 
			& 1.35  & 3.88 & 0.14 & 0.000 & \textbf{0.000} \\
			log of lymph nodes	& -0.06 & 0.94 & 0.11 & 0.587 & 1.000          
			& 0.14  & 1.14 & 0.12 & 0.267 & 1.000          \\
			log of estrogen+1	& 0.01  & 1.01 & 0.05 & 0.919 & 1.000          
			& -0.05 & 0.95 & 0.06 & 0.436 & 1.000          \\
			log of progesterone+1	& -0.02 & 0.98 & 0.05 & 0.731 & 1.000          
			& 0.11  & 1.11 & 0.06 & 0.076 & 0.685          \\
			Postmenopausal (vs. premenopausal)	& 0.10  & 1.10 & 0.46 & 0.838 & 1.000          
			& -0.31 & 0.73 & 0.63 & 0.624 & 1.000          \\
			Tumor size (ref. $<20$mm)\\
			\,\,\,\,\, 20-50mm 	& 0.18  & 1.20 & 0.17 & 0.305 & 1.000          
			& -0.10 & 0.91 & 0.24 & 0.682 & 1.000          \\
			\,\,\,\,\, $>50$mm 	& -0.18 & 0.83 & 0.21 & 0.381 & 1.000          
			& 0.17  & 1.19 & 0.30 & 0.557 & 1.000          \\
			Hormone therapy	& 0.28  & 1.32 & 0.21 & 0.179 & 1.000          
			& -0.27 & 0.76 & 0.24 & 0.262 & 1.000          \\
			Chemotherapy		& 0.06  & 1.06 & 0.37 & 0.881 & 1.000          
			& -0.20 & 0.82 & 0.55 & 0.714 & 1.000          \\
			Tumor grade 3 (vs. 2)	& 0.04  & 1.04 & 0.23 & 0.866 & 1.000          
			& 0.01  & 1.01 & 0.23 & 0.978 & 1.000         \\
			\midrule
			\textbf{Transition: relapse $\rightarrow$ death} \\
			\midrule
			Age at surgery*	& -0.10 & 0.90 & 0.06 & 0.081 & 0.407          
			& 0.12  & 1.13 & 0.05 & 0.017 & 0.137          \\
			Years from surgery to relapse*	& 1.08  & 2.94 & 0.26 & 0.000 & \textbf{0.000} 
			& -1.15 & 0.32 & 0.27 & 0.000 & \textbf{0.000} \\
			log of lymph nodes		& 0.04  & 1.04 & 0.06 & 0.527 & 0.854          
			& 0.08  & 1.08 & 0.04 & 0.091 & 0.548          \\
			log of estrogen+1		& 0.14  & 1.16 & 0.04 & 0.000 & \textbf{0.002} 
			& -0.01 & 0.99 & 0.02 & 0.529 & 1.000          \\
			log of progesterone+1	& 0.10  & 1.11 & 0.04 & 0.020 & 0.140          
			& -0.12 & 0.89 & 0.02 & 0.000 & \textbf{0.000} \\
			Postmenopausal (vs. premenopausal)	& -0.13 & 0.88 & 0.17 & 0.427 & 0.854          
			& -0.15 & 0.86 & 0.14 & 0.292 & 0.876          \\
			Tumor size (ref. $<20$mm)	\\
			\,\,\,\,\, 20-50mm 	& -0.24 & 0.79 & 0.11 & 0.029 & 0.176          
			& 0.24  & 1.27 & 0.10 & 0.012 & 0.111          \\
			\,\,\,\,\,  $>50$mm 		& -0.22 & 0.80 & 0.13 & 0.099 & 0.407          
			& 0.26  & 1.29 & 0.12 & 0.026 & 0.185          \\
			Hormone therapy	& 0.34  & 1.40 & 0.11 & 0.002 & \textbf{0.016} 
			& 0.04  & 1.04 & 0.10 & 0.672 & 1.000          \\
			Chemotherapy		& -0.17 & 0.85 & 0.15 & 0.267 & 0.800          
			& 0.17  & 1.18 & 0.11 & 0.129 & 0.644          \\
			Tumor grade 3 (vs. 2)	& -0.36 & 0.70 & 0.08 & 0.000 & \textbf{0.000} 
			& 0.12  & 1.13 & 0.09 & 0.200 & 0.800 \\ 
			\midrule
			\multicolumn{11}{l}{*devided by 10} \\
			\bottomrule
	\end{tabular}}
	\label{reg_no_frailty}
\end{table}

\newpage
\end{document}